\documentclass[times]{cnmauth}
\usepackage{bm}
\usepackage[dvips,colorlinks,bookmarksopen,bookmarksnumbered,citecolor=red,urlcolor=red]{hyperref}
\usepackage{moreverb}
\usepackage{paralist}
\usepackage{psfrag}

\def\volumeyear{2012}
\def\volumenumber{28}
\def\startpage{317}
\def\endpage{345}

\newcommand   \D  [2]{\frac{\partial   #1}{\partial #2}}
\newcommand   \DD [2]{\frac{\partial^2 #1}{\partial #2 ^2}}

\renewcommand \vec[1]{\bm{\mathrm{#1}}}

\def \Da{{\rm d} a}
\def \Dx{{\rm d} \x}
\def \Ds{{\rm d} \s}
\def \Eb{E_{\text{b}}}
\def \Es{E_{\text{s}}}
\def \F{\vec{F}}
\def \Fb{\F_{\text{b}}}
\def \Fs{\F_{\text{s}}}
\def \II{\mathbb{I}}
\def \RR{\mathbb{R}}
\def \X{\vec{X}}
\def \cEs{{\mathcal E}_{\text{s}}}
\def \cF{\vec{\mathcal F}}
\def \cb{c_{\text{b}}}
\def \f{\vec{f}}
\def \grad{\nabla}
\def \half{\frac{1}{2}}
\def \p{\partial}
\def \s{\vec{s}}
\def \u{\vec{u}}
\def \x{\vec{x}}

\def \OmegaAo{\Omega^{\text{Ao}}}
\def \OmegaLV{\Omega^{\text{LV}}}
\def \OmegaO{\Omega^{\text{O}}}
\def \PAo{P^{\text{Ao}}}
\def \PLV{P^{\text{LV}}}
\def \PWk{P^{\text{Wk}}}
\def \QAo{Q^{\text{Ao}}}
\def \QLV{Q^{\text{LV}}}
\def \Rc{R_{\text{c}}}
\def \Rp{R_{\text{p}}}

\def \Div{\nabla_h \cdot \mbox{}}
\def \Grad{\nabla_h}
\def \Lap{\nabla_h^2}
\def \dt{\Delta t}
\def \dx{\Delta x}
\def \ds{\Delta s}

\def \ellmax{\ell_{\text{max}}}
\def \hell{h^{\ell}}
\def \hellmax{h^{\ellmax}}
\def \hellminus{h^{\ell-1}}
\def \hellplus{h^{\ell+1}}
\def \nref{{\mathfrak n}}

\def \c{\vec{c}}

\def \cS{\vec{\mathcal S}}
\def \G{\mathcal G}

\begin{document}

\setcounter{page}{317}

\runningheads{B.~E.~Griffith}{Immersed boundary model of aortic heart
  valve dynamics}

\title{Immersed boundary model of aortic heart valve dynamics with
  physiological driving and loading conditions}

\author{Boyce E.~Griffith}

\address{Leon H.~Charney Division of Cardiology, Department of
  Medicine, and Program in Computational Biology, Sackler Institute of
  Graduate Biomedical Sciences, New York University School of
  Medicine, 550 First Avenue, New York, NY 10016 USA}

\corraddr{Boyce E.~Griffith, Leon H.~Charney Division of Cardiology,
  Department of Medicine, New York University School of Medicine, 550
  First Avenue, New York, NY 10016 USA.  E-mail:
  boyce.griffith@nyumc.org.}

\begin{abstract}
  The immersed boundary (IB) method is a mathematical and numerical
  framework for problems of fluid-structure interaction, treating the
  particular case in which an elastic structure is immersed in a
  viscous incompressible fluid.  The IB approach to such problems is
  to describe the elasticity of the immersed structure in Lagrangian
  form, and to describe the momentum, viscosity, and incompressibility
  of the coupled fluid-structure system in Eulerian form.  Interaction
  between Lagrangian and Eulerian variables is mediated by integral
  equations with Dirac delta function kernels.  The IB method provides
  a unified formulation for fluid-structure interaction models
  involving both thin elastic boundaries and also thick viscoelastic
  bodies.  In this work, we describe the application of an adaptive,
  staggered-grid version of the IB method to the three-dimensional
  simulation of the fluid dynamics of the aortic heart valve.  Our
  model describes the thin leaflets of the aortic valve as immersed
  elastic boundaries, and describes the wall of the aortic root as a
  thick, semi-rigid elastic structure.  A physiological
  left-ventricular pressure waveform is used to drive flow through the
  model valve, and dynamic pressure loading conditions are provided by
  a reduced (zero-dimensional) circulation model that has been fit to
  clinical data.  We use this model and method to simulate aortic
  valve dynamics over multiple cardiac cycles.  The model is shown to
  approach rapidly a periodic steady state in which physiological
  cardiac output is obtained at physiological pressures.  These
  realistic flow rates are not specified in the model, however.
  Instead, they emerge from the fluid-structure interaction
  simulation.
\end{abstract}


\maketitle

\section{Introduction}
\label{s:introduction}

The immersed boundary (IB) method is a mathematical and numerical
approach to problems of fluid-structure interaction that was
introduced by Peskin to model the fluid dynamics of heart valves
\cite{Peskin72,Peskin77}.  The IB methodology has subsequently been
used to model diverse problems in biological fluid dynamics
\cite{Peskin02} and other problems in which a rigid or elastic
structure is immersed in a fluid flow
\cite{Peskin02,GIaccarino03,RMittal05}.  The IB method for
fluid-structure interaction treats problems in which an elastic
structure is immersed in a viscous incompressible fluid, describing
the elasticity of the immersed structure in Lagrangian form, and
describing the momentum, viscosity, and incompressibility of the
coupled fluid-structure system in Eulerian form.  Integral equations
with Dirac delta function kernels couple the Lagrangian and Eulerian
variables.  When discretized for computer simulation, the IB method
approximates the Lagrangian equations on a curvilinear mesh,
approximates the Eulerian equations on a Cartesian grid, and
approximates the Lagrangian-Eulerian interaction equations by
replacing the singular delta function with a regularized version of
the delta function.

A key strength of the IB approach to fluid-structure interaction is
that it does not require conforming Lagrangian and Eulerian
discretizations.  Specifically, the IB method permits the Lagrangian
mesh to cut through the background Eulerian grid in an arbitrary
manner and does not require dynamically generated body-fitted meshes.
This attribute of the method greatly simplifies the task of grid
generation and facilitates simulations involving large deformations of
the elastic structure.  An additional feature of the IB formulation is
that it provides a unified approach to constructing models involving
both thin elastic boundaries (i.e., immersed structures that are of
codimension one with respect to the fluid) and also thick elastic
bodies (i.e., immersed structures that are of codimension zero with
respect to the fluid) \cite{Peskin02}.

In this work, we describe an adaptive version of the IB method and the
application of this method to the simulation of the fluid dynamics of
the aortic heart valve.
Each year, approximately $\mbox{250,000}$ procedures are performed to
repair or replace damaged or destroyed heart valves
\cite{Yoganathan04}.  Severe aortic valve disease is generally treated
by replacement with either a mechanical or a bioprosthetic valve
\cite{CarrSavage04}, and approximately $\mbox{50,000}$ aortic valve
replacements are performed annually to treat severe aortic stenosis
\cite{RVFreeman05,ROBonow06,IMSingh08}.  Because many of the
difficulties of prosthetic heart valves are related to the fluid
dynamics of the replacement valve \cite{Yoganathan04}, mathematical
and computational models that enable the study of the fluid-mechanical
mechanisms of valve function and dysfunction may ultimately aid in
improving treatment outcomes for the many patients suffering from
valvular heart diseases.

The aortic valve model employed herein is similar, but not identical,
to that described by Griffith et
al.~\cite{GriffithLuoMcQueenPeskin09}.  We model the thin leaflets of
the aortic valve as immersed boundaries comprised of systems of
elastic fibers that resist extension, compression, and bending, and we
model the aortic root and ascending aorta as a thick, semi-rigid
elastic structure.  To construct the model valve leaflets, we use the
mathematical theory of Peskin and McQueen \cite{PeskinMcQueen94},
which describes the architecture of the systems of collagen fibers
within the valve leaflets that allow the closed valve to support a
significant pressure load.  The geometry of the model aortic root is
based on the idealized description of Reul et al.~\cite{Reul90}, which
was derived from imaging data collected from healthy patients, and we
use dimensions that are based on measurements by Swanson and
Clark~\cite{SwansonClark74} of human aortic roots harvested after
autopsy.  A Windkessel model fit to human data by Stergiopulos et
al.~\cite{Stergiopulos99} provides physiological loading conditions
for the model valve.  Two limitations of our earlier model
\cite{GriffithLuoMcQueenPeskin09}, which are overcome in the present
work, are that it used only a highly idealized left-ventricular
driving pressure waveform, and that it considered only a single
cardiac cycle.  In this work, we use a physiological driving pressure
waveform that is based on human clinical data \cite{JPMurgo80}, and we
perform multibeat simulations of the fluid dynamics of the aortic
heart valve.  We emphasize that we do not prescribe the flow rate at
either the upstream or downstream boundaries of the model vessel.
Instead, we impose a realistic, periodic left-ventricular driving
pressure at the upstream boundary along with a dynamic circulatory
loading pressure at the downstream boundary.  With the driving and
loading conditions used in the present work, our model rapidly
approaches a periodic steady state in which physiological cardiac
output is obtained at physiological pressures.

There are also important differences between the numerical methods
used in the present study and those of our earlier simulations of
aortic valve dynamics \cite{GriffithLuoMcQueenPeskin09}.  Although
both use an adaptive version of the IB method for fluid-structure
interaction, our earlier study used a cell-centered IB method
\cite{GriffithHornungMcQueenPeskin07}, whereas herein we use a
staggered-grid discretization.  This is notable because we have
recently demonstrated that staggered-grid IB methods yield
substantially improved accuracy when compared to cell-centered
discretizations \cite{BEGriffithXX-ib_volume_conservation}.
Specifically, we have found that using a staggered-grid Eulerian
discretization improves the volume-conservation properties of the IB
method by one to two orders of magnitude in comparison to a
cell-centered discretization
\cite{BEGriffithXX-ib_volume_conservation}.  Staggered-grid IB methods
also yield improved resolution of pressure discontinuities
\cite{BEGriffithXX-ib_volume_conservation}.  In the present
application, such discontinuities occur along the thin heart valve
leaflets and are especially pronounced when the valve is closed and
supporting a significant, physiological pressure load.

The three-dimensional adaptive IB method used in our simulations is
similar to the two-dimensional adaptive IB method of Roma et
al.~\cite{RomaPeskinBerger99}.  Specifically, both schemes use a
globally second-order accurate staggered-grid (i.e., marker-and-cell
or MAC \cite{HarlowWelch65}) discretization of the incompressible
Navier-Stokes equations on block-structured adaptively refined
Cartesian grids, and both schemes implement formally second-order
accurate versions of the IB method (i.e., schemes that yield
second-order convergence rates for problems with sufficiently smooth
solutions \cite{LaiPeskin00,GriffithPeskin05}).  There are also
important differences between the present scheme and the scheme of
Roma et al.  For instance, the method of Roma et al.~uses centered
differencing to approximate the nonlinear advection terms of the
incompressible Navier-Stokes equations, whereas we use a
staggered-grid version \cite{Griffith09} of the xsPPM7 variant
\cite{RiderEtAl07} of the piecewise parabolic method (PPM)
\cite{ColellaWoodward84} that enables the application of the present
method to high Reynolds number flows.  Our scheme also uses the
projection method not as a fractional-step solver for the
incompressible Navier-Stokes equations, but rather as a preconditioner
for an iterative Krylov method applied to an unsplit discretization of
those equations \cite{Griffith09}.  Our approach eliminates the
timestep-splitting error associated with standard projection methods.
It also greatly simplifies the specification of physical boundary
conditions along the outer boundaries of the computational domain.  In
this work, such physical boundary conditions couple the detailed,
three-dimensional fluid-structure interaction model to the reduced
circulation models that provide realistic driving and loading
conditions.

\section{The continuous equations of motion}
\label{s:the_continuous_equations_of_motion}

The IB formulation of the equations of motion for a coupled
fluid-structure system describes the elasticity of the immersed
structure in Lagrangian form and describes the momentum, velocity, and
incompressibility of the fluid-structure system in Eulerian form.  Let
$\x = (x_1,x_2,x_3) \in \Omega$ denote Cartesian physical coordinates,
with $\Omega \subset \RR^3$ denoting the physical region that is
occupied by the fluid-structure system; let $\s = (s_1,s_2,s_3) \in U$
denote Lagrangian material coordinates that are attached to the
immersed elastic structure, with $U \subset \RR^3$ denoting the
Lagrangian coordinate domain; and let $\X(\s,t) \in \Omega$ denote the
physical position of material point $\s$ at time $t$.  We consider the
case in which the fluid possesses a uniform mass density $\rho$ and
dynamic viscosity $\mu$, and we assume that the structure is neutrally
buoyant and has the same viscous properties as the fluid in which it
is immersed.  These assumptions are not essential to the method,
however, and generalizations of the IB method have been developed to
permit the mass density of the structure to differ from that of the
fluid \cite{Peskin02,ZhuPeskin02,ZhuPeskin03,KimEtAl03,Kim06,Kim07}.
Work is also underway to develop new extensions of the IB method that
permit the viscosity of the structure to differ from that of the
fluid.

The IB formulation of the equations of fluid-structure interaction is
\cite{Peskin02}:
\begin{align}
  \rho \left(\D{\u}{t}(\x,t) + \u(\x,t) \cdot \grad \u(\x,t)\right)
                       &= -\grad p(\x,t) + \mu \grad^2 \u(\x,t) + \f(\x,t),     \label{e:momentum}          \\
  \grad \cdot \u(\x,t) &= 0,                                                    \label{e:incompressibility} \\
  \f(\x,t)             &= \int_U \F(\s,t) \, \delta(\x - \X(\s,t)) \, \Ds,      \label{e:force_density}     \\
  \D{\X}{t}(\s,t)      &= \int_\Omega \u(\x,t) \, \delta(\x - \X(\s,t)) \, \Dx, \label{e:interp}            \\
  \F(\s,t)             &= \cF[\X(\cdot,t)],                                     \label{e:lagrangian_force}
\end{align}
in which $\u(\x,t) = (u_1(\x,t),u_2(\x,t),u_3(\x,t))$ is the Eulerian
velocity field, $p(\x,t)$ is the Eulerian pressure, $\f(\x,t) =
(f_1(\x,t),f_2(\x,t),f_3(\x,t))$ is the Eulerian elastic force density
(i.e., the elastic force density with respect to the physical
coordinate system, so that $\f(\x,t) \, \Dx$ has units of force),
$\F(\s,t) = (F_1(\s,t),F_2(\s,t),F_3(\s,t))$ is the Lagrangian elastic
force density (i.e., the elastic force density with respect to the
material coordinate system, so that $\F(\s,t) \, \Ds$ has units of
force), $\cF:\X\mapsto\F$ is a functional that specifies the
Lagrangian elastic force density in terms of the deformation of the
immersed structure, and $\delta(\x) = \delta(x_1) \, \delta(x_2) \,
\delta(x_3)$ is the three-dimensional Dirac delta function.  In this
formulation, eqs.~\eqref{e:force_density} and \eqref{e:interp} are the
interaction equations that couple the Lagrangian and Eulerian
variables.  Eq.~\eqref{e:force_density} converts the Lagrangian force
density $\F(\s,t)$ into the equivalent Eulerian force density
$\f(\x,t)$.  Eq.~\eqref{e:interp} states that the physical position of
each Lagrangian material point $\s$ moves with velocity
$\u(\X(\s,t),t)$, thereby implying that there is no fluid slip at
fluid-structure interfaces.  Notice, however, that the no-slip
condition of a viscous fluid does not appear in the equations as a
constraint on the fluid motion.  Instead, the no-slip condition
determines the motion of the immersed structure.  See, e.g., Peskin
\cite{Peskin02} for further discussion of these equations.

We next describe the form of the Lagrangian elastic force density
functional $\cF:\X\mapsto\F$ used in our model.  Like our earlier
study of cardiac valve dynamics \cite{GriffithLuoMcQueenPeskin09}, we
model the flexible leaflets of the aortic valve as thin elastic
boundaries, and we model the vessel wall as a thick, semi-rigid
elastic structure.  The elasticity of these structures is described in
terms of families of fibers that resist extension, compression, and
bending.  We identify the model fibers of the valve leaflets with the
collagen fibers that enable the real valve to support a significant
pressure load when closed.  In the case of the vessel wall, we do not
identify the model fibers with particular physiological features;
instead, these fibers are used simply to fix the geometry of the
vessel.

As is frequently done in IB models \cite{Peskin02}, we define the
fiber elasticity in terms of a strain-energy functional $E =
E[\X(\cdot,t)]$.  The corresponding Lagrangian elastic force density
may be expressed in terms of the Fr\'{e}chet derivative of $E$.
Specifically, $\F$ is defined by
\begin{equation}
  \vec{F} = - \frac{\delta E}{\delta \X}, \label{e:frechet1}
\end{equation}
which is shorthand for
\begin{equation}
  \delta E[\X(\cdot,t)] = - \int_U \vec{F}(\s,t) \cdot \delta \X(\s,t) \, \Ds. \label{e:frechet2}
\end{equation}
Notice that in eqs.~\eqref{e:frechet1} and \eqref{e:frechet2},
$\delta$ denotes the perturbation operator, not the Dirac delta
function.  To specify $E$, it is convenient to choose the Lagrangian
coordinates $\s = (s_1,s_2,s_3) \in U$ so that each fixed value of
$(s_1,s_2)$ labels a particular fiber.  This implies that the mapping
$s_3 \mapsto \X(s_1^0,s_2^0,s_3)$ is a parametric representation of
the fiber labeled by $(s_1,s_2) = (s_1^0,s_2^0)$.  The curvilinear
coordinate $s_3$ need not correspond to arc length along the fiber,
however, and even if $s_3$ were to correspond to arc length in an
initial or reference configuration, notice that it generally will not
remain arc length as the structure deforms.

As we have done previously \cite{GriffithLuoMcQueenPeskin09}, we
describe the total elastic energy functional $E$ as the sum of a
stretching energy $\Es$ and a bending energy $\Eb$, so that $E = \Es +
\Eb$.  In turn, these elastic energy functionals determine a
stretching force density $\Fs$ and a bending force density $\Fb$, so
that $\F = \Fs + \Fb$.  The stretching energy is
\begin{equation}
  \Es = \int_{\Omega} \cEs\left(\left|\D{\X}{s_3}\right|;\s\right) \, \Ds, \label{e:E_s}
\end{equation}
in which $\cEs$ is a local stretching energy.  The corresponding
stretching force is given by \cite{Peskin02}
\begin{equation}
  \Fs = \D{}{s_3} \left( \cEs'\left(\left|\D{\X}{s_3}\right|;\s\right) \frac{\p\X/\p s_3}{\left|\p\X/\p s_3\right|} \right),
\end{equation}
in which $\cEs'$ indicates the derivative of $\cEs$ with respect to
its first argument.  By identifying $T = \cEs'\left(\left|\p\X/\p
    s_3\right|;\s\right)$ as the fiber tension and $\vec{\tau} =
\p\X/\p s_3 / \left| \p\X / \p s_3\right|$ as the fiber-aligned unit
tangent vector, we may rewrite $\Fs$ as
\begin{equation}
  \Fs = \D{}{s_3}\left(T \vec{\tau}\right).  \label{e:F_s}
\end{equation}
The bending energy used in our model is
\begin{equation}
  \Eb = \half \int_{\Omega} \cb(\s) \left|\DD{\X}{s_3} - \DD{\bar{\X}}{s_3}\right|^{2} \, \Ds, \label{e:E_b}
\end{equation}
in which $\cb = \cb(\s)$ is the spatially inhomogeneous bending
stiffness, and $\bar{\X} = \bar{\X}(\s)$ is the reference
configuration of the structure.  The corresponding bending-resistant
force is given by \cite{GriffithLuoMcQueenPeskin09}
\begin{equation}
  \Fb = \DD{}{s_3}\left(\cb(\s) \left(\DD{\bar{\X}}{s_3} - \DD{\X}{s_3}\right)\right).
\end{equation}
We take the reference configuration to be the initial configuration,
i.e., $\bar{\X}(\s) = \X(\s,0)$.  We remark that we use
bending-resistant forces only within the model valve leaflets, i.e.,
$\cb(\s) \neq 0$ only for those fibers that comprise the valve
leaflets.  Because we model the valve leaflets as thin elastic
surfaces immersed in fluid, including bending-resistant forces allows
the model leaflets to account for the thickness of real valve
leaflets, which are thin but, of course, not infinitely thin.  In our
model, we increase $\cb$ near the tips of the free edges of the valve
leaflets to account for the fibrous noduli arantii.

Next, we specify the boundary conditions imposed along the outer
boundary of the physical domain $\Omega$.  We take $\Omega$ to be a
$\mbox{10 cm} \times \mbox{10 cm} \times \mbox{15 cm}$ rectangular
box, and we employ a combination of solid-wall and prescribed-pressure
boundary conditions along $\p\Omega$.  Solid-wall boundary conditions
are simply homogeneous Dirichlet conditions for the velocity field
$\u(\x,t)$.  At solid-wall boundaries, a boundary condition for the
pressure is neither needed nor permitted.  By prescribed-pressure
boundary conditions, we mean a combination of normal-traction and
zero-tangential-slip boundary conditions.  For a viscous
incompressible fluid, it is easy to show that combining
normal-traction and zero-tangential-slip boundary conditions along a
flat boundary allows for the pointwise specification of the pressure
$p(\x,t)$ on that boundary.  To see this, recall that the Cauchy
stress tensor of a viscous incompressible fluid is
\begin{equation}
  \vec{\sigma} =  -p \II + \mu \left[ \grad\vec{u} + \left(\grad\vec{u}\right)^{T} \right].
\end{equation}
Let the outward unit normal at a position $\x \in \p\Omega$ be denoted
by $\vec{n} = \vec{n}(\x)$, and let a unit tangent vector at a
position $\x \in \p\Omega$ be denoted by $\vec{t} = \vec{t}(\x)$.  By
prescribing the normal traction at the boundary, we are prescribing
the value of the normal component of the normal stress, i.e.,
\begin{equation}
  \vec{n} \cdot \vec{\sigma} \cdot \vec{n} = -p + 2 \mu \D{}{n}(\vec{u}\cdot\vec{n}).
\end{equation}
The zero-tangential-slip condition imposed on $\u$ implies that $\u
\cdot \vec{t} \equiv 0$ along $\p\Omega$, and combining this condition
with the incompressibility constraint implies that
$\D{}{n}(\vec{u}\cdot\vec{n}) \equiv 0$ along $\p\Omega$.  Therefore,
along $\p\Omega$, the normal component of the normal stress reduces to
\begin{equation}
  \vec{n} \cdot \vec{\sigma} \cdot \vec{n} = -p.
\end{equation}
Thus, combining normal-traction and zero-tangential-slip boundary
conditions allows us to prescribe the value of the pressure pointwise
along the boundary.

\begin{figure}
  \centering
  \input{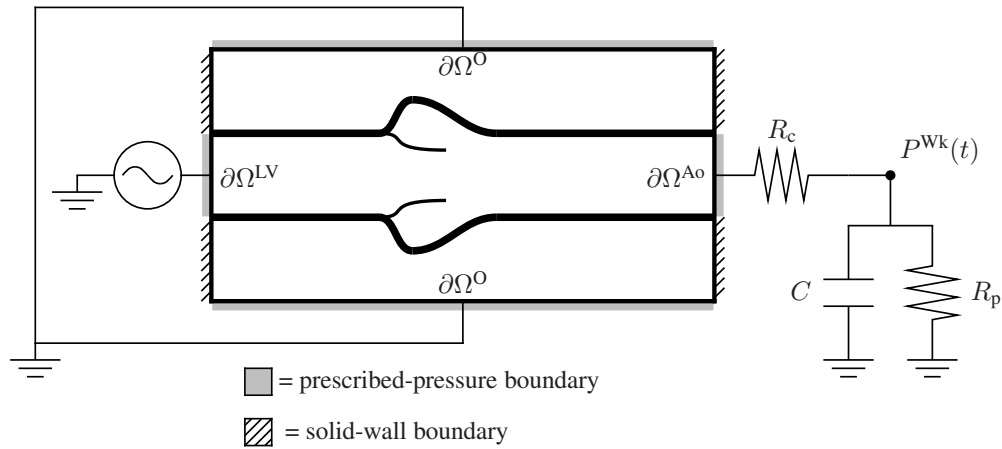}
  \caption{Schematic diagram of the model showing the coupling between
    the detailed fluid-structure interaction model and the reduced
    circulation models that provide physiological driving and loading
    conditions.  The model vessel is coupled via the upstream boundary
    $\p\OmegaLV$ to a prescribed left-ventricular pressure source that
    drives flow through the valve.  At the downstream boundary
    $\p\OmegaAo$, the model vessel is coupled to a circulation model
    that provides a realistic pressure load.  The open boundary
    $\p\OmegaO$ provides a zero-pressure fluid reservoir that balances
    any changes in the volume of the vessel, which is modeled as a
    semi-rigid elastic structure.}
  \label{f:physical_bc2d}
\end{figure}

The model vessel attaches directly to $\p \Omega$, the outermost
boundary of the physical domain.  A schematic diagram is provided in
fig.~\ref{f:physical_bc2d}.  Along $\p\OmegaLV$, the upstream boundary
of the vessel, a time-dependent left-ventricular pressure waveform
$\PLV(t)$ is prescribed to drive flow through the model valve, so that
\begin{equation}
  p(\x,t) = \PLV(t), \ \x \in \p\OmegaLV.
\end{equation}
The specific left-ventricular pressure waveform used in the present
model is adapted from the study of Murgo et al.~\cite{JPMurgo80}.  The
rate of flow entering the model vessel via the upstream boundary is
\begin{equation}
  \QLV(t) = - \int_{\p\OmegaLV} \u(\x,t) \cdot \vec{n} \, \Da,
\end{equation}
in which $\vec{n}$ is the outward unit normal along $\p\Omega$, and
$\Da = \Da(\x)$ is the area element in the Cartesian coordinate
system.

On $\p\OmegaAo$, the downstream boundary of the vessel, we use a
reduced (i.e., ordinary differential equation) circulation model to
determine the pressure $\PAo(t)$ that provides dynamic loading
conditions for the model valve, so that
\begin{equation}
  p(\x,t) = \PAo(t), \ \x \in \p\OmegaAo.
\end{equation}
The reduced circulation model used in this work is a three-element
Windkessel model with characteristic resistance $\Rc$, peripheral
resistance $\Rp$, and arterial compliance $C$; see
fig.~\ref{f:physical_bc2d}.  The rate of flow leaving the model vessel
via the downstream boundary is
\begin{equation}
  \QAo(t) = \int_{\p\OmegaAo} \u(\x,t) \cdot \vec{n} \, \Da.
\end{equation}
Because we specify the pressure along $\p\OmegaAo$, the value of
$\QAo(t)$ is not known in advance; instead, it must be determined by
the coupled model.  The flow leaving the fluid-structure interaction
model via $\p\OmegaAo$ is exactly the flow through the circulation
model, so that
\begin{align}
  \frac{1}{\Rc}\left(\PAo(t) - \PWk(t)\right) &= \QAo(t), \label{wk1} \\
  C \frac{\text{d}\PWk}{\text{d}t}(t) + \frac{1}{\Rp} \PWk(t) &= \QAo(t), \label{wk2}
\end{align}
in which $\PWk(t)$ is the stored pressure in the Windkessel model.
Notice that the value of $\PAo(t)$ is completely determined by
$\PWk(t)$ and $\QAo(t)$.  In our simulations, we set $\Rc = 0.033$
(mmHg ml$^{-1}$ s), $\Rp = 0.79$ (mmHg ml$^{-1}$ s), and $C = 1.75$
(ml mmHg$^{-1}$), corresponding to the human ``Type A'' beat
characterized by Stergiopulos et al.~\cite{Stergiopulos99}.

Although we have found that coupling the detailed and reduced models
via prescribed-pressure boundary conditions works well in practice,
other choices of boundary conditions are possible.  For instance, it
is straightforward to devise boundary conditions for the
incompressible Navier-Stokes equations that prescribe the \emph{mean}
pressure along a portion of a flat boundary; see ch.~3 sec.~8 of
Gresho and Sani \cite{GreshoSani98} for details.  Alternatively, one
may wish to couple the detailed and reduced models by prescribing
boundary conditions for the normal component of the velocity.  A
drawback of this approach is that it would require determining an
appropriate velocity profile at the boundary.  A more serious
limitation of this alternative approach is that prescribing the flow
rate as a boundary condition, at either the upstream or the downstream
boundary, makes it impossible to impose a realistic pressure
difference across the model valve during the diastolic phase of the
cardiac cycle.  By using pressure boundary conditions at both the
upstream and downstream boundaries of the vessel, we allow the normal
component of the velocity profile at the boundary to be determined by
the model, and we are able to impose realistic pressure loads on the
model valve throughout the cardiac cycle.

Along $\p\OmegaO$, the outermost portion of $\p\Omega$ exterior to the
model vessel, we set the pressure to equal zero.  This external
boundary condition provides an open boundary that acts to couple the
fluid-structure interaction model to a zero-pressure fluid reservoir.
This constant-pressure reservoir allows the vessel to change volume
during the course of the simulation, i.e., it permits a mismatch
between the instantaneous flow rates at the inflow and outflow
boundaries of the vessel.  Because we model the vessel wall as a
semi-rigid elastic structure, we have that $\QLV(t) \approx \QAo(t)$.
Of course, once the model reaches periodic steady state, the
time-integrated inflow and outflow volumes must match.  The real
aortic root is a flexible structure with significant compliance,
however, and a model vessel that accounts for the physiological
compliance of the aortic root would generally result in instantaneous
differences between the inflow rate through $\p\OmegaLV$ and the
outflow rate through $\p\OmegaAo$.

All that remains is to specify the initial conditions.  At time $t =
0$, we set $\u(\x,t) = 0$ along with $\PLV(t) = \PWk(t) = 0$, so that
all prescribed normal traction along the boundary are equal to zero.
During a brief initialization period lasting 12.8 ms, we increase the
left-ventricular driving pressure to a value of approximately 10 mmHg,
and we increase the stored pressure in the Windkessel model to 85
mmHg, thereby establishing a realistic pressure load on the closed
valve.  During this initialization period, $\PWk(t)$ is treated as a
boundary condition and not as a state variable.  That is to say,
$\PWk(t)$ does not satisfy eq.~\eqref{wk2} for $t \leq \mbox{12.8
  ms}$; rather, the value of $\PWk(t)$ is prescribed.  Once the model
is initialized, however, $\PWk(t)$ is treated as a state variable, the
dynamics of which are determined by eq.~\eqref{wk2}.

\section{The discrete equations of motion}
\label{s:the_discrete_equations_of_motion}

\subsection{Lagrangian and Eulerian spatial discretizations}

As in our earlier simulation studies of cardiac fluid dynamics
\cite{GriffithLuoMcQueenPeskin09,GriffithHornungMcQueenPeskin07}, we
discretize the Lagrangian equations on a fiber-aligned curvilinear
mesh, and we discretize the Eulerian equations on a block-structured
locally refined Cartesian grid that is adaptively generated to conform
to the moving fiber mesh.  The curvilinear mesh spacings are $\ds_1$,
$\ds_2$, and $\ds_3$, and we use the indices $(l,m,n)$ to label the
nodes of the Lagrangian mesh, so that $\X_{l,m,n}$ and $\F_{l,m,n}$
are the position and Lagrangian elastic force density associated with
curvilinear mesh node $(l,m,n)$.  The nodal values of $\F_{l,m,n}$ are
computed from the physical positions of the nodes of the curvilinear
mesh via standard second-order accurate finite difference
approximations to $\D{}{s_3}$ and to $\DD{}{s_3}$.  This approach is
equivalent to describing the elasticity of the discretized model in
terms of systems of springs and beams.

\begin{figure}
  \centering
  \input{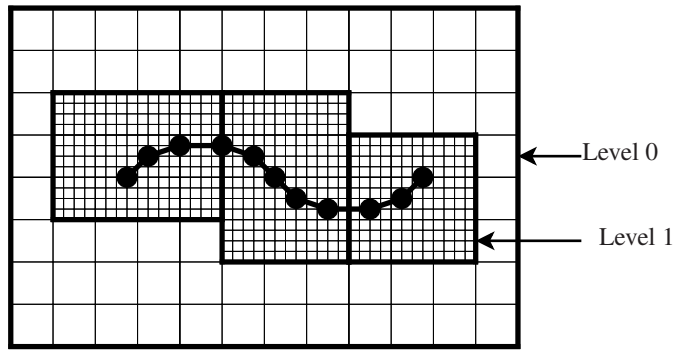}
  \caption{A two-dimensional locally refined hierarchical Cartesian
    grid comprised of two levels with refinement ratio $\nref=4$.
    Cartesian grid patch boundaries are indicated by bold lines.  A
    single immersed fiber is embedded in level 1 of patch hierarchy.
    To simplify the discretization of the interaction equations that
    couple the Lagrangian and Eulerian variables, the Cartesian grid
    patches are constructed in a manner that ensures that the
    curvilinear mesh nodes are well separated from interfaces in
    Cartesian grid resolution.}
  \label{f:valid_grid}
\end{figure}

The locally refined Cartesian grid is organized as a hierarchy of
nested grid levels that are labeled $\ell = 0,\ldots,\ellmax$, with
$\ell = 0$ denoting the coarsest level of the hierarchical grid and
with $\ell = \ellmax$ denoting the finest level.  Each grid level
$\ell$ is comprised of the union of rectangular Cartesian grid
patches.  All grid patches in a given level $\ell$ of the grid
hierarchy share the same uniform grid spacing $\hell$, and the grid
spacings are chosen so that $\hell = \hellminus / \nref$, in which
$\nref>\mbox{1}$ is an integer refinement ratio.  The patch levels are
constructed to satisfy the proper nesting condition
\cite{BergerColella89}, which generally requires that the union of the
level $\ell+1$ grid patches be strictly contained within the union of
the level $\ell$ grid patches.  The proper nesting condition is
relaxed at the outermost boundary of the physical domain, thereby
allowing high Eulerian spatial resolution all the way up to $\p\Omega$
in cases in which such resolution is needed.  The patch levels are
generated so that the faces of the grid patches that comprise level
$\ell > 0$ are coincident with the faces of the Cartesian grid cells
that comprise level $\ell-1$, the next coarser level of the grid.
This construction simplifies the development of composite-grid
discretization methods that couple the levels of the locally refined
grid.  Except where noted, in our simulations, we employ a two-level
adaptive grid, so that $\ellmax = 1$, and we use $\nref=4$.  An
example two-dimensional grid is shown in fig.~\ref{f:valid_grid}.

\begin{figure}
  \centering
  \includegraphics[width=0.6\textwidth]{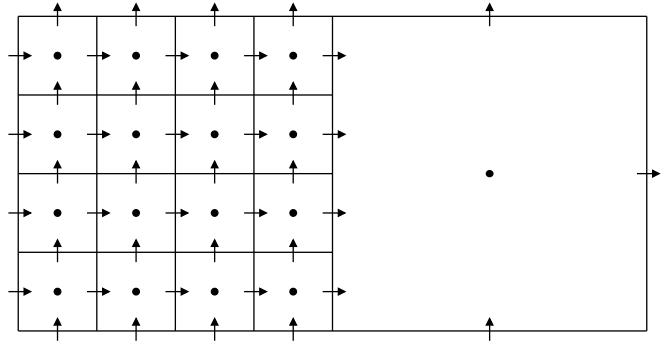}
  \caption{A two-dimensional locally refined staggered-grid
    discretization.  The velocity field $\vec{u}$ is approximated in
    terms of those vector components that are normal to the faces of
    the grid cells, and the pressure $p$ is approximated at the
    centers of the grid cells.}
  \label{f:mac_grid}
\end{figure}

To discretize the Eulerian incompressible Navier-Stokes equations in
space, we employ a locally refined version of a three-dimensional
staggered-grid finite difference scheme; see fig.~\ref{f:mac_grid}.
The computational domain $\Omega$ is a rectangular box, $\Omega =
[0,L_1] \times [0,L_2] \times [0,L_3]$, and the coarsest level of the
locally refined Cartesian grid is a uniform discretization of
$\Omega$, so that the union of the level $\ell = 0$ grid patches form
a regular $N_1 \times N_2 \times N_3$ Cartesian grid with grid
spacings $\dx_1 = \frac{L_1}{N_1}$, $\dx_2 = \frac{L_2}{N_2}$, and
$\dx_3 = \frac{L_3}{N_3}$.  For simplicity, we assume that $\dx_1 =
\dx_2 = \dx_3 = h^0$.  On each level $\ell$ of the locally refined
grid, $\mbox{$(i,j,k)$}$ labels a particular Cartesian grid cell, and
$\x_{i,j,k} = \mbox{$\left( \left(i+\half\right)\hell,
    \left(j+\half\right)\hell, \left(k+\half\right)\hell \right)$}$
denotes the physical location of the center of that cell.  The
physical region covered by Cartesian grid cell $(i,j,k)$ on level
$\ell$ is denoted by $\c_{i,j,k}^{\ell}$, and the set of Cartesian
grid cell indices associated with level $\ell$ is denoted by
$\G^{\ell}$.  The components of the Eulerian velocity field $\u =
(u_1,u_2,u_3)$ are respectively approximated at the centers of the
$x_1$, $x_2$, and $x_3$ faces of the Cartesian grid cells, i.e., at
positions $\x_{i-\half,j,k} = \mbox{$\left( i\hell,
    \left(j+\half\right)\hell, \left(k+\half\right)\hell \right)$}$,
$\x_{i,j-\half,k} = \mbox{$\left( \left(i+\half\right)\hell, j\hell,
    \left(k+\half\right)\hell \right)$}$, and $\x_{i,j,k-\half} =
\mbox{$\left( \left(i+\half\right)\hell, \left(j+\half\right)\hell,
    k\hell \right)$}$.  The pressure $p$ is approximated at the
centers of the Cartesian grid cells.  We use $(u_1)_{i-\half,j,k}$,
$(u_2)_{i,j-\half,k}$, $(u_3)_{i,j,k-\half}$, and $p_{i,j,k}$ to
denote the values of $\u$ and $p$ that are stored on the grid.  A
staggered discretization is also used for the Eulerian body force $\f
= (f_1,f_2,f_3)$, so that $f_1$, $f_2$, and $f_3$ are respectively
approximated at the centers of the $x_1$, $x_2$, and $x_3$ faces of
the Cartesian grid cells.

Let $\Omega^{\ell} \subseteq \Omega$ denote the physical region
covered by the union of the level $\ell$ grid patches.  By
construction, $\Omega^0 = \Omega$, and $\Omega^{\ell+1} \subseteq
\Omega^{\ell}$.  Moreover, away from $\p\Omega$, $\Omega^{\ell+1}$ is
strictly contained within $\Omega^{\ell}$.  Notice that $\Omega^{\ell}
= \cup_{(i,j,k) \in \G^{\ell}} \c_{i,j,k}^{\ell}$.  The coarse-fine
interface between levels $\ell$ and $\ell+1$ is $\p\Omega^{\ell+1}
\setminus \p\Omega^{\ell}$.  Because the grid levels are constructed
to satisfy the proper nesting condition, however, $\p\Omega^{\ell+1}
\cap \p\Omega^{\ell} \subset \p\Omega$, so that the coarse-fine
interface between levels $\ell$ and $\ell+1$ is $\p\Omega^{\ell+1}
\setminus \p\Omega$.

The refined region of level $\ell < \ellmax$, denoted by
$\Omega^{\ell,\text{ref}}$, consists of the portion of $\Omega^{\ell}$
that is covered by $\Omega^{\ell+1}$, i.e., $\Omega^{\ell,\text{ref}}
= \Omega^{\ell} \cap \Omega^{\ell+1} = \Omega^{\ell+1}$.  The grid
values of any quantity stored on the Cartesian grid that are
physically located in $\Omega^{\ell,\text{ref}}$ are referred to as
invalid values.  The remaining values, i.e., those that are located in
$\Omega^{\ell} \setminus \Omega^{\ell,\text{ref}}$, are referred to as
valid values.  The invalid values of level $\ell <\ellmax$ are
constrained to be the restriction of the overlying level $\ell+1$
values.  Notice that these overlying values could be either valid or
invalid values, depending on the configuration of the grid hierarchy.
The simplest restriction procedure is to define the coarse-grid
invalid values to be the averages of the overlying fine-grid values;
however, other restriction procedures are possible and, in fact,
necessary to obtain higher-order accuracy at coarse-fine interfaces.

Each Cartesian grid patch is augmented by a layer of ghost cells that
provide the values at the patch boundary that are needed to evaluate
the approximations to the Eulerian spatial differential operators in
the patch interior.  We consider three types of ghost cells for the
level $\ell$ grid patches:
\begin{inparaenum}[(1)]
\item ghost cells $(i,j,k)$ that overlap the interior of another level
  $\ell$ grid patch, so that $\c_{i,j,k}^{\ell} \subset
  \Omega^{\ell}$;
\item ghost cells $(i,j,k)$ that overlap the interior of some level
  $\ell-1$ grid patch but not the interior of any of the level $\ell$
  grid patches, so that $\c_{i,j,k}^{\ell} \subset \Omega^{\ell-1}$
  but $\c_{i,j,k}^{\ell} \not\subset \Omega^{\ell}$; and
\item ghost cells that are exterior to $\Omega$, so that
  $\c_{i,j,k}^{\ell} \not\subset \Omega$.
\end{inparaenum}
For level $\ell$ ghost cells that overlap the interior of a
neighboring level $\ell$-grid patch, the values in that ghost cell are
simply copies of the neighboring interior values.  If a level $\ell$
ghost cell overlaps the interior of a neighboring level $\ell-1$ grid
patch but does not overlap the interior of any level $\ell$ grid
patch, that ghost cell is said to be located on the coarse side of a
coarse-fine interface.  Ghost values located on the coarse side of a
coarse-fine interface are defined by an interpolation procedure that
uses both fine-~and coarse-grid values.  As we have done previously
\cite{GriffithLuoMcQueenPeskin09}, we use a specialized quadratic
interpolation procedure
\cite{Minion96b,MartinColella00,MartinColellaGraves08} to compute
cell-centered ghost values at coarse-fine interfaces.  For
face-centered quantities, we use a generalization of this procedure
that employs a combination of quadratic and cubic interpolation at
coarse-fine interfaces.  (Details of these coarse-fine interface
discretizations are provided in the appendix.)  Finally, ghost values
that are exterior to the physical domain are determined via the
physical boundary conditions in a manner that is analogous to the
uniform-grid scheme for the incompressible Navier-Stokes equations of
Griffith \cite{Griffith09}.

We denote by $\Div \u \approx \grad \cdot \u$, $\Grad p \approx \grad
p$, and $\Lap \u \approx \grad^2 \u$ composite-grid finite difference
approximations to the divergence, gradient, and Laplace operators,
respectively.  We employ a standard staggered-grid (MAC)
discretization approach, in which $\Div \u$ is computed at cell
centers, whereas both $\Grad p$ and $\Lap \u$ are computed on cell
faces.  Briefly stated, we use standard second-order accurate
staggered-grid approximations to these operators within each Cartesian
grid patch \cite{Griffith09}.  These uniform-grid patch-based
discretizations are coupled to form composite-grid discretizations via
restriction and prolongation operators.  Although our composite-grid
finite difference discretization of the incompressible Navier-Stokes
equations is globally second-order accurate, this discretization does
not retain the pointwise second-order accuracy of the basic
uniform-grid approximation because of localized reductions in accuracy
at coarse-fine interfaces.

To compute $\Div \u$ on the composite grid, we first use simple
averaging to restrict $\u$ from finer levels of the grid to coarser
levels of the grid.  We then compute for each level $\ell$ grid cell
\begin{align}
  \left( \Div \u \right)_{i,j,k} = & \frac{
    \left(u_1\right)_{i+\half,j,k} - \left(u_1\right)_{i-\half,j,k}
  }{\hell} + \frac{
    \left(u_2\right)_{i,j+\half,k} - \left(u_2\right)_{i,j-\half,k}
  }{\hell} + \mbox{} \nonumber \\ & \frac{
    \left(u_3\right)_{i,j,k+\half} - \left(u_3\right)_{i,j,k-\half}
  }{\hell}. \label{e:div_h}
\end{align}
To compute $\Grad p$ on the composite grid, we first use cubic
interpolation to restrict $p$ from finer levels of the grid to coarser
levels of the grid, and we then compute the composite-grid
cell-centered interpolation of $p$ in the ghost cells along the coarse
side of the coarse-fine interface.  This approach ensures that the
ghost values are at least third-order accurate interpolations of $p$.
With the coarse-fine interface ghost values so determined, we then
compute for each level $\ell$ cell face
\begin{align}
  (\Grad p)_{i-\half,j,k} &= \frac{p_{i,j,k} - p_{i-1,j,k}}{\hell}, \label{e:grad_h_1} \\
  (\Grad p)_{i,j-\half,k} &= \frac{p_{i,j,k} - p_{i,j-1,k}}{\hell}, \label{e:grad_h_2} \\
  (\Grad p)_{i,j,k-\half} &= \frac{p_{i,j,k} - p_{i,j,k-1}}{\hell}. \label{e:grad_h_3}
\end{align}
Finally, to compute $\Lap \u$, we first use cubic interpolation to
restrict $\u$ from finer levels of the grid to coarser levels of the
grid, and we then compute coarse-fine interface ghost values of $\u$
using a composite-grid face-centered interpolation scheme, again
ensuring that the ghost values are at least third-order accurate.  We
then compute for each $x_1$ face on level $\ell$
\begin{align}
  (\Lap u_1)_{i-\half,j,k} =
  &\frac{\left(u_1\right)_{i+\half,j,k} - 2 \left(u_1\right)_{i-\half,j,k} + \left(u_1\right)_{i-\frac{3}{2},j,k}}{\left(\hell\right)^2} + \mbox{} \nonumber \\
  &\frac{\left(u_1\right)_{i-\half,j+1,k} - 2 \left(u_1\right)_{i-\half,j,k} + \left(u_1\right)_{i-\half,j-1,k}}{\left(\hell\right)^2} + \mbox{} \nonumber \\
  &\frac{\left(u_1\right)_{i-\half,j,k+1} - 2 \left(u_1\right)_{i-\half,j,k} + \left(u_1\right)_{i-\half,j,k-1}}{\left(\hell\right)^2}.  \label{e:Lap_h}
\end{align}
Similar formulae are used to evaluate $\Lap u_2$ and $\Lap u_3$ on the
$x_2$ and $x_3$ faces of the composite grid.

In the case of a uniform-grid discretization, notice that $\Div \Grad$
is the usual 7-point cell-centered finite difference approximation to
the Laplacian.  This property facilities the construction of efficient
preconditioners based on the projection method \cite{Griffith09} and
approximate Schur complement methods \cite{ElmanEtAl06,ElmanEtAl08}.
In the presence of local mesh refinement, $\Div \Grad$ is a relatively
standard composite-grid generalization of the 7-point Laplacian
\cite{GriffithLuoMcQueenPeskin09,GriffithHornungMcQueenPeskin07,Minion96b,MartinColella00,MartinColellaGraves08}
for which efficient solution algorithms, such as FAC (the fast
adaptive composite-grid method)
\cite{McCormick89,McCormickEtAl89,McCormick92}, have been developed.

\subsection{Lagrangian-Eulerian interaction}

To approximate the Lagrangian-Eulerian interaction equations,
eqs.~\eqref{e:force_density} and \eqref{e:interp}, we replace the
delta function $\delta(\x)$ with a regularized version of the delta
function $\delta_h(\x)$.  The regularized delta function that we use
is of the tensor-product form $\delta_h(\x) = \delta_h(x_1) \,
\delta_h(x_2) \, \delta_h(x_3)$, and we use a one-dimensional
regularized delta function that is of the form $\delta_h(x) =
\frac{1}{h} \phi\left(\frac{x}{h}\right)$.  We take $\phi = \phi(r)$
to be the four-point delta function of Peskin \cite{Peskin02}.

We have found that fluid-structure interfaces generally require the
highest available spatial resolution in an IB simulation.  Therefore,
we construct the Cartesian grid hierarchy so that the curvilinear mesh
is embedded in the finest level of the grid.  Additionally, because we
use a regularized delta function with a support of four Cartesian
meshwidths in each coordinate direction, we ensure that the locally
refined grid is generated so that each curvilinear mesh point is
physically located at least two level $\ellmax$ grid cells away from
$\p\Omega^{\ellmax} \setminus \p\Omega$, the coarse-fine interface
between levels $\ellmax-1$ and $\ellmax$.  With these constraints on
the configuration of the locally refined Cartesian grid, the nodes of
the curvilinear mesh are directly coupled only to valid values defined
on the finest level of the locally refined grid.  This construction
therefore allows us to discretize the Lagrangian-Eulerian interaction
equations as if we were using a uniformly fine grid with resolution
$\hellmax$.

With $\f = (f_1,f_2,f_3)$ and $\F = (F_1,F_2,F_3)$, we approximate
eq.~\eqref{e:force_density} componentwise by
\begin{align}
  (f_1)_{i-\half,j,k} &= \sum_{(l,m,n)} (F_1)_{l,m,n} \, \delta_{\hellmax}(\x_{i-\half,j,k} - \X_{l,m,n}) \, \ds_1 \, \ds_2 \, \ds_3, \label{e:discrete_force_density1} \\
  (f_2)_{i,j-\half,k} &= \sum_{(l,m,n)} (F_2)_{l,m,n} \, \delta_{\hellmax}(\x_{i,j-\half,k} - \X_{l,m,n}) \, \ds_1 \, \ds_2 \, \ds_3, \label{e:discrete_force_density2} \\
  (f_3)_{i,j,k-\half} &= \sum_{(l,m,n)} (F_3)_{l,m,n} \, \delta_{\hellmax}(\x_{i,j,k-\half} - \X_{l,m,n}) \, \ds_1 \, \ds_2 \, \ds_3, \label{e:discrete_force_density3}
\end{align}
for $(i,j,k) \in \G^{\ellmax}$.  The valid values of $\f$ are set to
equal zero on all coarser levels of the grid hierarchy.  Similarly,
with $\u = (u_1,u_2,u_3)$ and with $\X = (X_1,X_2,X_3)$, we
approximate eq.~\eqref{e:interp} by
\begin{align}
  \frac{{\rm d}}{{\rm d}t} (X_1)_{l,m,n} &= \sum_{(i,j,k)\in\G^{\ellmax}} (u_1)_{i-\half,j,k} \, \delta_{\hellmax}(\x_{i-\half,j,k} - \X_{l,m,n}) \, \left(\hellmax\right)^3, \label{e:discrete_interp1} \\
  \frac{{\rm d}}{{\rm d}t} (X_2)_{l,m,n} &= \sum_{(i,j,k)\in\G^{\ellmax}} (u_2)_{i,j-\half,k} \, \delta_{\hellmax}(\x_{i,j-\half,k} - \X_{l,m,n}) \, \left(\hellmax\right)^3, \label{e:discrete_interp2} \\
  \frac{{\rm d}}{{\rm d}t} (X_3)_{l,m,n} &= \sum_{(i,j,k)\in\G^{\ellmax}} (u_3)_{i,j,k-\half} \, \delta_{\hellmax}(\x_{i,j,k-\half} - \X_{l,m,n}) \, \left(\hellmax\right)^3, \label{e:discrete_interp3}
\end{align}
in which we again consider only Cartesian grid cells on the finest
level of the hierarchical grid.  For those curvilinear mesh nodes that
are in the vicinity of physical boundaries, we use the modified
regularized delta function formulation of Griffith et
al.~\cite{GriffithLuoMcQueenPeskin09}.  This approach ensures that
force and torque are conserved during Lagrangian-Eulerian interaction,
even near $\p\Omega$.  To simplify the description of our timestepping
algorithm, we use the shorthand $\f = \cS[\X] \, \F$ and $\frac{{\rm
    d}}{{\rm d}t} \X = \cS^*[\X] \, \u$, in which the force-spreading
and velocity-interpolation operators, $\cS[\X]$ and $\cS^*[\X]$, are
implicitly defined by
eqs.~\eqref{e:discrete_force_density1}--\eqref{e:discrete_force_density3}
and \eqref{e:discrete_interp1}--\eqref{e:discrete_interp3},
respectively.

\subsection{Temporal discretization}

We employ a simple timestep-splitting scheme to discretize the
equations in time.  Briefly, during each timestep, we first solve the
fluid-structure interaction equations, treating the values of the
upstream and downstream pressure boundary conditions as fixed.  We
then update the state variables of the circulation model, treating the
fluid-structure interaction model state variables as fixed.  This
approach amounts to a first-order timestep splitting of the equations.

We discretize the fluid-structure interaction equations in time using
truncated fixed-point iteration.  We treat the linear terms in the
incompressible Navier-Stokes equations implicitly, and we treat all
other terms explicitly.  Let $\X^{n+1,k}$, $\u^{n+1,k}$, and
$p^{n+\half,k}$ denote the approximations to the values of $\X$ and
$\u$ at time $t^{n+1} = (n+1)\dt$ and to the value of $p$ at time
$t^{n+\half} = (n+\half)\dt$ obtained after $k$ steps of fixed-point
iteration, with $\X^{n+1,0} = \X^n$, $\u^{n+1,0} = \u^n$, and
$p^{n+\half,0} = p^{n-\half}$.  Letting $\X^{n+\half,k} =
\half\left(\X^n + \X^{n+1,k}\right)$ and $\u^{n+\half,k} =
\half\left(\u^n + \u^{n+1,k}\right)$, we obtain $\X^{n+1,k+1}$,
$\u^{n+1,k+1}$, and $p^{n+\half,k+1}$ by solving the linear system of
equations
\begin{align}
  \rho \left(\frac{\u^{n+1,k+1} - \u^n}{\dt} + \vec{N}^{n+\half,k}\right) &= - \Grad p^{n+\half,k+1} + \mu \Lap \u^{n+\half,k+1} + \vec{f}^{n+\half,k}, \label{e:staggered1} \\
  \Div \u^{n+1,k+1} &= 0, \label{e:staggered2} \\
  \f^{n+\half,k} &= \cS[\X^{n+\half,k}] \, \F^{n+\half,k}, \label{e:staggered3} \\
  \frac{\X^{n+1,k+1} - \X^n}{\dt} &= \cS^*[\X^{n+\half,k}] \, \u^{n+\half,k+1}, \label{e:staggered4} \\
  \F^{n+\half,k} &= \Fs[\X^{n+\half,k}] + \Fb[\X^{n+\half,k}], \label{e:staggered5}
\end{align}
in which $\vec{N}^{n+\half,k} \approx \left[\u \cdot \grad
  \u\right]^{n+\half}$ is an explicit approximation to the advection
term that uses the xsPPM7 variant \cite{RiderEtAl07} of the piecewise
parabolic method (PPM) \cite{ColellaWoodward84} to discretize the
nonlinear advection terms; see Griffith~\cite{Griffith09} for details.
We use two cycles of fixed-point iteration per timestep to obtain a
second-order accurate timestepping scheme.  When solving for
$\X^{n+1,k+1}$, $\u^{n+1,k+1}$, and $p^{n+\half,k+1}$, we fix the
pressure at the upstream boundary $\p\OmegaLV$ to be $p = \PLV(t^n)$,
and we fix the pressure at the downstream boundary $\p\OmegaAo$ to be
$p = \PAo(t^n)$.

Next, having computed $\X^{n+1}$, $\u^{n+1}$, and $p^{n+\half}$, we
use $\u^{n+1}$ to compute $\QAo\left(t^{n+1}\right)$, the
instantaneous rate of flow leaving the vessel through $\p\OmegaAo$ at
time $t^{n+1}$.  We then update the value of $\PWk$ via a second-order
accurate explicit Runge-Kutta method, so that
\begin{align}
  C \frac{\tilde{P}^{\text{Wk}}\left(t^{n+1}\right) - \PWk(t^n)}{\dt} + \frac{1}{\Rp} \PWk\left(t^{n}\right) &= \QAo\left(t^{n+1}\right), \\
  C \frac{P^{\text{Wk}}\left(t^{n+1}\right) - P^{\text{Wk}}\left(t^{n}\right)}{\dt} + \frac{1}{\Rp} \frac{\tilde{P}^{\text{Wk}}\left(t^{n+1}\right) + P^{\text{Wk}}\left(t^{n}\right)}{2} &= \QAo\left(t^{n+1}\right).
\end{align}
We then set
\begin{equation}
  \PAo\left(t^{n+1}\right) = \Rc \QAo\left(t^{n+1}\right) + \PWk\left(t^{n+1}\right),
\end{equation}
which serves as the downstream pressure boundary condition for the
fluid-structure interaction model during the subsequent timestep.  For
the initial timestep, we set $\PLV(0) = \PAo(0) = 0$, values that are
consistent with the initial conditions of the continuous system.

\subsection{Cartesian grid adaptive mesh refinement}

The locally refined grid is constructed in a recursive fashion: First,
level 0 is constructed to cover the entire physical domain $\Omega$.
Next, having constructed levels $0,\ldots,\ell < \ellmax$, level
$\ell+1$ is generated by
\begin{inparaenum}[(1)]
\item tagging cells on level $\ell$ for refinement,
\item covering the tagged level $\ell$ grid cells by rectangular boxes
  generated by the Berger-Rigoutsos point-clustering algorithm
  \cite{BergerRigoutsos91}, and
\item refining the generated boxes by the integer refinement ratio
  $\nref$ to form the level $\ell+1$ grid patches.
\end{inparaenum}
Our cell-tagging criteria are simple rules that ensure that the
immersed structure remains covered throughout the simulation by the
grid cells that comprise the finest level of the hierarchical grid,
and that attempt to ensure that flow features requiring enhanced
resolution, such as vortices shed from the free edges of the valve
leaflets, remain covered by grid cells of an appropriate resolution.
Specifically, we tag grid cell $(i,j,k)$ on level $\ell < \ellmax$ for
refinement whenever there exists some curvilinear mesh node $(l,m,n)$
such that $\X_{l,m,n} \in \c_{i,j,k}^{\ell}$, or whenever the local
magnitude of the vorticity $\|\vec{\omega}\|_{i,j,k} = \|\grad_h
\times \u\|_{i,j,k}$ exceeds a relative threshold.  Additional cells
are added to the finest level of the grid to ensure that the
coarse-fine interface between levels $\ellmax-1$ and $\ellmax$ is
sufficiently far away from each of the curvilinear mesh nodes to
prevent complicating the discretization of the Lagrangian-Eulerian
interaction equations, as discussed previously.  We emphasize that the
positions of the curvilinear mesh nodes are not constrained to conform
in any way to the locally refined Cartesian grid.  Instead, the
Cartesian grid patch hierarchy is adaptively updated to conform to the
evolving configuration of the immersed elastic structure.

To prevent the immersed structure from ``escaping'' from the finest
level of the grid, it is necessary to regenerate the locally refined
grid at an interval that is determined by the CFL number of the flow.
In our simulations, we choose the timestep size $\dt$ to satisfy a CFL
condition of the form
\begin{equation}
  \dt \leq \frac{1}{5} \min_{0 \leq \ell \leq \ellmax} \min_{(i,j,k) \in \G^{\ell}} \frac{\hell}{\|\u(\x_{i,j,k})\|_{\infty}}. \label{e:cfl_condition}
\end{equation}
This condition implies that each curvilinear mesh point moves at most
$\frac{1}{5}$ fractional meshwidths per timestep.  Therefore, to
ensure that the immersed structure remains covered by
$\Omega^{\ellmax}$, we must adaptively regenerate the grid hierarchy
at least every five timesteps.  In our simulations, we actually
regenerate the grid every four timesteps.  In practice, we could
generally postpone regridding because the actual timestep size is
generally smaller than that required by eq.~\eqref{e:cfl_condition} as
a consequence of an additional stability restriction on $\dt$ that is
of the form $\dt = O\left(\left(\hellmax\right)^4\right)$.  This
severe stability restriction results from our time-explicit treatment
of the bending-resistant elastic force.  For a model that includes
only extension-~and compression-resistant elastic elements, the
stability restriction is reduced to $\dt =
O\left(\left(\hellmax\right)^2\right)$.

Each time that the locally refined Cartesian grid is regenerated,
Eulerian quantities must be transferred from the old grid hierarchy to
the new one.  In newly refined regions of the physical domain, the
velocity field is prolonged from coarser levels of the old grid via a
specialized conservative interpolation scheme that preserves the
discrete divergence and curl of the staggered-grid velocity field
\cite{TothRoe02}.  (The basic divergence-~and curl-preserving
interpolation scheme \cite{TothRoe02} considers only the case in which
$\nref=2$; however, this procedure is easily generalized to cases in
which $\nref$ is a power of two via recursion.)  The pressure, which
is not a state variable of the system and which is used in the
subsequent timestep only as an initial approximation to the updated
pressure computed during that timestep (see below), is prolonged by
simple linear interpolation.  In newly coarsened regions of the
domain, the values of the velocity and pressure are set to be the
averages of the overlying fine-grid values from the old grid
hierarchy.

\section{Solution methodology}
\label{s:solution_methodology}

Solving for $\X^{n+1,k+1}$, $\vec{u}^{n+1,k+1}$, and $p^{n+\half,k+1}$
in eqs.~\eqref{e:staggered1}--\eqref{e:staggered5} requires the
solution of the linear system of equations,
\begin{equation}
  \left(
    \begin{array}{cc}
      \frac{\rho}{\dt} I - \frac{\mu}{2} \Lap & \Grad \\
      -\Div & 0
    \end{array}
  \right)
  \left(
    \begin{array}{c}
      \u^{n+1,k+1} \\
      p^{n+\half,k+1}
    \end{array}
  \right)
  =
  \left(
    \begin{array}{c}
      \left(\frac{\rho}{\dt} I + \frac{\mu}{2} \Lap\right) \u^{n} - \rho \vec{N}^{n+\half,k} + \f^{n+\half,k} \\
      0
    \end{array}
  \right)
  \label{e:linear_system}
\end{equation}
We solve eq.~\eqref{e:linear_system} via the FGMRES algorithm
\cite{Saad93}, using $\vec{u}^{n+1,k}$ and $p^{n+\half,k}$ as initial
approximations to $\vec{u}^{n+1,k+1}$ and $p^{n+\half,k+1}$, and using
the projection method as a preconditioner.  (Recall that we define
$\vec{u}^{n+1,0} = \vec{u}^{n}$ and $p^{n+\half,0} = p^{n-\half}$.)
Letting $A$ denote the matrix corresponding to this block system,
i.e.,
\begin{equation}
  A =
  \left(
    \begin{array}{cc}
      \frac{\rho}{\dt} I - \frac{\mu}{2} \Lap & \Grad \\
      -\Div & 0
    \end{array}
  \right)
\end{equation}
we write the corresponding projection method-based preconditioner
matrix $B$ as \cite{Griffith09}
\begin{align}
  B = &
  \left(
    \begin{array}{cc}
      I & -\frac{\dt}{\rho} \Grad \\
      0 & I - \frac{\dt}{\rho} \frac{\mu}{2} \Div \Grad
    \end{array}
  \right)
  \left(
    \begin{array}{cc}
      I & 0 \\
      0 & (\Div \Grad)^{-1}
    \end{array}
  \right) \cdot \mbox{} \nonumber \\ &
  \left(
    \begin{array}{cc}
      I & 0 \\
      -\frac{\rho}{\dt} \Div & -\frac{\rho}{\dt} I
    \end{array}
  \right)
  \left(
    \begin{array}{cc}
      \left(\frac{\rho}{\dt}I - \frac{\mu}{2}\Lap\right)^{-1} & 0 \\
      0 & I
    \end{array}
  \right)
\end{align}
Because we use a preconditioned Krylov method to solve for
$\vec{u}^{n+1,k+1}$ and $p^{n+\half,k+1}$, it is not necessary to form
explicitly the matrices corresponding to $A$ and $B$.  Instead, we
need only to be able to compute the application of these operators to
arbitrary velocity-~and pressure-like quantities.  Computing the
action of $A$ requires implementations of the finite difference
approximations to the divergence, gradient, and Laplace operators
described previously.  Computing the action of $B$ additionally
requires solvers for cell-centered and face-centered Poisson-type
problems.  It is not necessary, however, to employ exact solvers for
these subdomain problems.  In fact, these subdomain solvers may be
quite approximate.  At least in the present application, we have found
that the performance of our implementation is optimized by using a
single multigrid V-cycle of a cell-centered FAC preconditioner for the
pressure subsystem, corresponding to the block $(\Div \Grad)^{-1}$ of
the preconditioner $B$, and by applying two iterations of the
conjugate gradient method for the velocity subsystem, corresponding to
the block $\left(\frac{\rho}{\dt}I - \frac{\mu}{2}\Lap\right)^{-1}$ of
the preconditioner.  We are able to avoid using a multigrid method for
the velocity subsystem because the Reynolds number of the flow is
relatively large and the timestep size is relatively small, so that
the linear system $\left(\frac{\rho}{\dt}I - \frac{\mu}{2}\Lap\right)$
is well conditioned.

We remark that although the projection method can be an extremely
effective preconditioner \cite{Griffith09}, it does not appear to be
widely used in this manner in practice.  Instead, it is generally used
as an approximate solver for the incompressible Navier-Stokes
equations
\cite{MartinColella00,MartinColellaGraves08,BellColellaGlaz89,AlmgrenBellEtAl98,BrownCortezMinion01}.
As a solver for the incompressible Navier-Stokes equations, the
projection method is a fractional-step scheme that first solves the
momentum equation over a time interval $[t^{n},t^{n+1}]$ for an
``intermediate'' velocity field without imposing the constraint of
incompressibility, and then projects that intermediate velocity field
onto the space of discretely divergence-free vector fields to obtain
an approximation to the incompressible velocity field at time
$t^{n+1}$.  These two steps correspond to the two subdomain solves of
our projection preconditioner, and each step requires the imposition
of ``artificial'' physical boundary conditions.  When the projection
method is used as a solver, the artificial boundary conditions must be
chosen carefully to yield a stable and accurate approximation to the
true boundary conditions that are to be imposed on the coupled
equations \cite{BrownCortezMinion01,YangProsperetti06}.  Obtaining
high-order accuracy may not be possible in all cases, such as for
problems involving outflow boundaries \cite{GuermondMinevShen06}, and
constructing discretizations that are both stable and accurate can be
difficult in practice \cite{GuyFogelson05}.

We prefer to use the projection method as a preconditioner rather than
as a solver.  First, doing so permits the Krylov method to eliminate
the timestep-splitting error of the basic projection method.  Second,
solving eq.~\eqref{e:linear_system}, an unsplit discretization of the
incompressible Stokes equations, permits us to impose directly the
true boundary conditions on $\u$ and $p$ in a coupled manner.
Specifically, the form of the artificial boundary conditions required
of the basic projection method does \emph{not} affect the accuracy of
the overall solver.  This greatly simplifies the development of
higher-order accurate discretiations of various types of boundary
conditions.  See Griffith \cite{Griffith09} for details on the
construction of accurate discretizations of various combinations of
normal and tangential velocity and traction boundary conditions.

\section{Implementation}
\label{s:implementation}

Our adaptive IB method is implemented in the IBAMR software framework,
a freely available C++ library for developing fluid-structure
interaction models that use the IB method \cite{IBAMR-web-page}.
IBAMR provides support for distributed-memory parallelism via MPI and
Cartesian grid adaptive mesh refinement.  IBAMR relies upon the SAMRAI
\cite{samrai-web-page,HornungKohn02,HornungWissinkKohn06}, PETSc
\cite{petsc-web-page,petsc-user-ref,petsc-efficient}, and \emph{hypre}
\cite{hypre-web-page,FalgoutYang02} libraries for much of its
functionality.

\section{Computational results}
\label{s:computational_results}

\subsection{Model results}

\begin{figure}
  \centering
  \begin{tabular}{lcclc}
    \vspace{-10pt} {\bf A.} & & & {\bf B.} & \\
    & \includegraphics[width=0.35\textwidth]{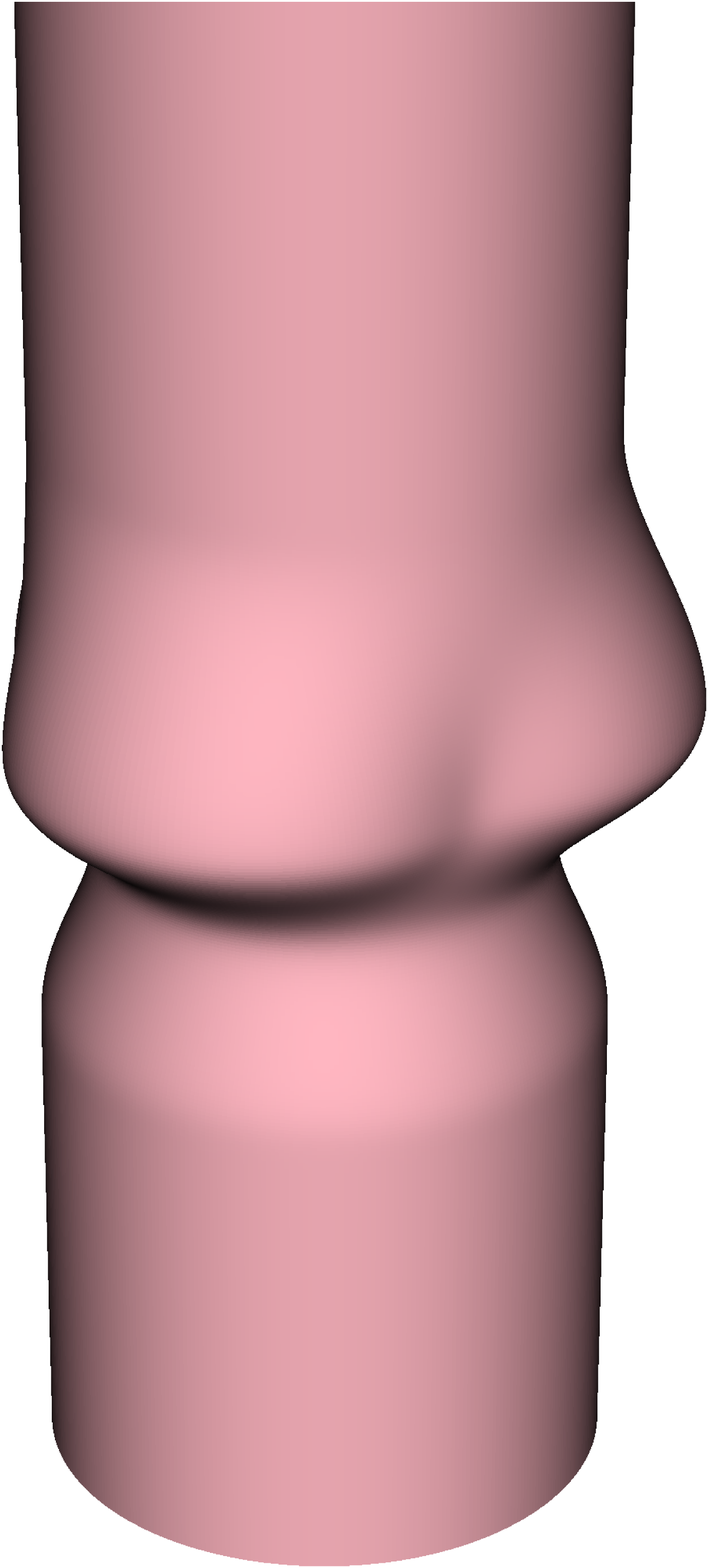} & \hspace{24pt} &
    & \includegraphics[width=0.35\textwidth]{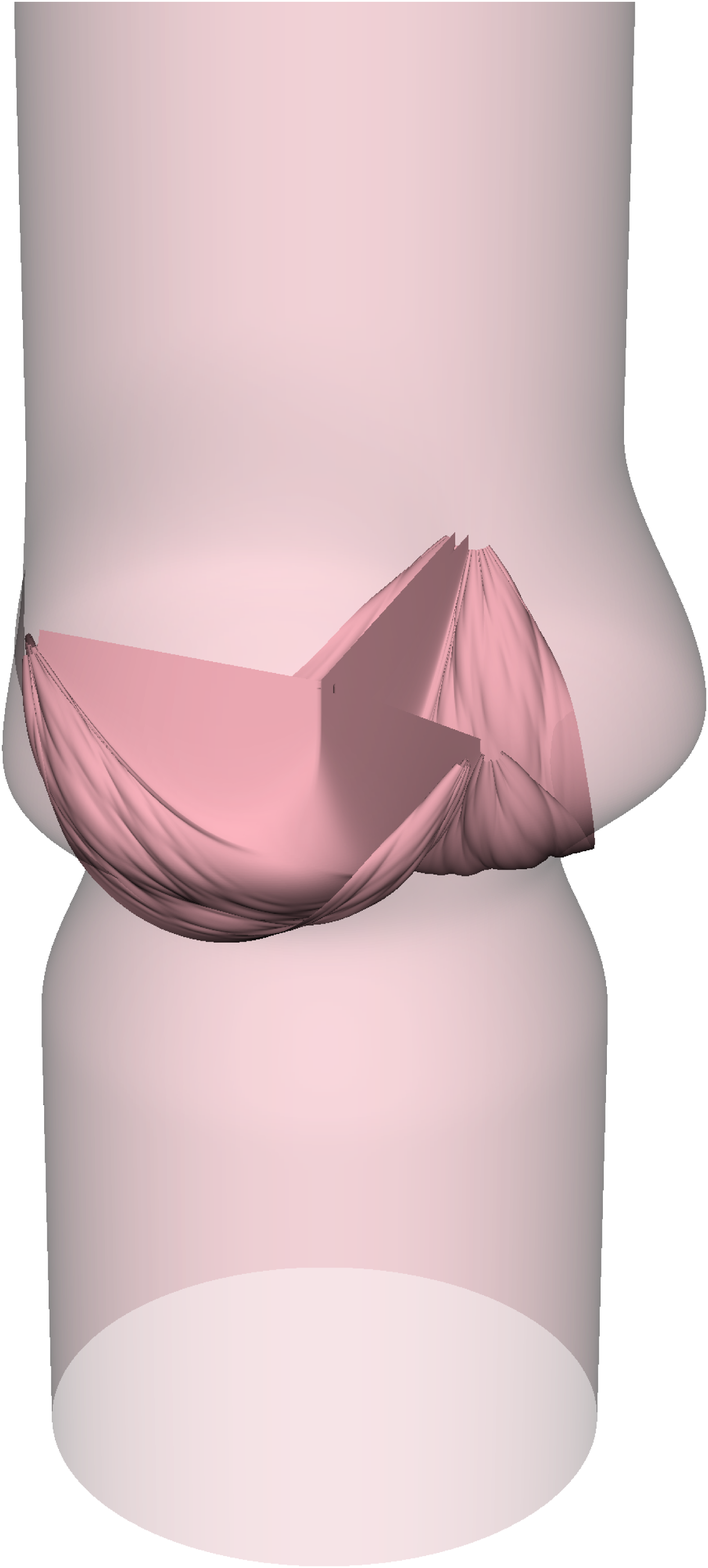}
  \end{tabular}
  \caption{The initial configuration of the model aortic root and
    valve leaflets.  In this and subsequent figures, the upstream
    boundary $\p\OmegaLV$ is generally located at the bottom of each
    panel, and the downstream boundary $\p\OmegaAo$ is generally
    located at the top of each panel. {\bf A.}~The geometry of the
    model aortic root is based on the idealized description of Reul et
    al.~\cite{Reul90}, using dimensions based on measurements by
    Swanson and Clark \cite{SwansonClark74}.  {\bf B.}~The geometry
    and fiber architecture of the model aortic valve leaflets are
    determined by the mathematical theory of Peskin and McQueen
    \cite{PeskinMcQueen94}.}
  \label{f:valve_model}
\end{figure}

\begin{figure}
  \centering
  \begin{tabular}{lccc}
    \vspace{-10pt} {\bf A.} & & & \\
    \vspace{3pt}
    & \includegraphics[height=135pt]{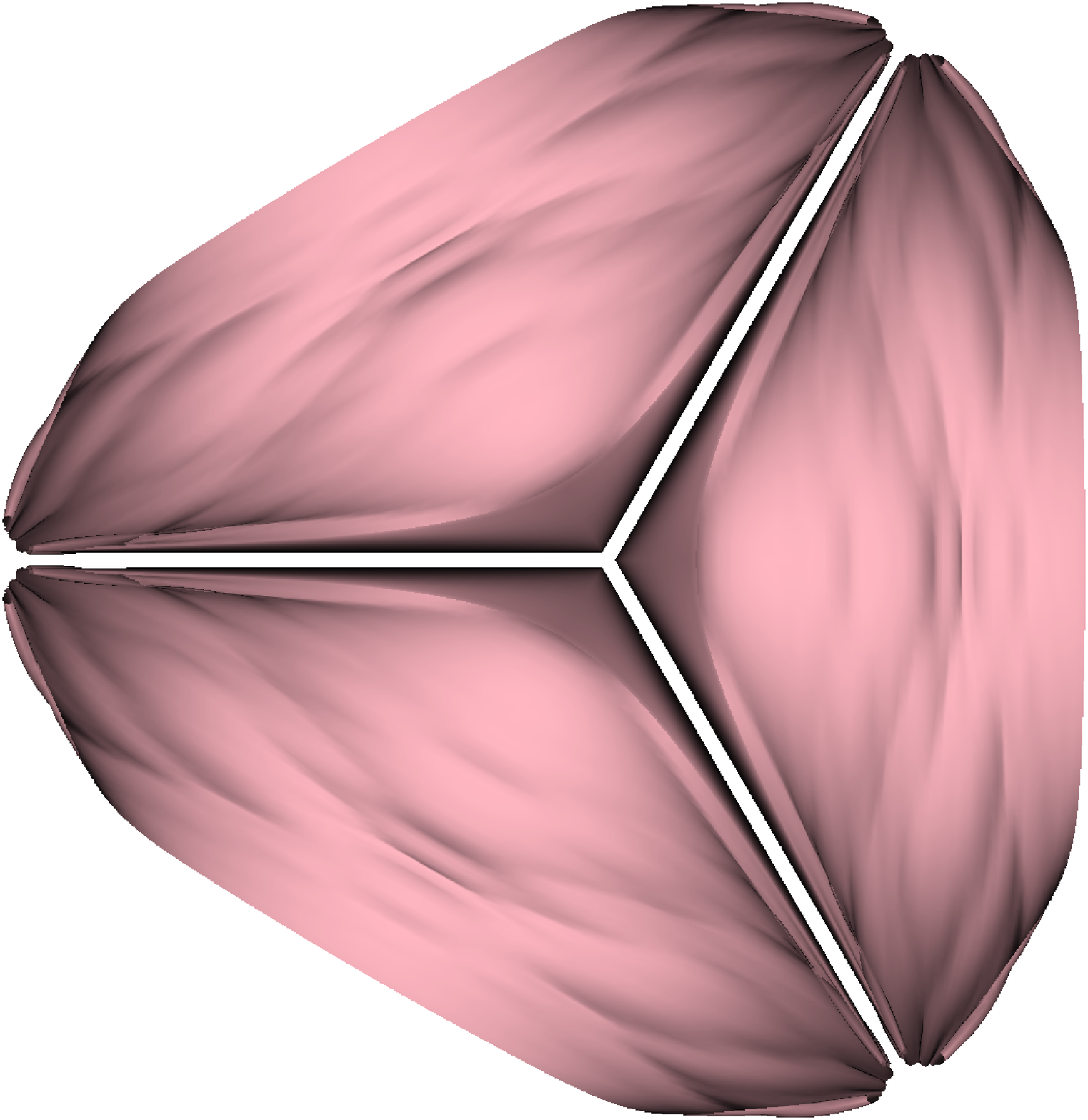} & \hspace{24pt}
    & \includegraphics[height=135pt]{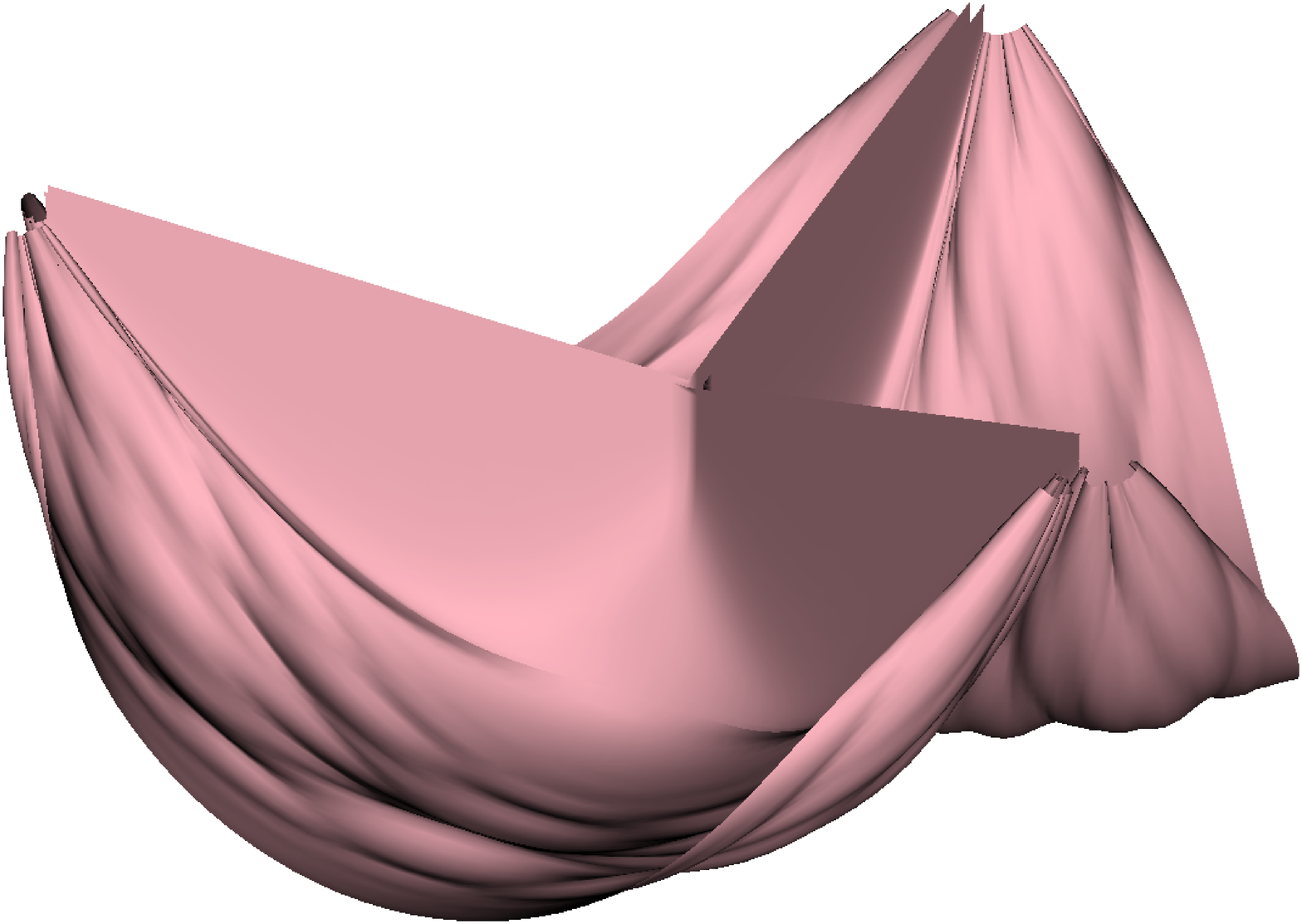} \\
    \vspace{-10pt} {\bf B.} & & & \\
    & \includegraphics[height=135pt]{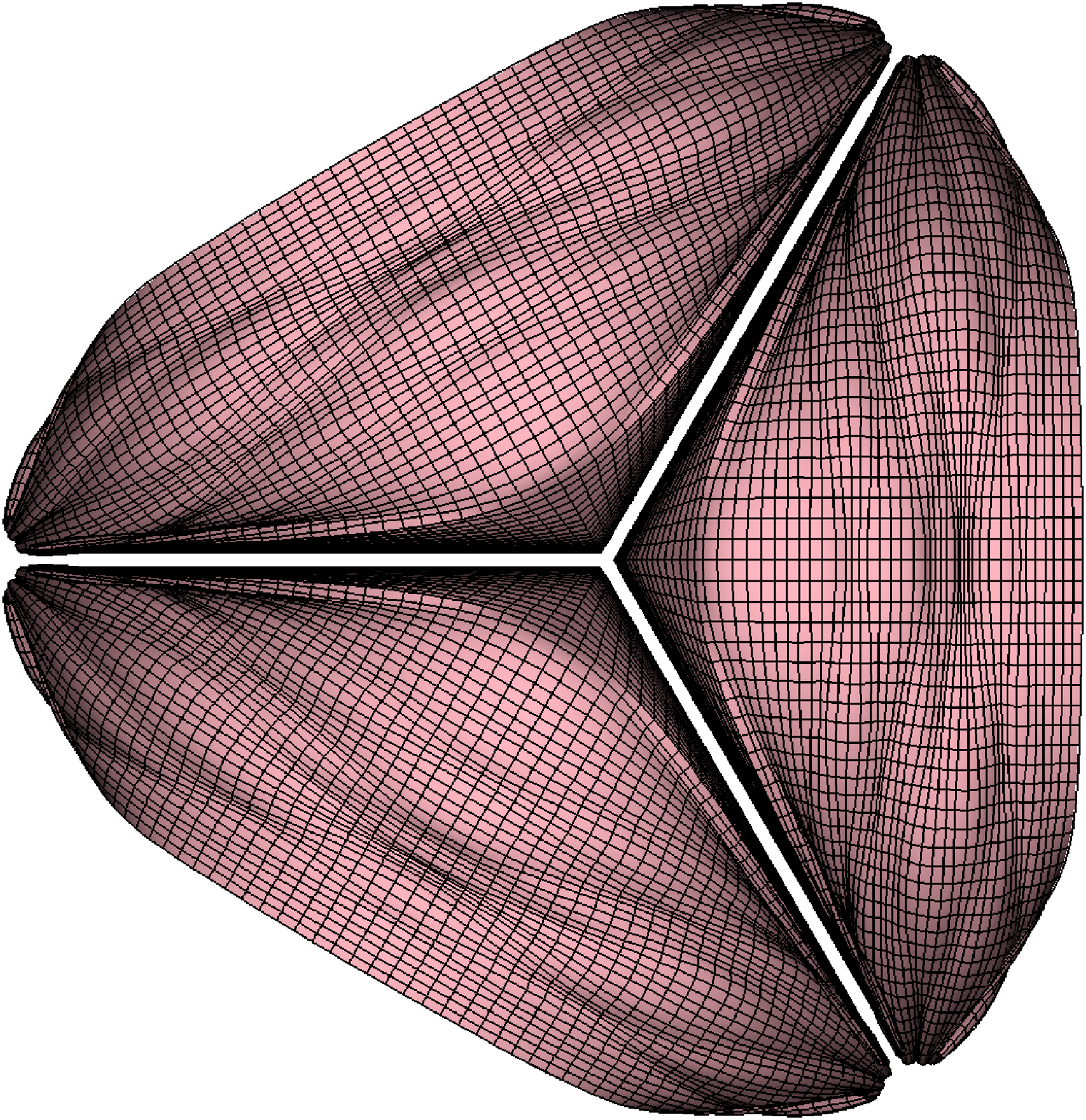} & \hspace{24pt}
    & \includegraphics[height=135pt]{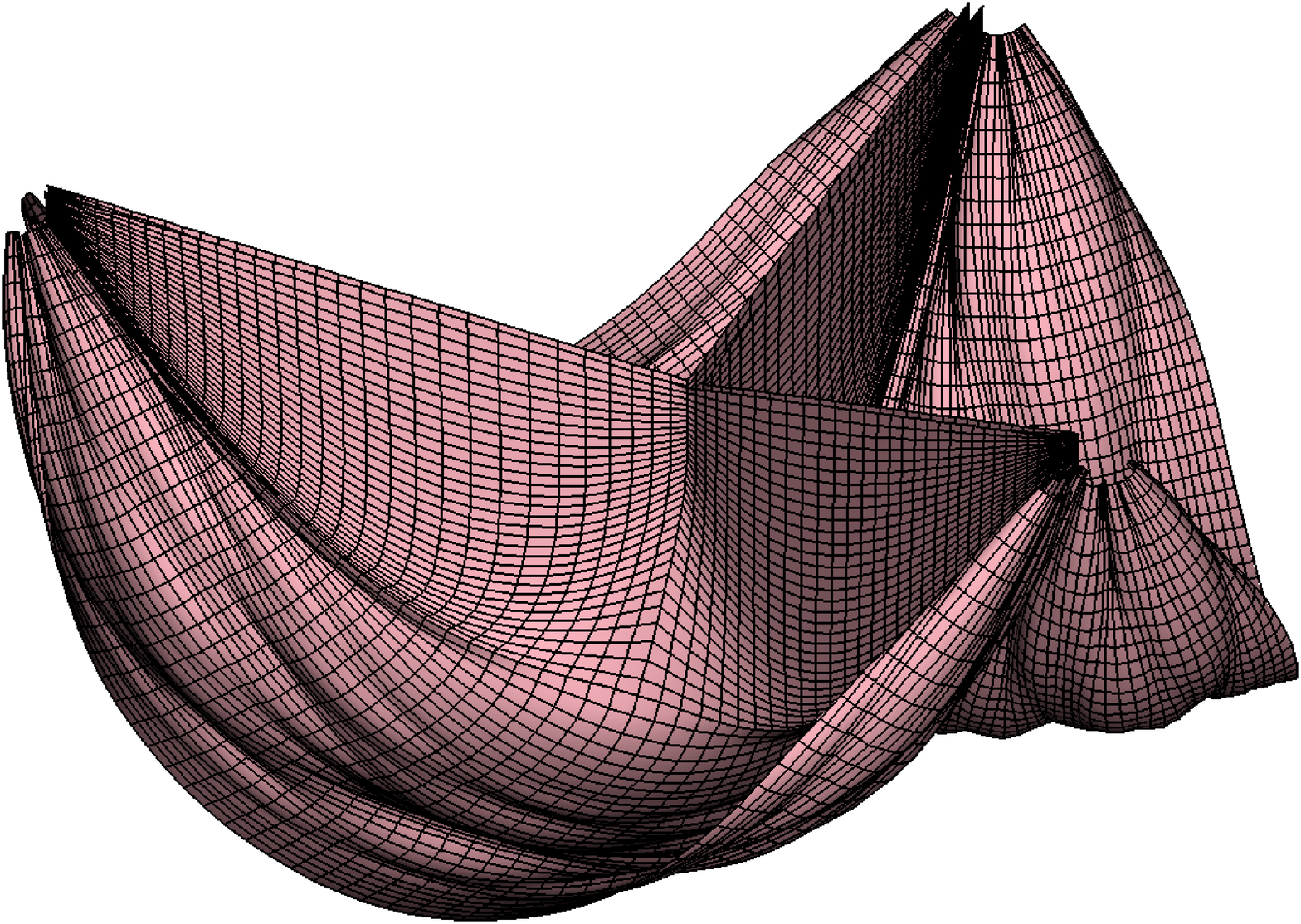}
  \end{tabular}
  \caption{The model valve leaflets in their initial, loaded
    configuration.  {\bf A.}~Surface renderings of the valve leaflets.
    {\bf B.}~Similar to A, but also showing a subset of the model
    fibers that comprise the leaflets.}
  \label{f:valve_fibers}
\end{figure}

Using the methods described herein, we have simulated the fluid
dynamics of the aortic valve over multiple cardiac cycles, using a
time-periodic left-ventricular driving pressure and dynamic loading
conditions.  In these simulations, the physical domain $\Omega$ is a
$\mbox{10 cm} \times \mbox{10 cm} \times \mbox{15 cm}$ rectangular box
that we discretize using a two-level adaptively refined Cartesian grid
with refinement ratio $\nref=4$ between grid levels.  The coarse-grid
resolution is $h^0 = \mbox{3.125 mm}$, which corresponds to that of a
$32 \times 32 \times 48$ uniform discretization of $\Omega$, and the
fine-grid resolution is $h^1 = \hellmax = \mbox{0.78125 mm}$, which
corresponds to that of a $128 \times 128 \times 192$ uniform
discretization of $\Omega$.  We model the valve leaflets as thin
elastic surfaces because real aortic valve leaflets are only
$\mbox{0.2 mm}$ thick \cite{ClarkFinke74}, which is somewhat thinner
than the Cartesian grid spacing on the finest level of the locally
refined Cartesian grid.  In contrast, we model the vessel as a thick
elastic structure because the thickness of the aortic wall is $\mbox{1
  mm}$ \cite{ShunkEtAl01}, which is relatively thick at the scale of
the Cartesian grid.  In our simulations, we use a uniform timestep
size of $\dt = \mbox{6.25 $\mu$s}$, thereby requiring 128,000
timesteps per $\mbox{0.8 s}$ cardiac cycle.  This value of $\dt$ was
empirically determined to be approximately the largest stable timestep
permitted by the present model and semi-explicit numerical method.

As discussed earlier, the elasticity of the valve leaflets and vessel
wall is modeled using systems of fibers.  In our model, each valve
leaflet is spanned by two families of fibers.  The first family of
fibers runs from commissure to commissure, and the second family runs
orthogonal to the commissural fibers.  We view the commissural fibers
as corresponding to the collagen fibers that allow the real valve
leaflets to support a significant pressure load, and we use the
mathematical theory of Peskin and McQueen \cite{PeskinMcQueen94} to
determine the architecture of these fibers.  We remark that the
construction of Peskin and McQueen yields discretized fibers that form
an orthogonal net, thereby facilitating the construction of the second
family of fibers.  Because this mathematical theory of the valve fiber
architecture describes the geometry of the closed, loaded valve, we
assume that the commissural fibers are initialized in a condition of
10\% strain, and we choose the commissural fiber stiffness so that the
closed valve supports a pressure load of approximately $\mbox{80
  mmHg}$.  The second family of fibers is 10\% as stiff as the
commissural fibers, thereby approximating the anisotropic material
properties of real aortic valve leaflets \cite{Sauren81}.  Notice that
although the valve leaflets are initialized in a strained
configuration, the boundary conditions at time $t=0$ do not provide
the closed valve with any pressure load.  This initial strain thereby
induces oscillations in the valve leaflets, as if the valve had been
struck like a drumhead at time $t = 0$.  These oscillations are
rapidly damped by the viscosity of the fluid but nonetheless render
the results of the first simulated beat somewhat atypical.  The
initial configuration of the model vessel and valve leaflets are shown
in figs.~\ref{f:valve_model} and \ref{f:valve_fibers}.

The geometry of the aortic root and ascending aorta is based on the
geometric description of Reul et al.~\cite{Reul90}, and the dimensions
of the model vessel are based on measurements by Swanson and Clark of
human aortic roots harvested post autopsy that were pressurized to
120~mmHg \cite{SwansonClark74}.  The stiffnesses of the fibers that
comprise the vessel wall are empirically determined to keep the vessel
essentially fixed in place.  Our model therefore neglects the
significant compliance of the real aortic root, which increases in
volume by approximately 35\% during ejection \cite{Lansac02}.  The
incorporation of a realistic description of the elasticity of the
aortic root and ascending aorta into our fluid-structure interaction
model remains important future work.

\begin{figure}
  \centering
  \begin{tabular}{lc}
    \vspace{-15.5pt} {\bf A.} & \\
    \vspace{3pt} & 
%
%
\begin{psfrags}%
\psfragscanon%
%
\psfrag{s03}[t][t]{\color[rgb]{0,0,0}\setlength{\tabcolsep}{0pt}\begin{tabular}{c}time (s)\end{tabular}}%
\psfrag{s04}[b][b]{\color[rgb]{0,0,0}\setlength{\tabcolsep}{0pt}\begin{tabular}{c}flow rate (ml/s)\end{tabular}}%
%
\psfrag{x01}[t][t]{0}%
\psfrag{x02}[t][t]{0.5}%
\psfrag{x03}[t][t]{1}%
\psfrag{x04}[t][t]{1.5}%
\psfrag{x05}[t][t]{2}%
\psfrag{x06}[t][t]{2.5}%
%
\psfrag{v01}[r][r]{-200}%
\psfrag{v02}[r][r]{-100}%
\psfrag{v03}[r][r]{0}%
\psfrag{v04}[r][r]{100}%
\psfrag{v05}[r][r]{200}%
\psfrag{v06}[r][r]{300}%
\psfrag{v07}[r][r]{400}%
\psfrag{v08}[r][r]{500}%
%
\resizebox{8cm}{!}{\includegraphics{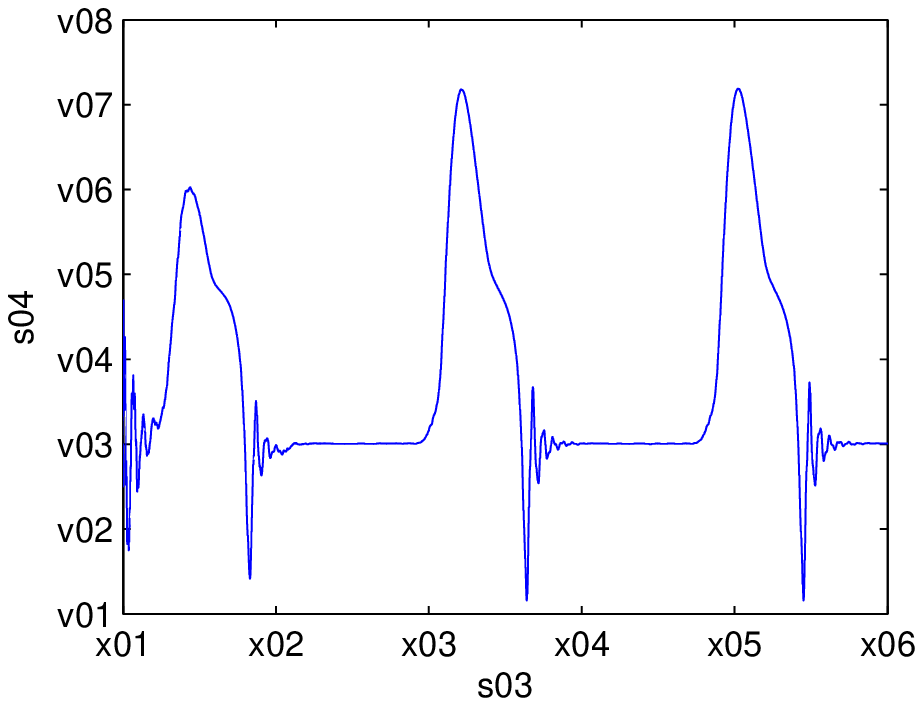}}%
\end{psfrags}%
%
 \\
    \vspace{-15.5pt} {\bf B.} & \\
    \vspace{3pt} & 
%
%
\begin{psfrags}%
\psfragscanon%
%
\psfrag{s05}[t][t]{\color[rgb]{0,0,0}\setlength{\tabcolsep}{0pt}\begin{tabular}{c}time (s)\end{tabular}}%
\psfrag{s06}[b][b]{\color[rgb]{0,0,0}\setlength{\tabcolsep}{0pt}\begin{tabular}{c}pressure (mmHg)\end{tabular}}%
\psfrag{s10}[][]{\color[rgb]{0,0,0}\setlength{\tabcolsep}{0pt}\begin{tabular}{c} \end{tabular}}%
\psfrag{s11}[][]{\color[rgb]{0,0,0}\setlength{\tabcolsep}{0pt}\begin{tabular}{c} \end{tabular}}%
\psfrag{s12}[l][l]{\color[rgb]{0,0,0}$\PAo$}%
\psfrag{s13}[l][l]{\color[rgb]{0,0,0}$\PLV$}%
\psfrag{s14}[l][l]{\color[rgb]{0,0,0}$\PAo$}%
%
\psfrag{x01}[t][t]{0}%
\psfrag{x02}[t][t]{0.5}%
\psfrag{x03}[t][t]{1}%
\psfrag{x04}[t][t]{1.5}%
\psfrag{x05}[t][t]{2}%
\psfrag{x06}[t][t]{2.5}%
%
\psfrag{v01}[r][r]{0}%
\psfrag{v02}[r][r]{20}%
\psfrag{v03}[r][r]{40}%
\psfrag{v04}[r][r]{60}%
\psfrag{v05}[r][r]{80}%
\psfrag{v06}[r][r]{100}%
\psfrag{v07}[r][r]{120}%
%
\resizebox{8cm}{!}{\includegraphics{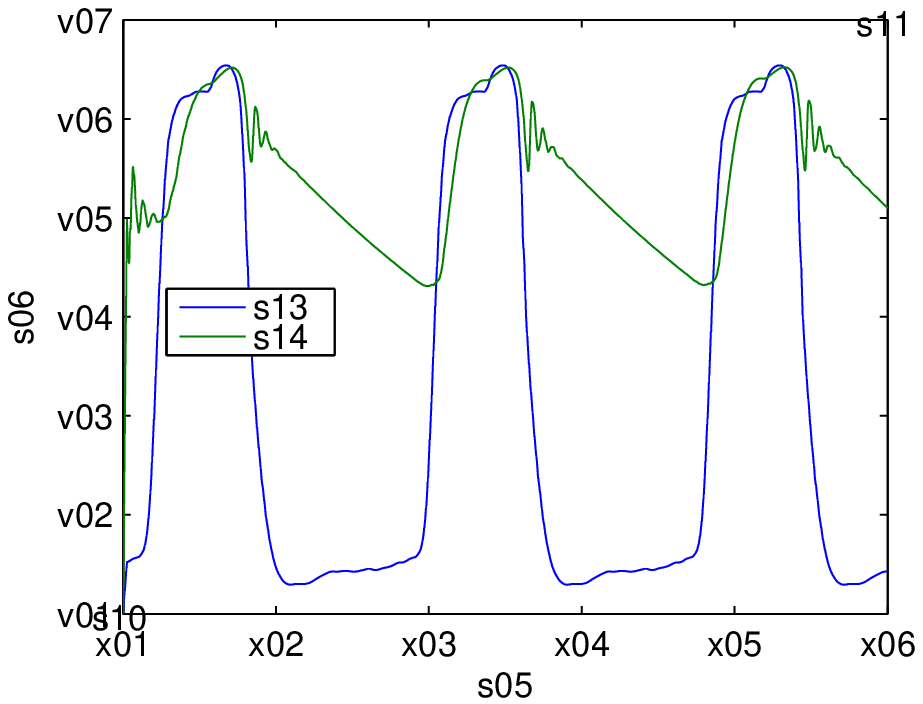}}%
\end{psfrags}%
%
 \\
    \vspace{-15.5pt} {\bf C.} & \\
    \vspace{3pt} & 
%
%
\begin{psfrags}%
\psfragscanon%
%
\psfrag{s05}[t][t]{\color[rgb]{0,0,0}\setlength{\tabcolsep}{0pt}\begin{tabular}{c}time (s)\end{tabular}}%
\psfrag{s06}[b][b]{\color[rgb]{0,0,0}\setlength{\tabcolsep}{0pt}\begin{tabular}{c}pressure (mmHg)\end{tabular}}%
\psfrag{s10}[][]{\color[rgb]{0,0,0}\setlength{\tabcolsep}{0pt}\begin{tabular}{c} \end{tabular}}%
\psfrag{s11}[][]{\color[rgb]{0,0,0}\setlength{\tabcolsep}{0pt}\begin{tabular}{c} \end{tabular}}%
\psfrag{s12}[l][l]{\color[rgb]{0,0,0}$\PWk$}%
\psfrag{s13}[l][l]{\color[rgb]{0,0,0}$\PAo$}%
\psfrag{s14}[l][l]{\color[rgb]{0,0,0}$\PWk$}%
%
\psfrag{x01}[t][t]{0}%
\psfrag{x02}[t][t]{0.5}%
\psfrag{x03}[t][t]{1}%
\psfrag{x04}[t][t]{1.5}%
\psfrag{x05}[t][t]{2}%
\psfrag{x06}[t][t]{2.5}%
%
\psfrag{v01}[r][r]{60}%
\psfrag{v02}[r][r]{70}%
\psfrag{v03}[r][r]{80}%
\psfrag{v04}[r][r]{90}%
\psfrag{v05}[r][r]{100}%
\psfrag{v06}[r][r]{110}%
\psfrag{v07}[r][r]{120}%
%
\resizebox{8cm}{!}{\includegraphics{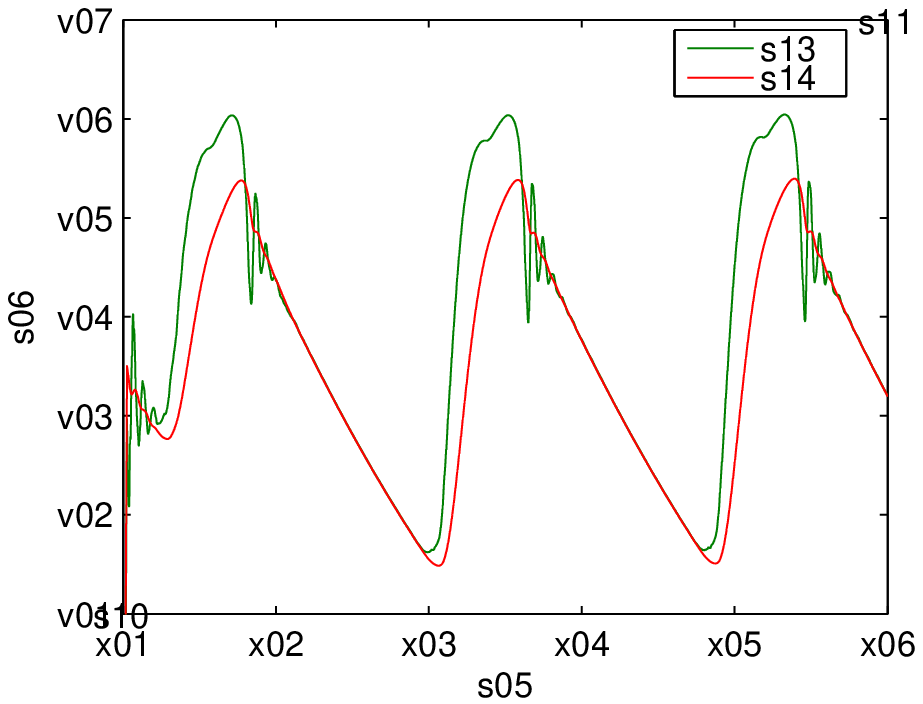}}%
\end{psfrags}%
%

  \end{tabular}
  \caption{Results from a multibeat simulation of aortic valve
    dynamics.  The familiar S$_{\text{2}}$ (``dup'') heart sound is
    clearly visible in the flow trace (panel A) and the aortic loading
    pressure (panels B and C).  Notice that the model, which is driven
    by a periodic left-ventricular pressure waveform, rapidly
    approaches a periodic steady state.  {\bf A.}~The flow rate
    through the valve as a function of time.  Mean stroke volume is
    approximately $\mbox{65 ml}$ for both the second and the third
    beat, and the peak flow rate is approximately $\mbox{420 ml
      s$^{-1}$}$.  We emphasize that the flow rate is \emph{not}
    specified in the model; rather, it \emph{emerges} from the
    fluid-structure interaction simulation.  {\bf B.}~The prescribed
    left-ventricular driving pressure $\PLV(t)$ and the computed
    loading aortic pressure $\PAo(t)$.  {\bf C.}~The pressures
    $\PAo(t)$ and $\PWk(t)$ determined by the coupled Windkessel
    model.}
  \label{f:bulk_flow_properties}
\end{figure}

\begin{figure}
  \centering
  \begin{tabular}{lrcccc}
    \vspace{-10pt} {\bf A.} & & & & & \\
    & \includegraphics[width=60.5pt]{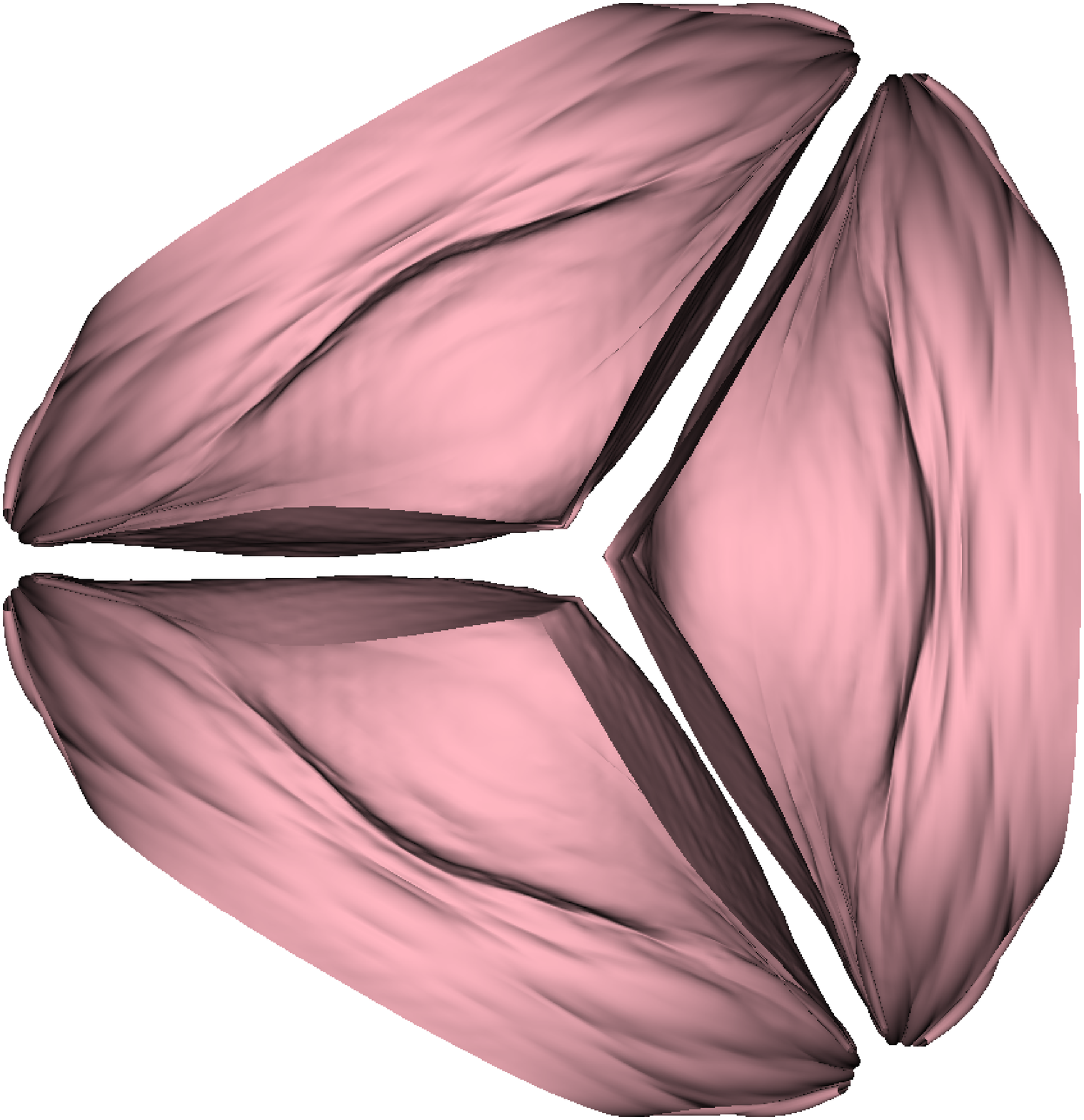}
    & \includegraphics[width=60.5pt]{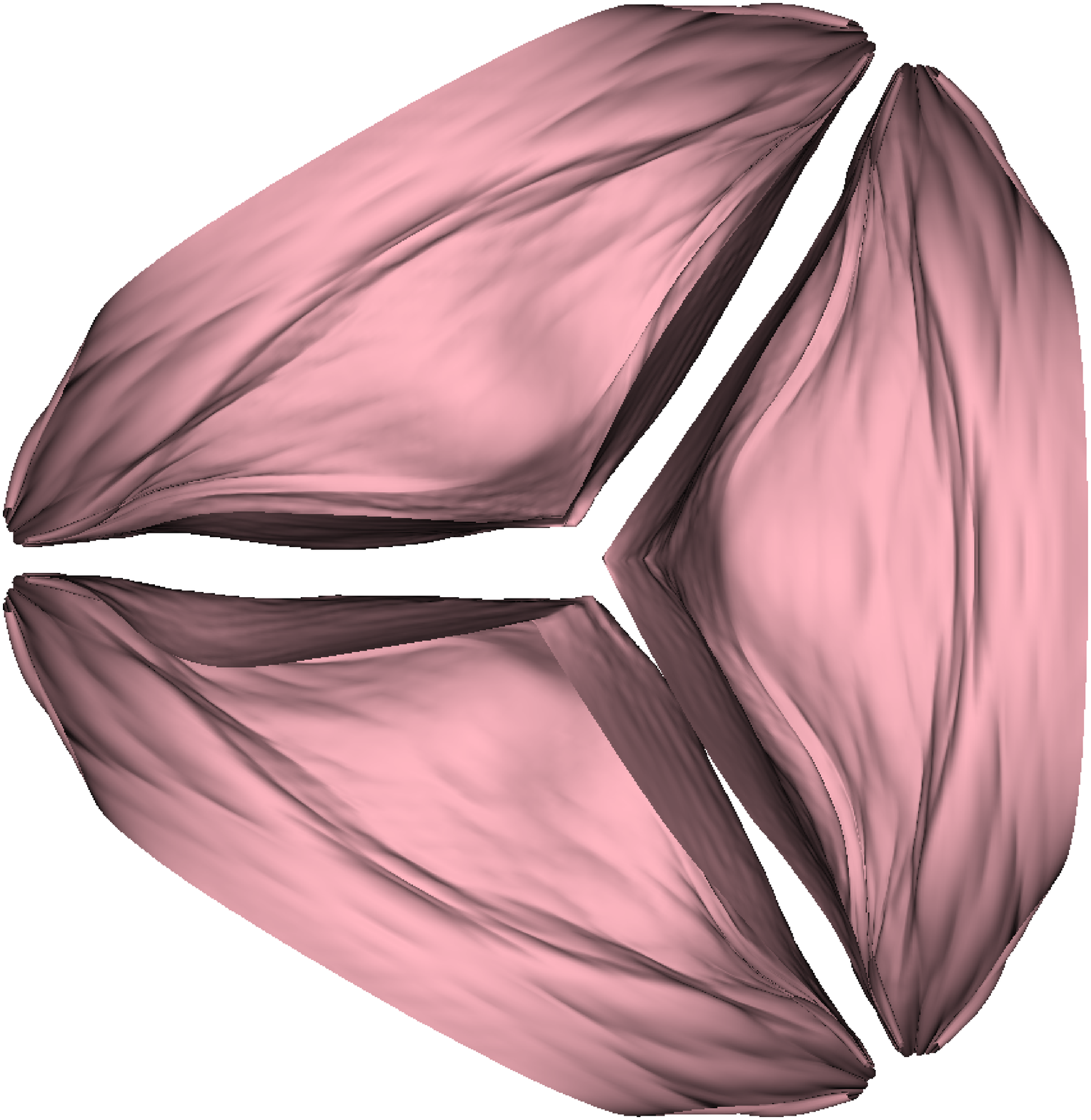}
    & \includegraphics[width=60.5pt]{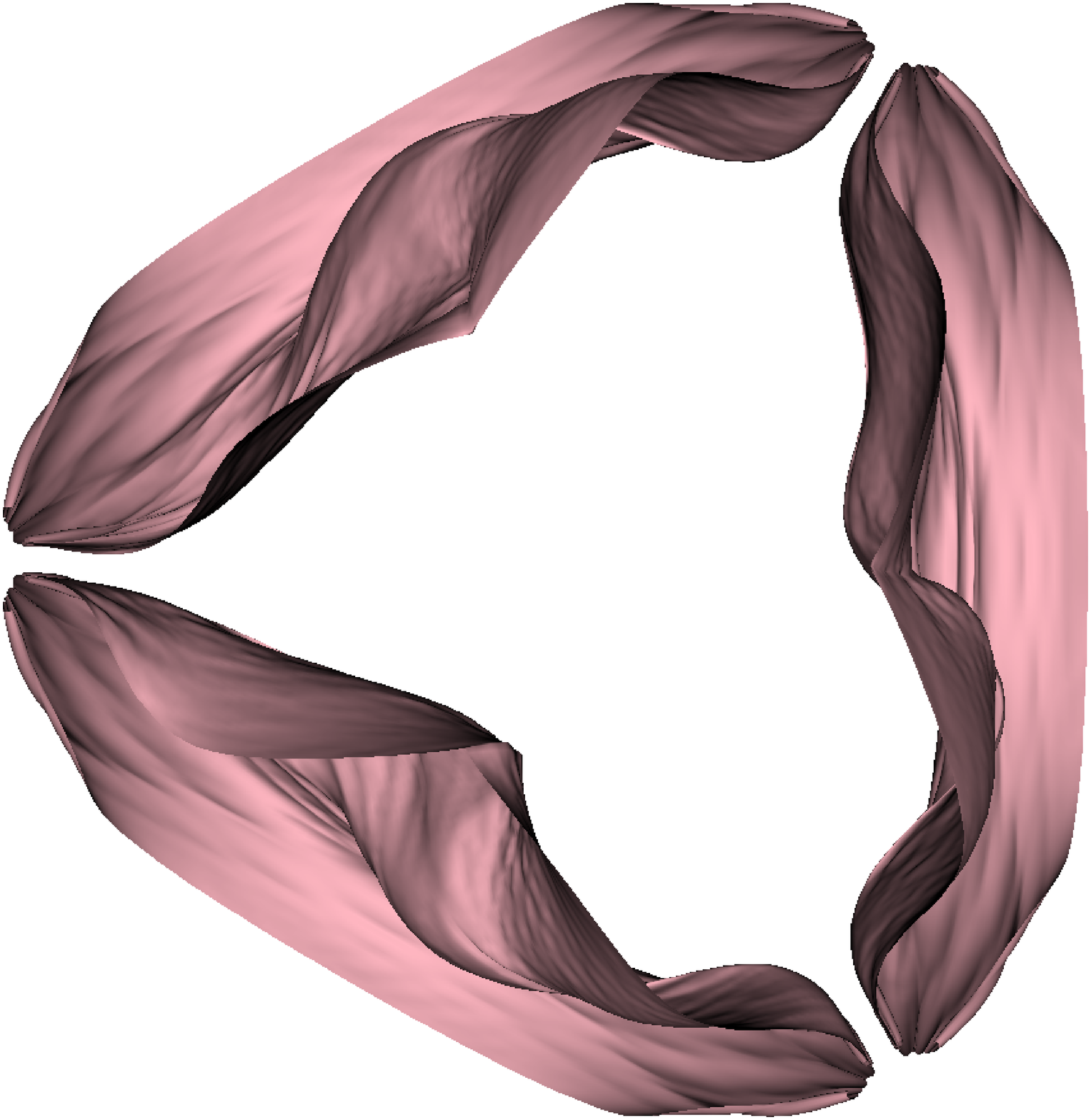}
    & \includegraphics[width=60.5pt]{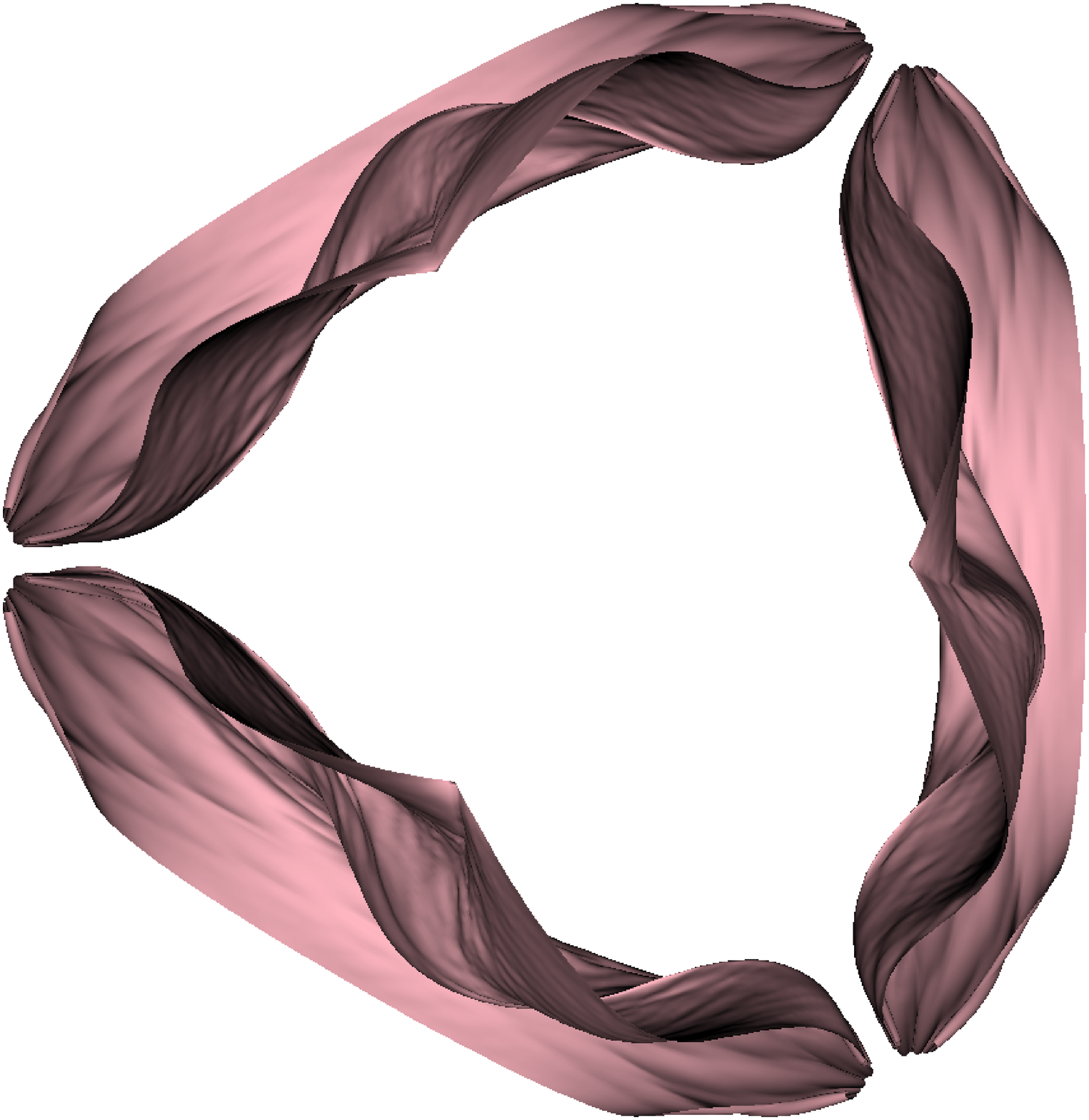}
    & \includegraphics[width=60.5pt]{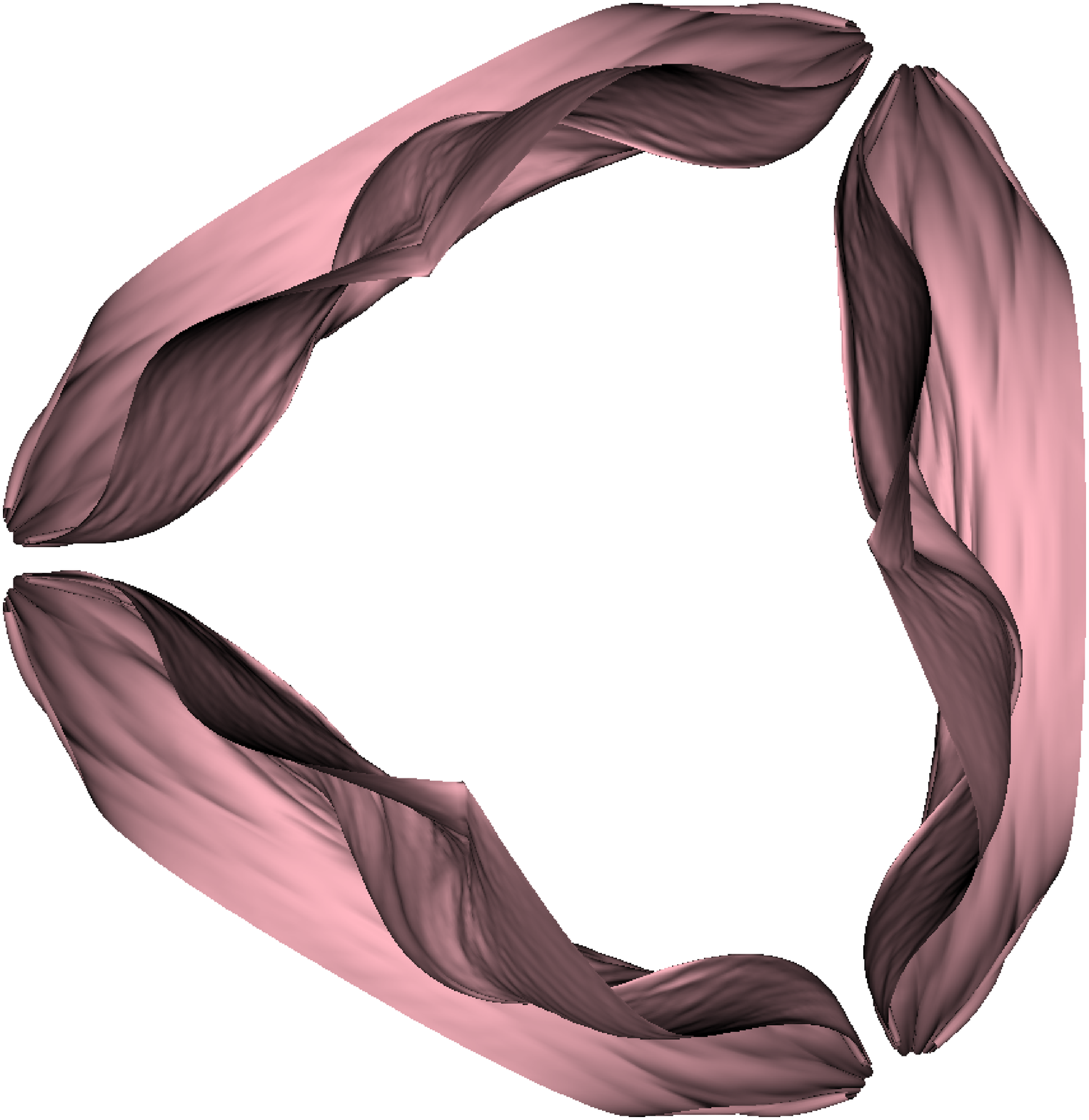} \\
    \vspace{-10pt} {\bf B.} & & & & & \\
    & \includegraphics[height=91.75pt]{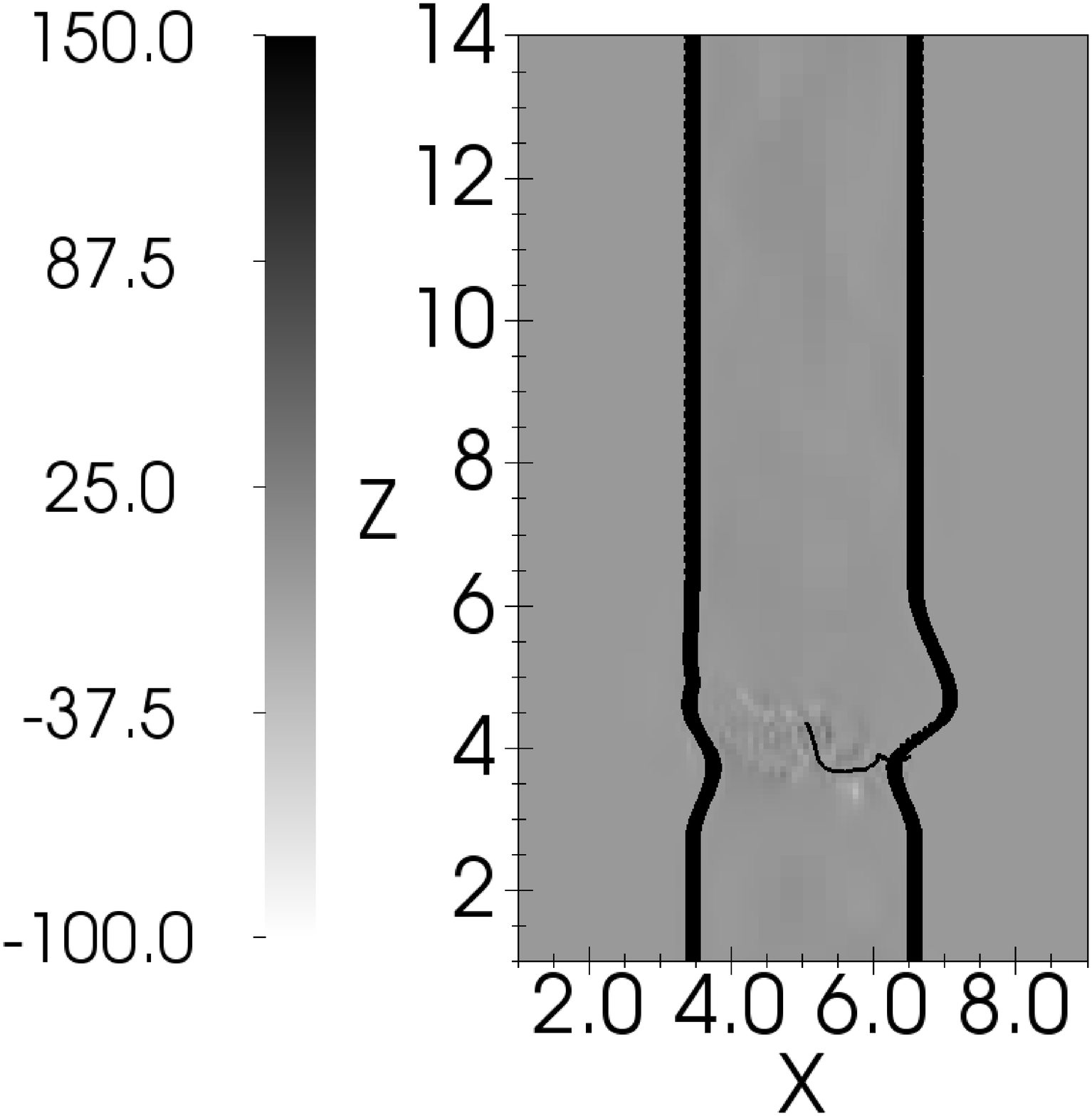}
    & \includegraphics[height=91.75pt]{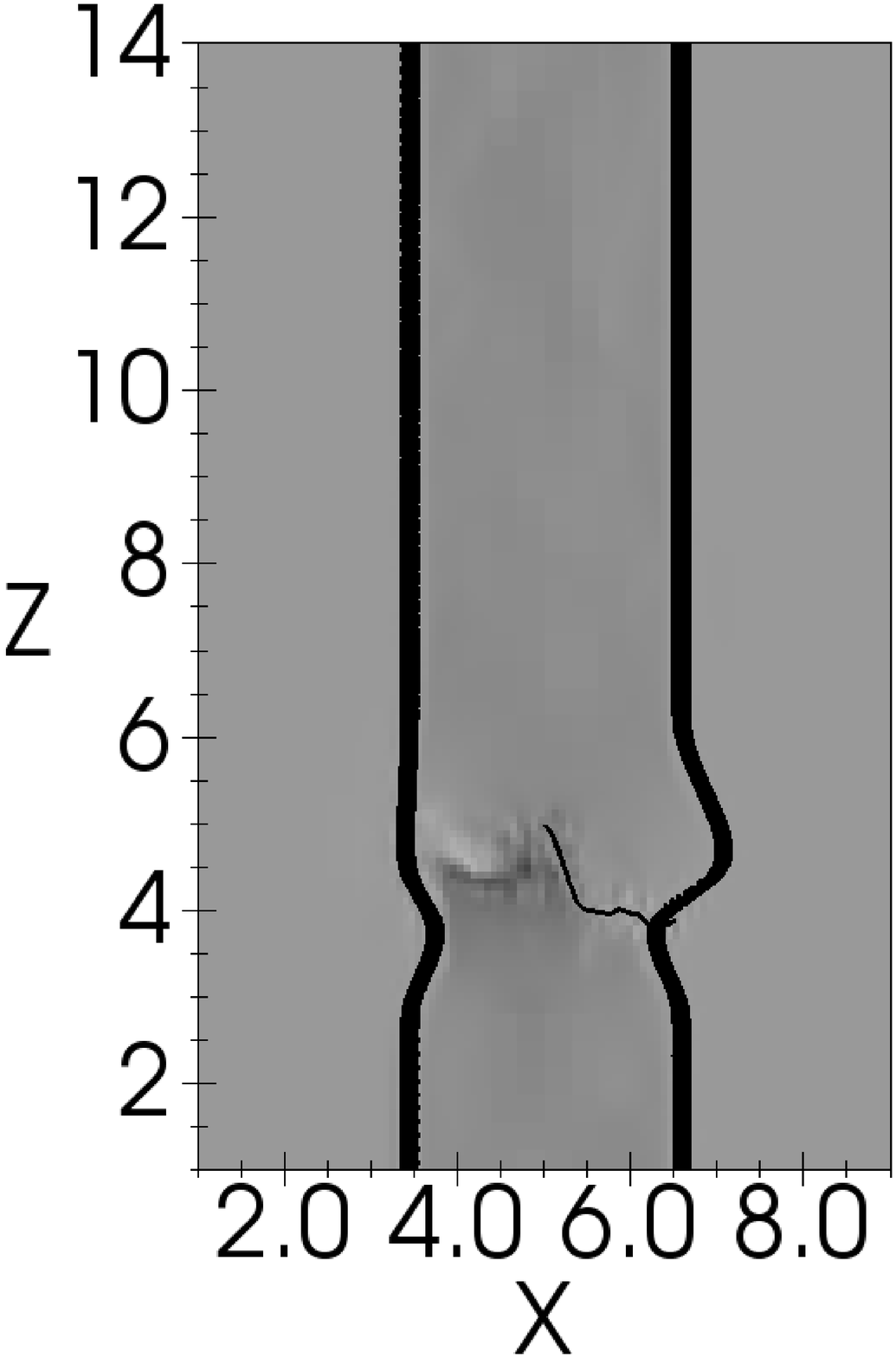}
    & \includegraphics[height=91.75pt]{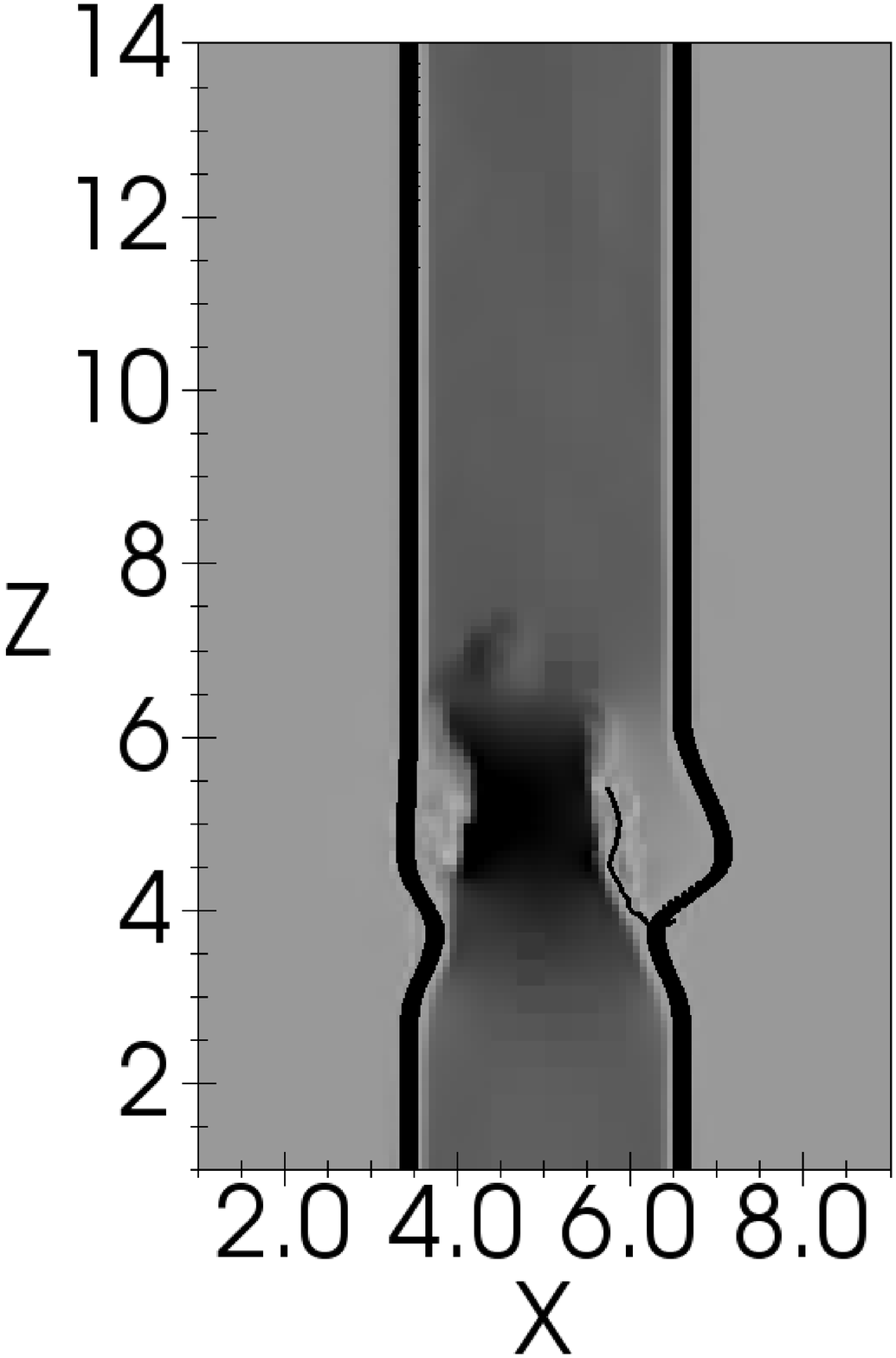}
    & \includegraphics[height=91.75pt]{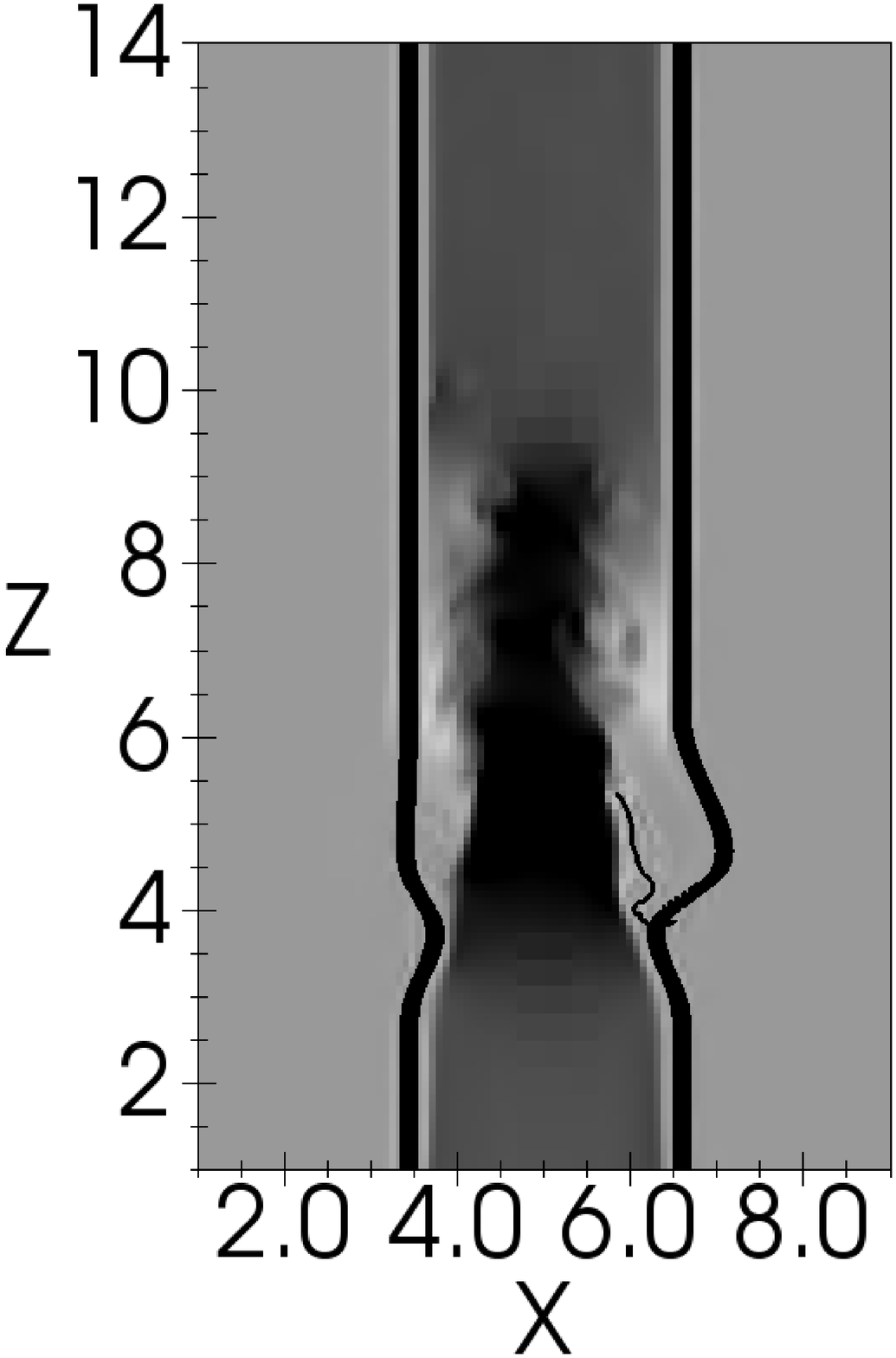}
    & \includegraphics[height=91.75pt]{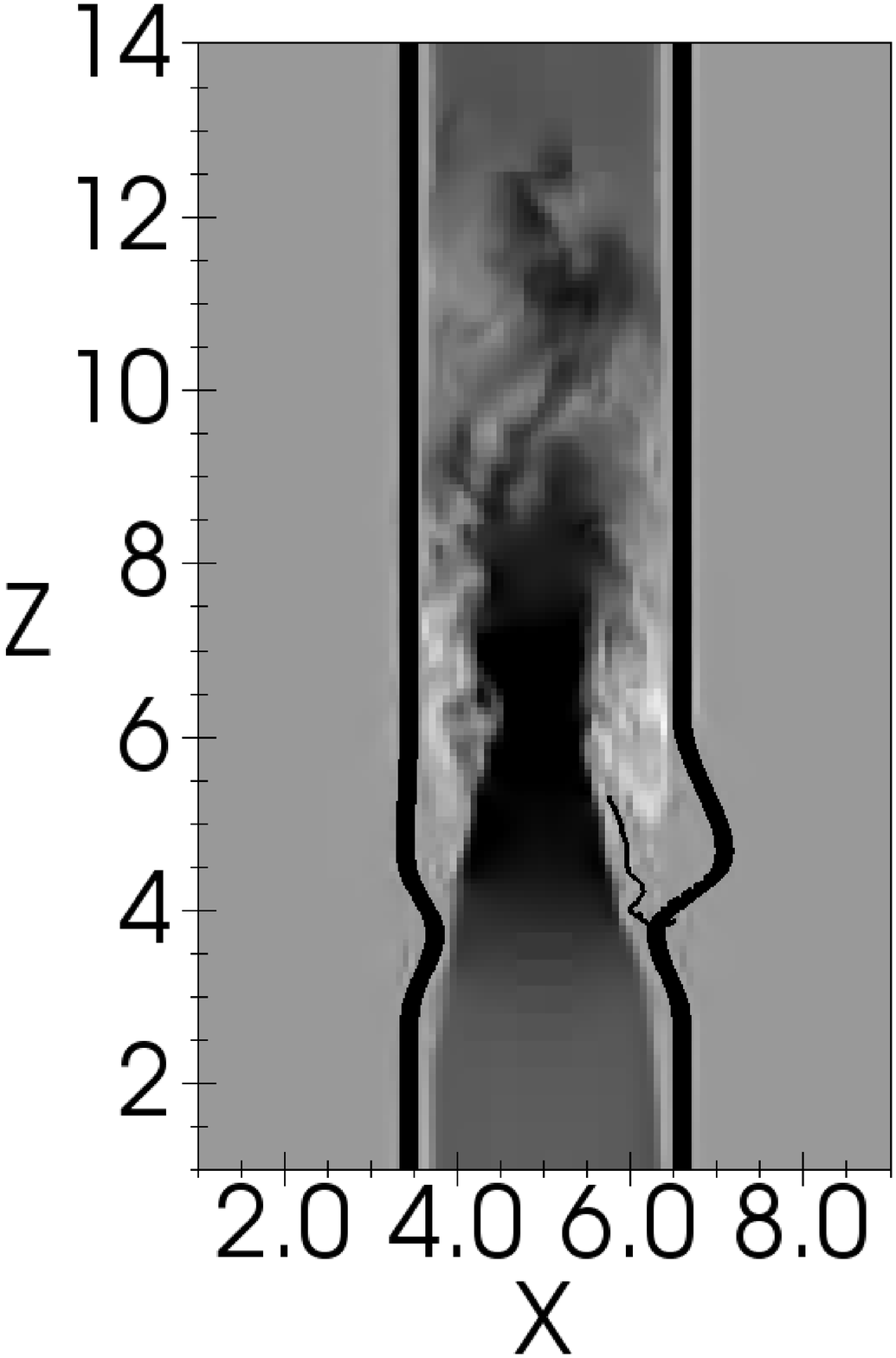} \\
    \vspace{-10pt} {\bf C.} & & & & & \\
    & \includegraphics[height=91.75pt]{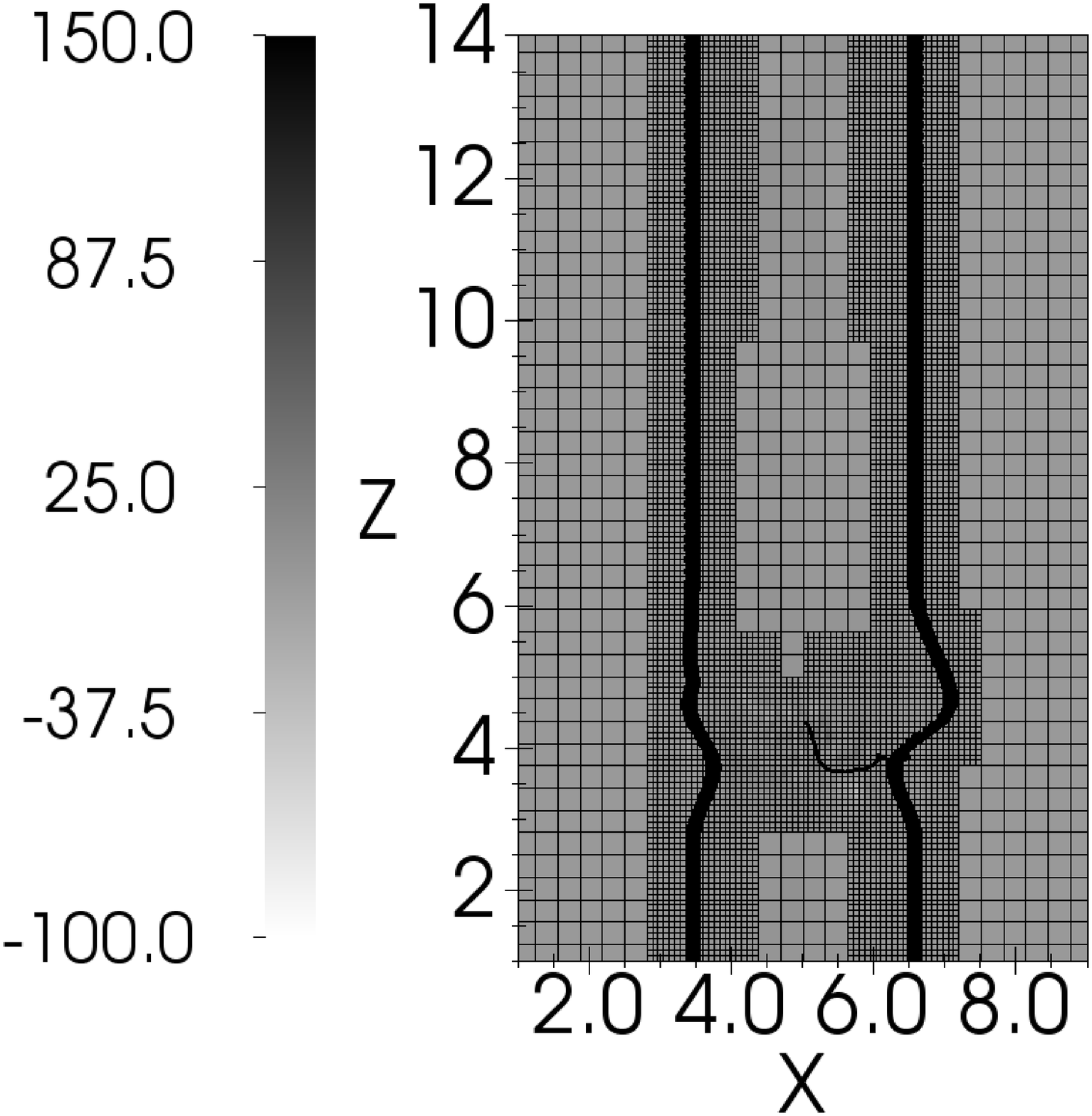}
    & \includegraphics[height=91.75pt]{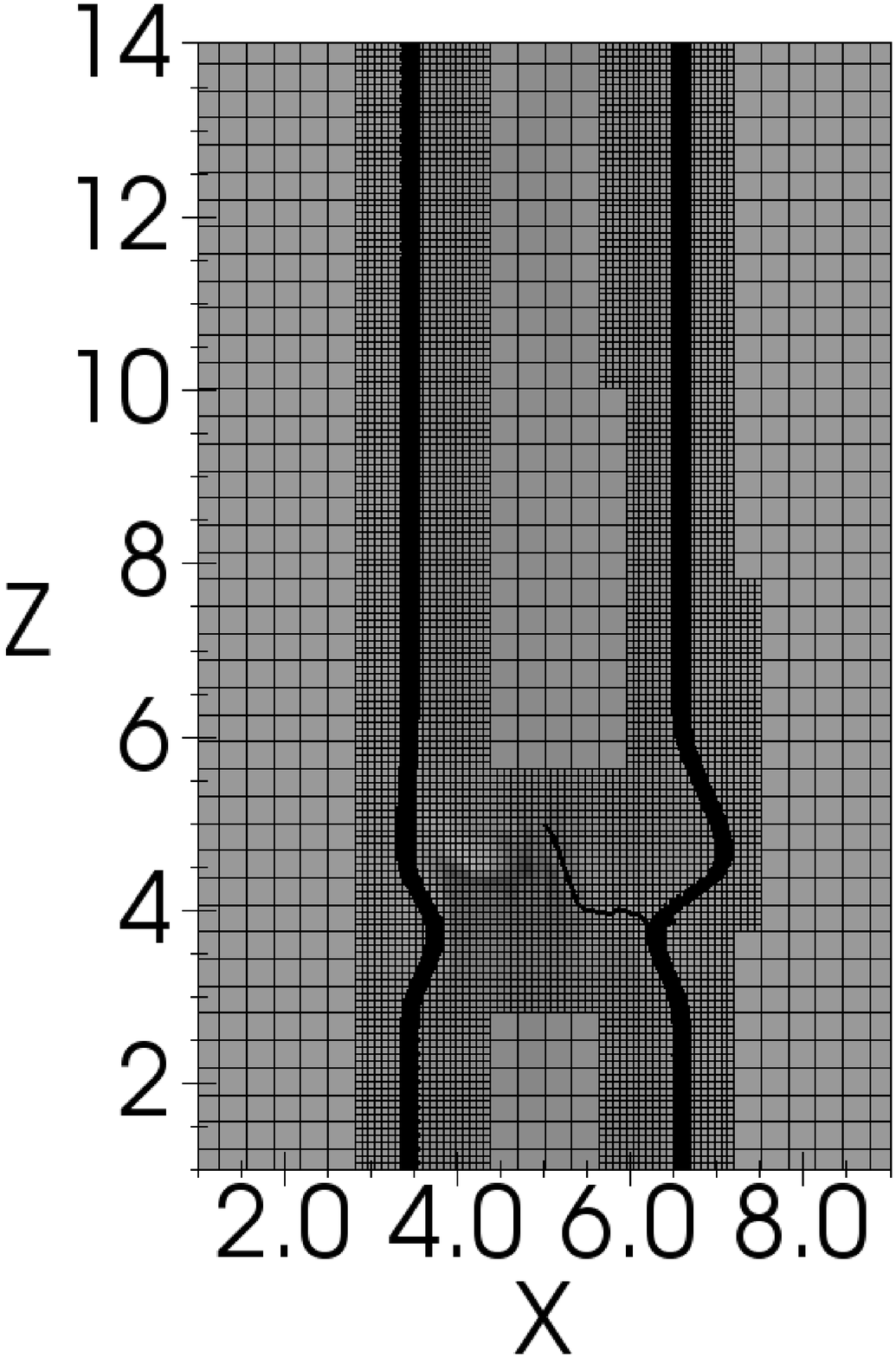}
    & \includegraphics[height=91.75pt]{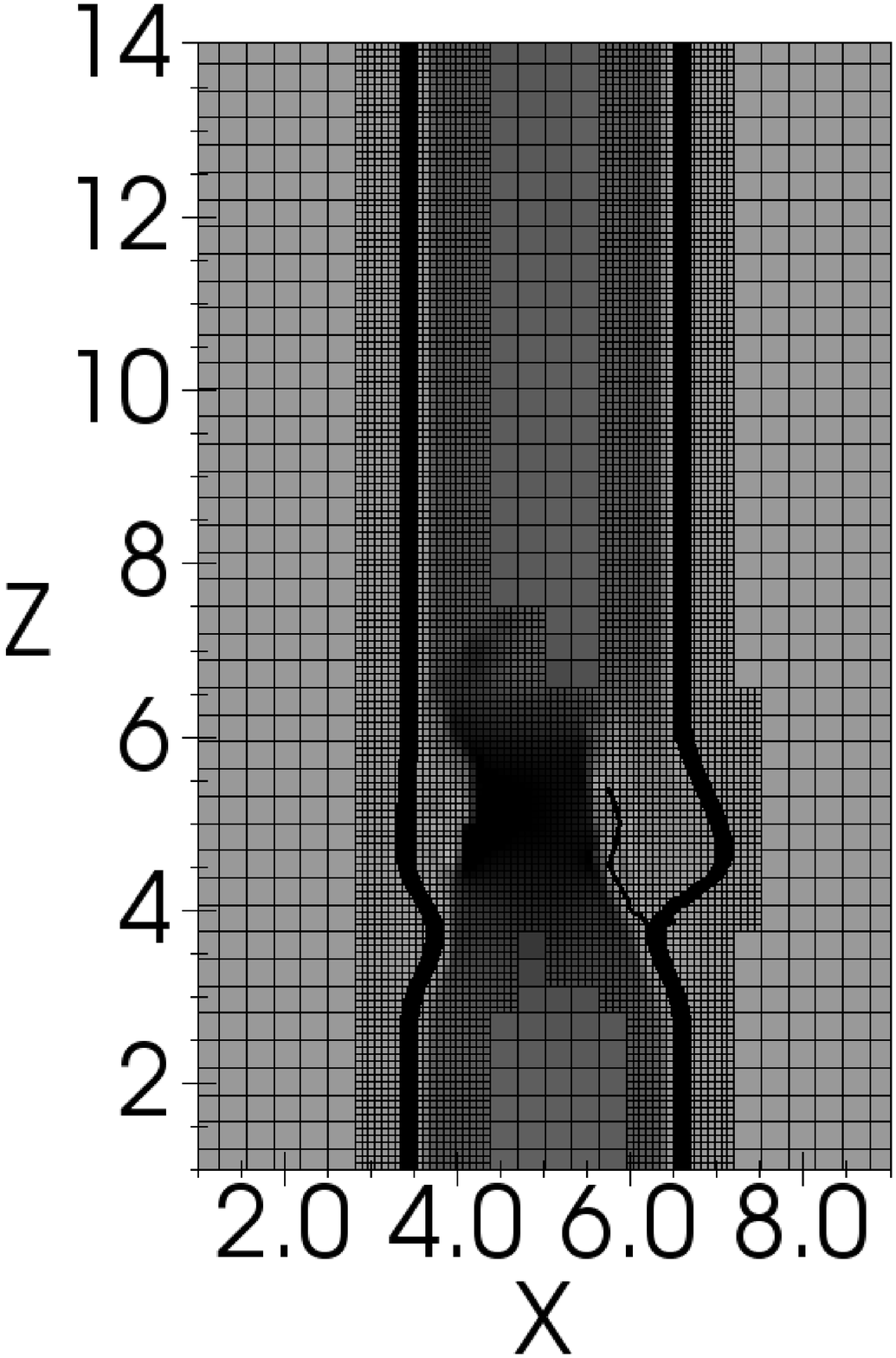}
    & \includegraphics[height=91.75pt]{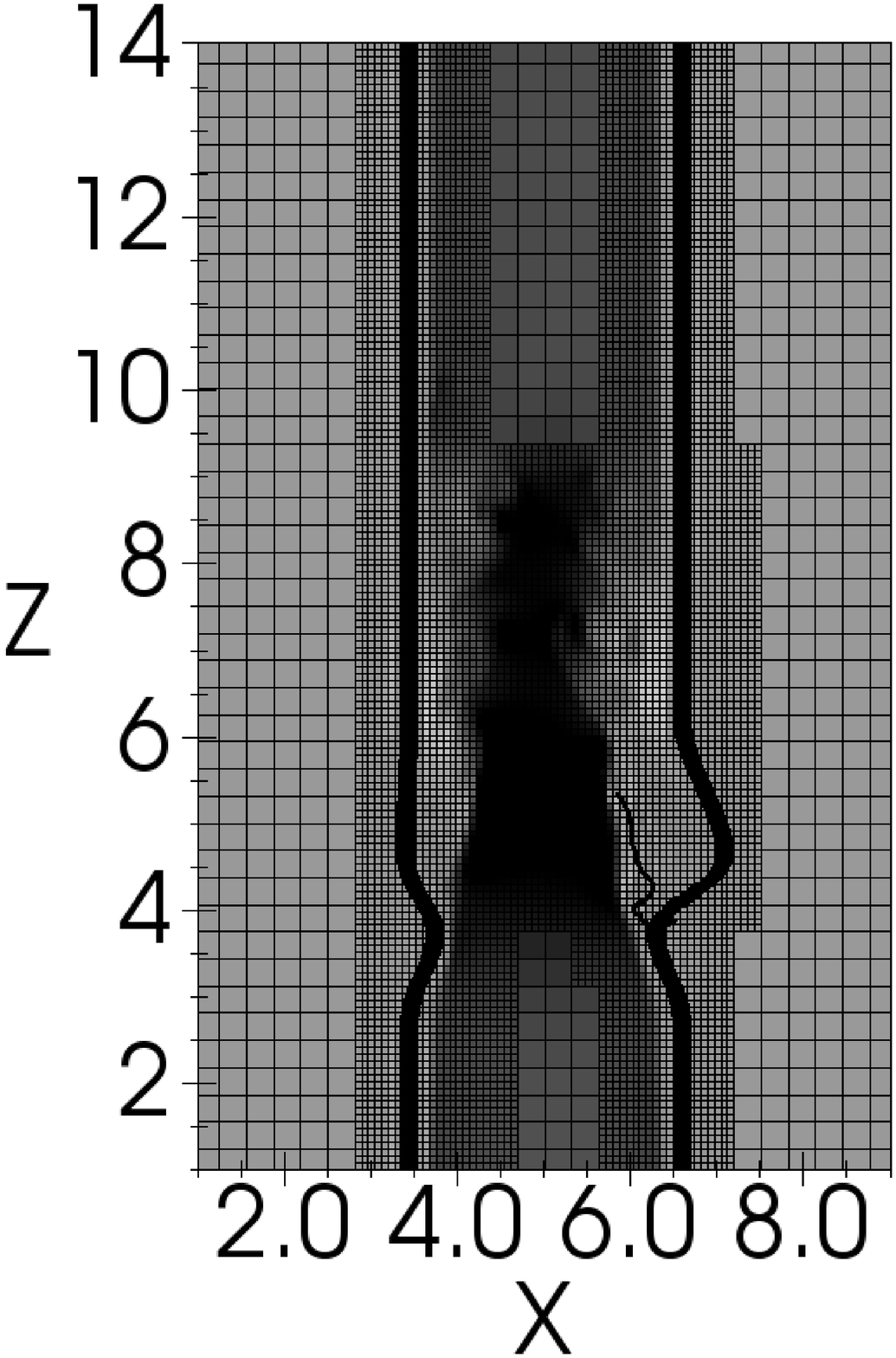}
    & \includegraphics[height=91.75pt]{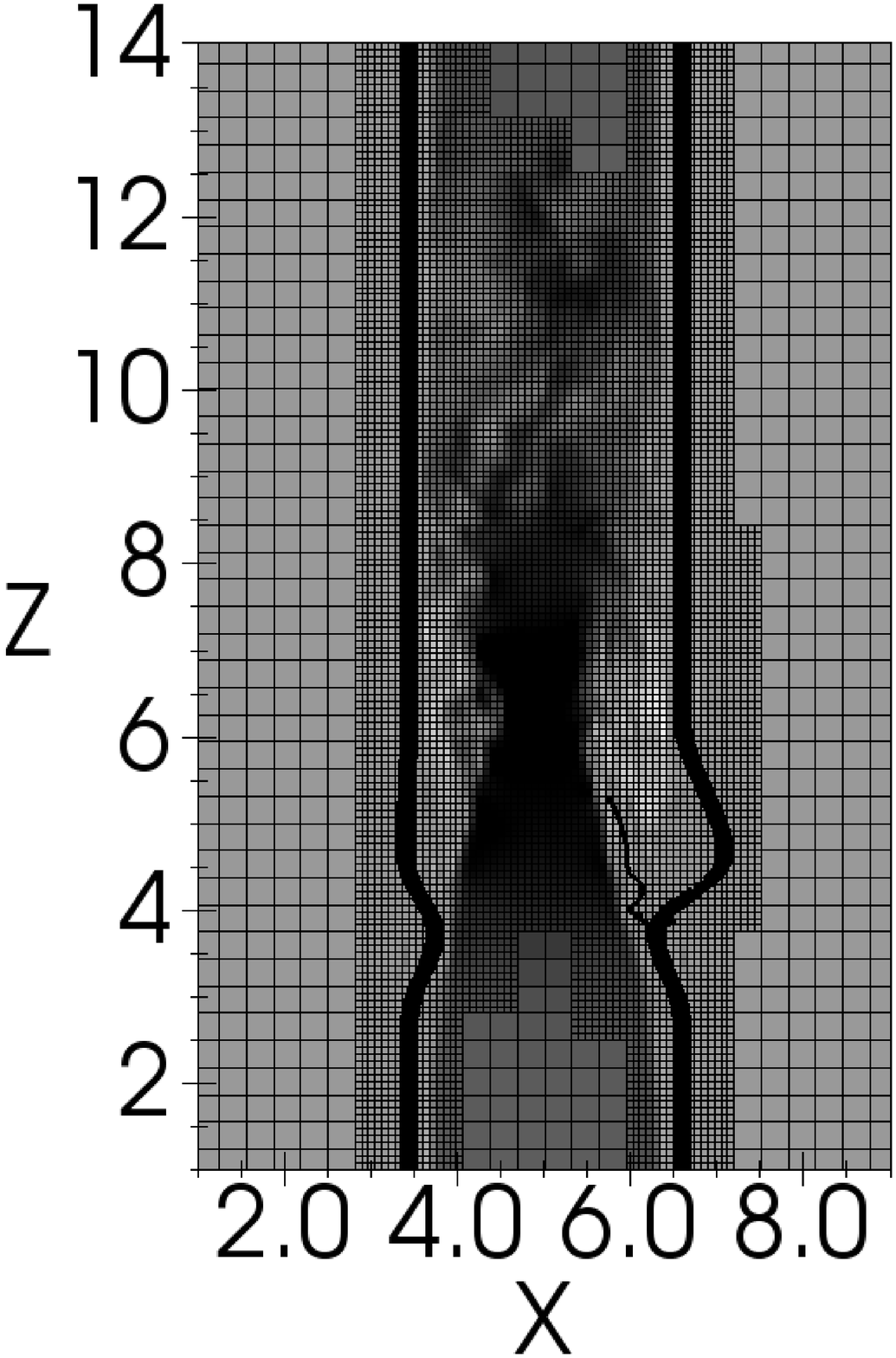}
  \end{tabular}
  \caption{The opening dynamics of the valve during the second cardiac
    cycle, shown at equally spaced time intervals.  {\bf A.}~The valve
    leaflets, as seen from the downstream boundary of the model.  {\bf
      B.}~The model valve and vessel and the streamwise ($u_3)$
    component of the flow velocity along a plane that bisects one of
    the valve leaflets.  {\bf C.}~Similar to B, but also showing the
    adaptively refined Cartesian grid.}
  \label{f:opening_dynamics}
\end{figure}

\begin{figure}
  \centering
  \begin{tabular}{lrcccc}
    \vspace{-10pt} {\bf A.} & & & & & \\
    & \includegraphics[width=60.5pt]{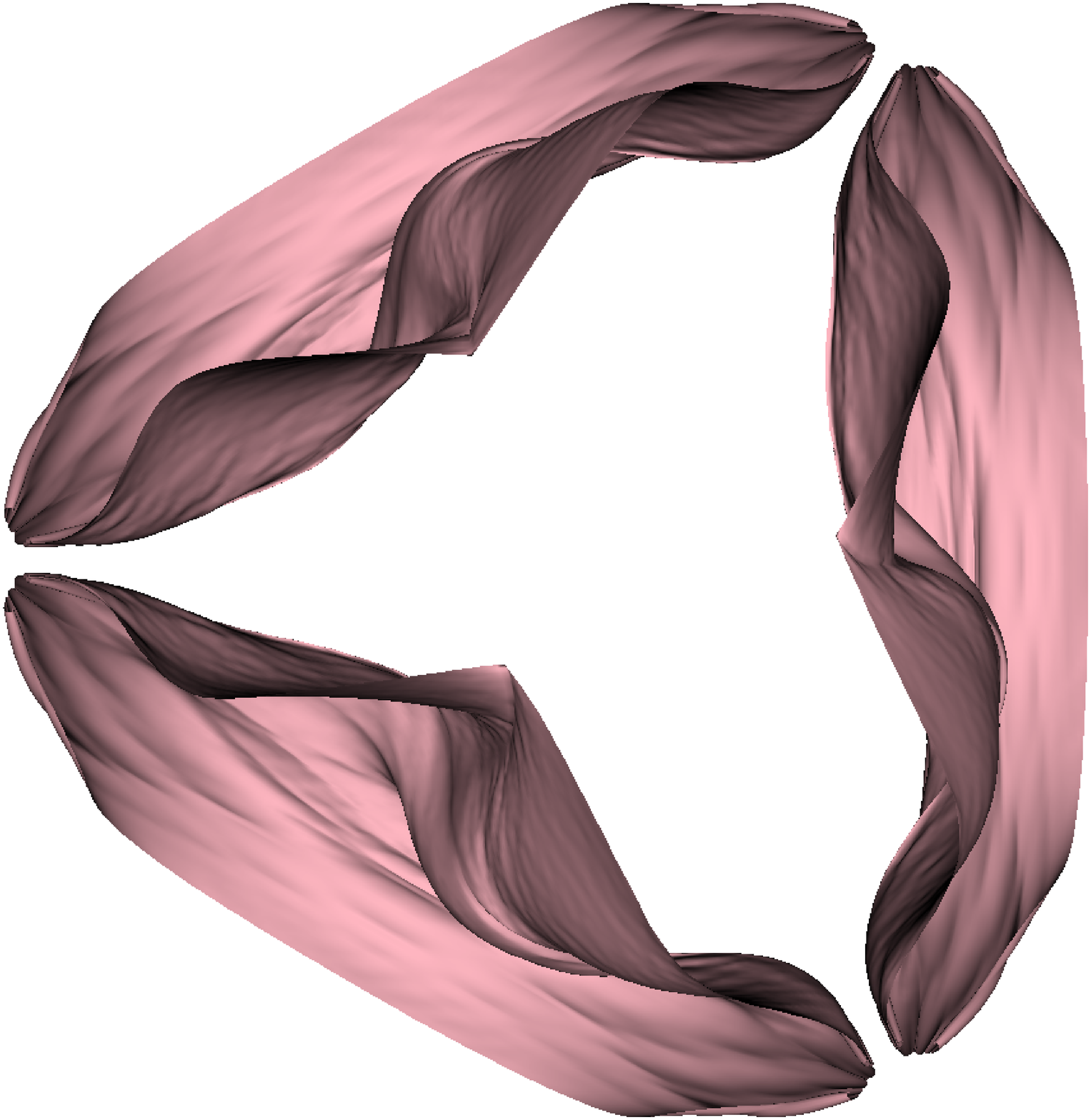}
    & \includegraphics[width=60.5pt]{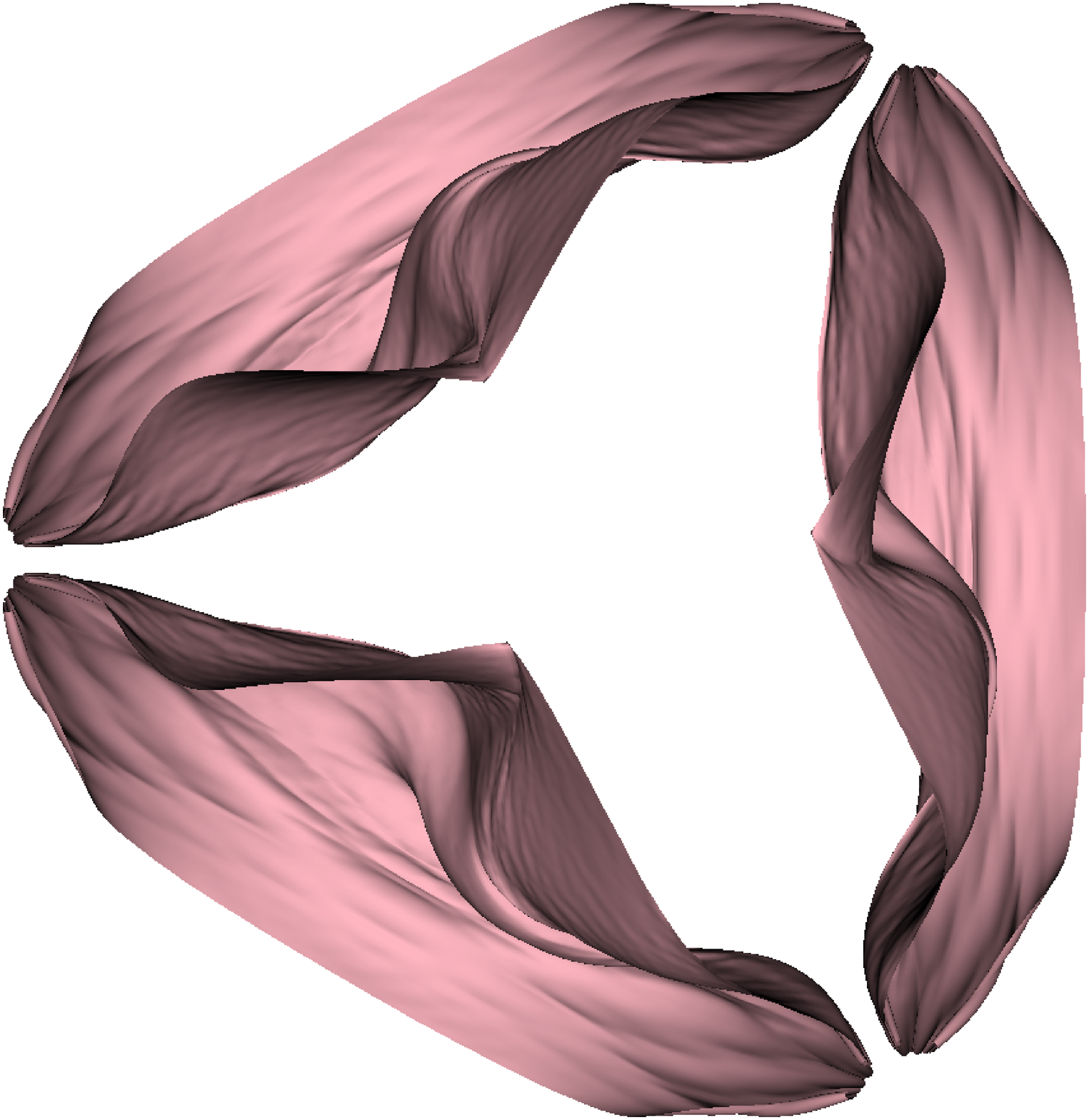}
    & \includegraphics[width=60.5pt]{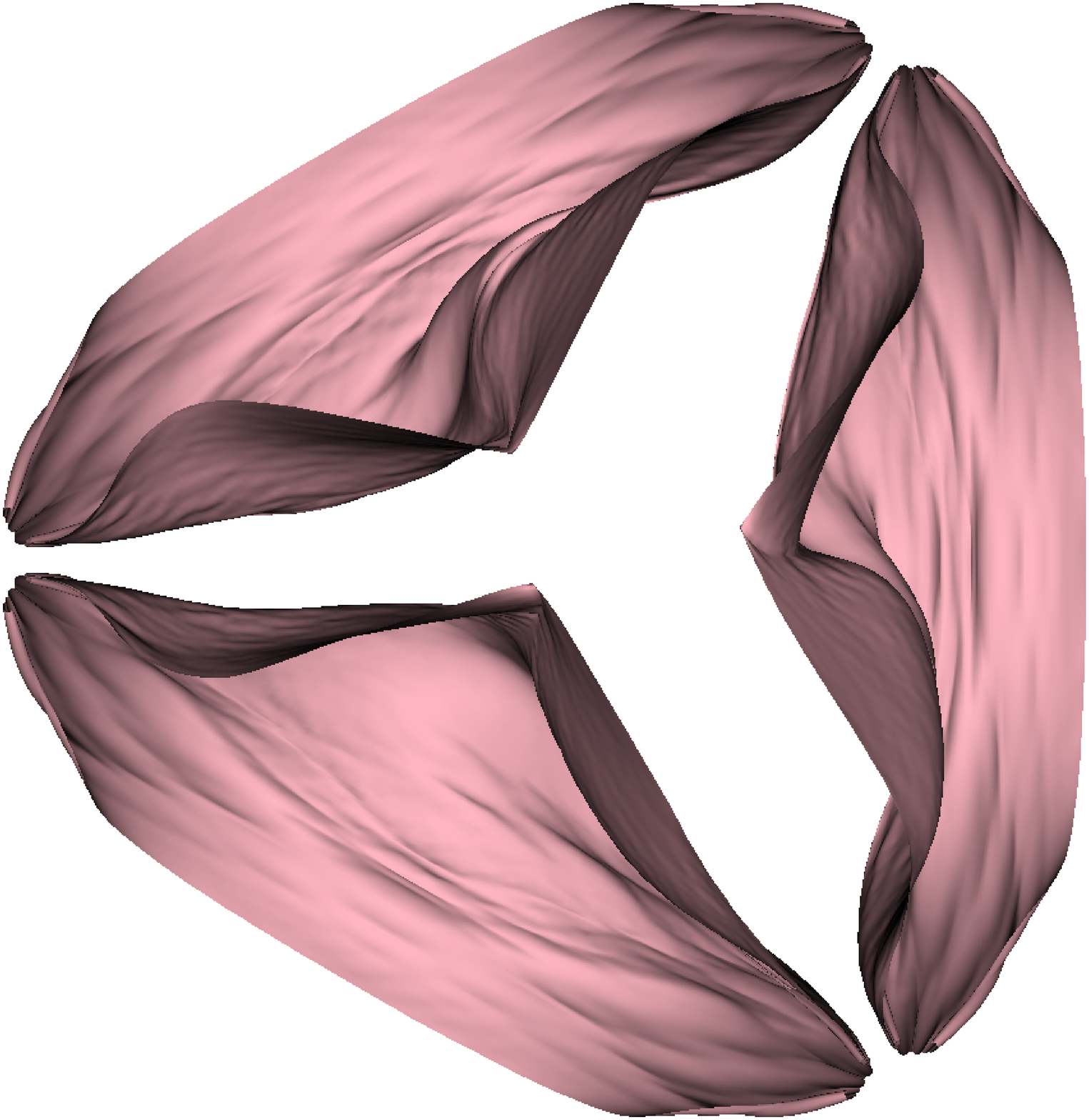}
    & \includegraphics[width=60.5pt]{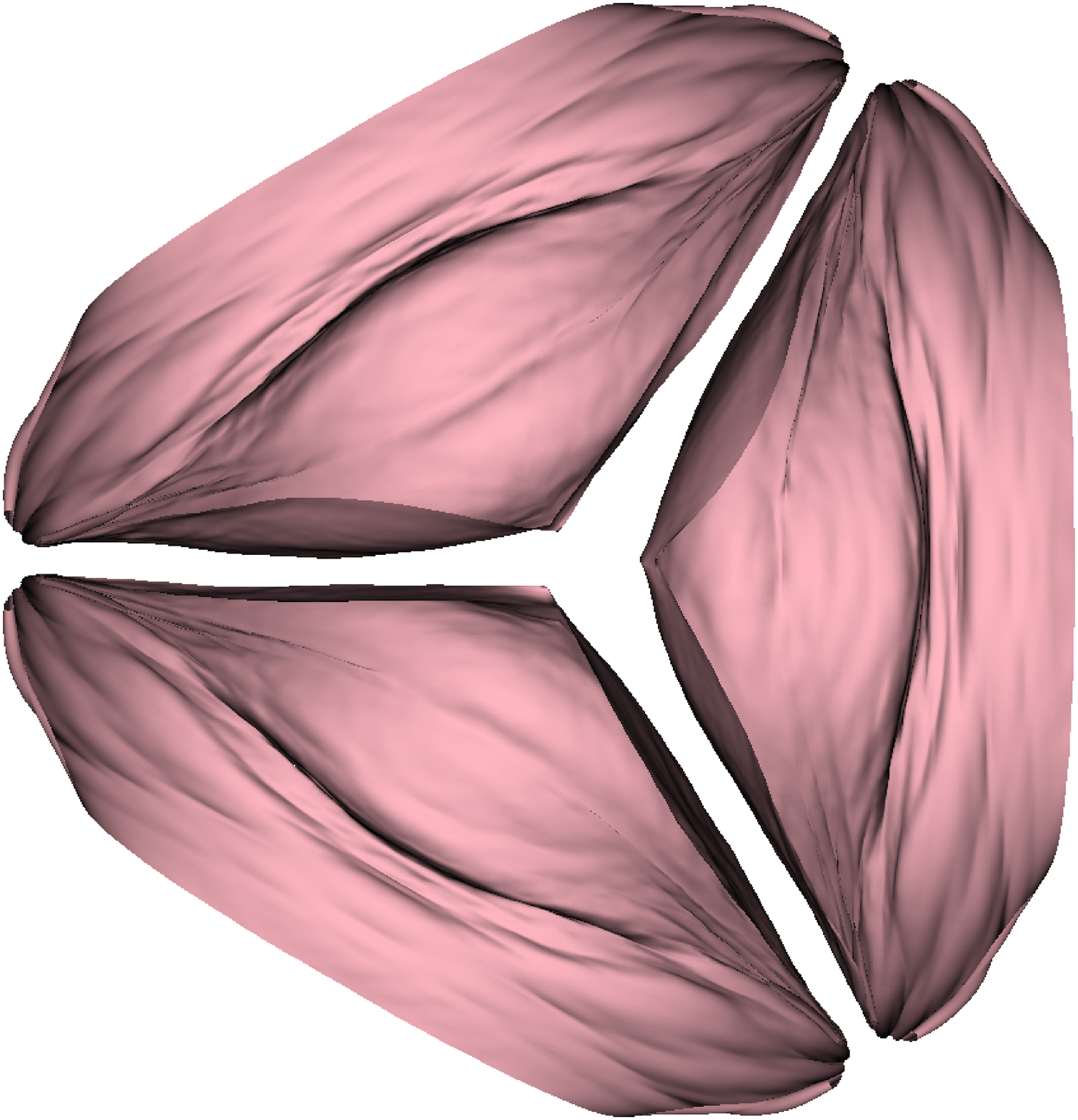}
    & \includegraphics[width=60.5pt]{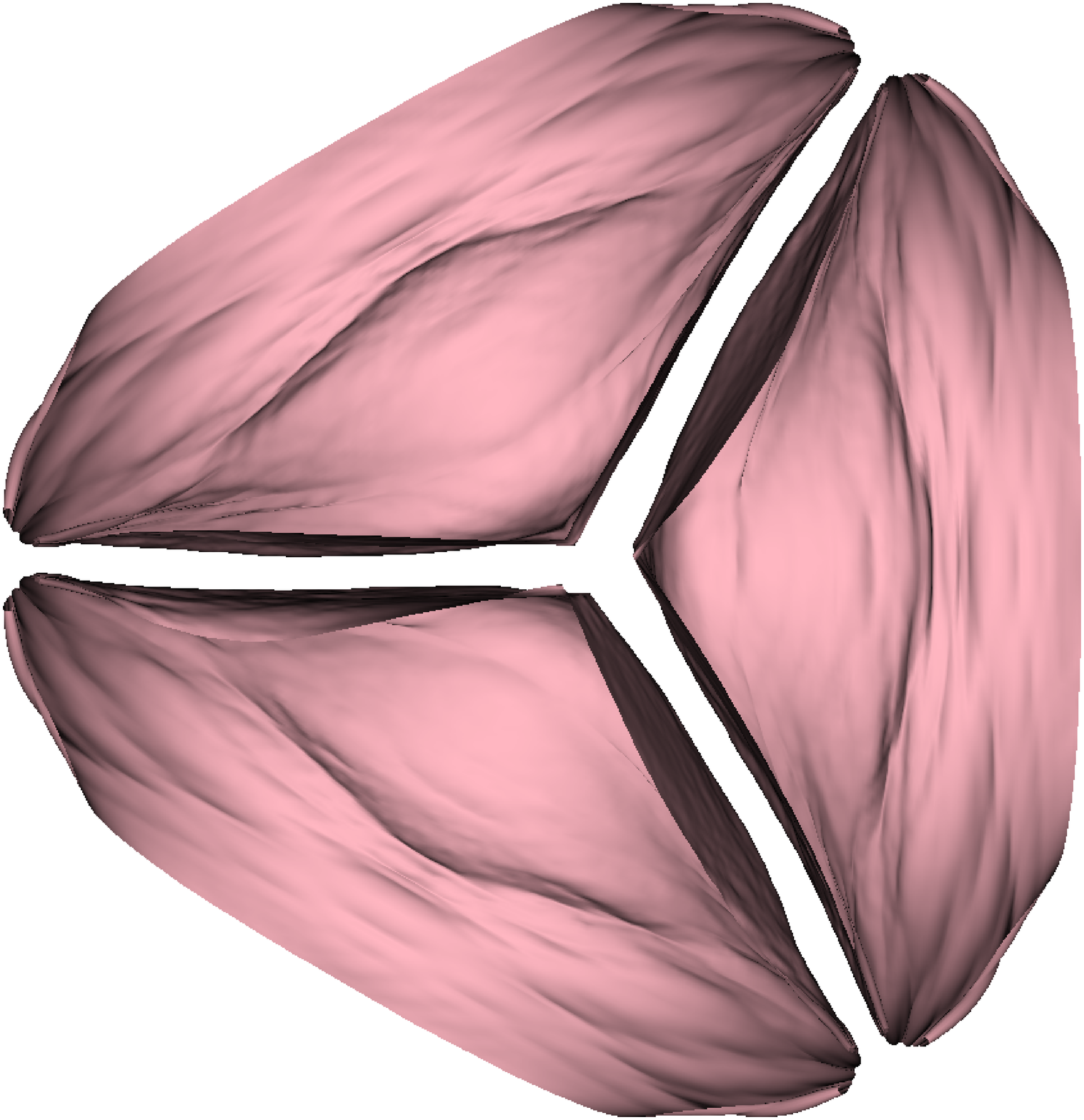} \\
    \vspace{-10pt} {\bf B.} & & & & & \\
    & \includegraphics[height=91.75pt]{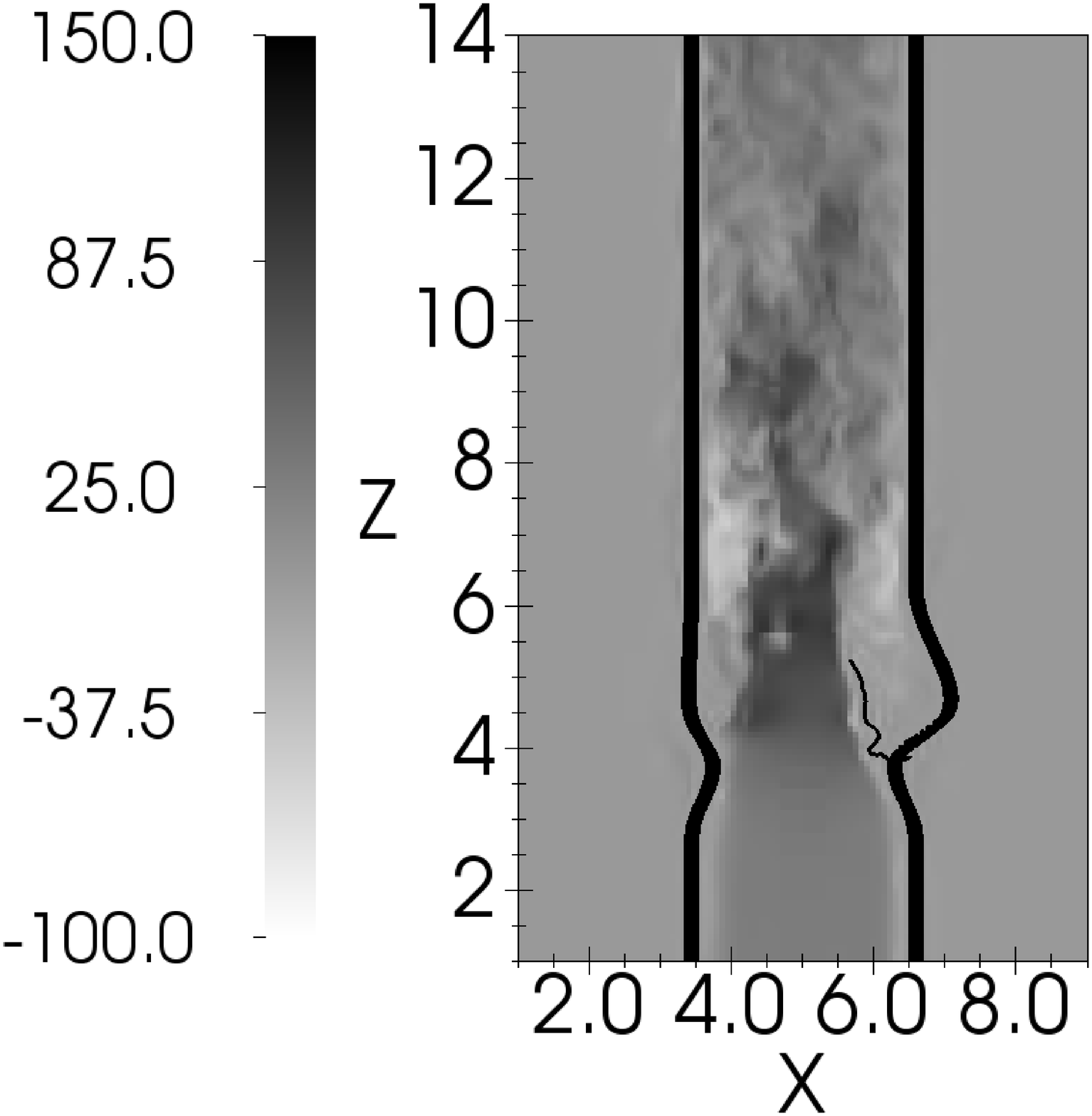}
    & \includegraphics[height=91.75pt]{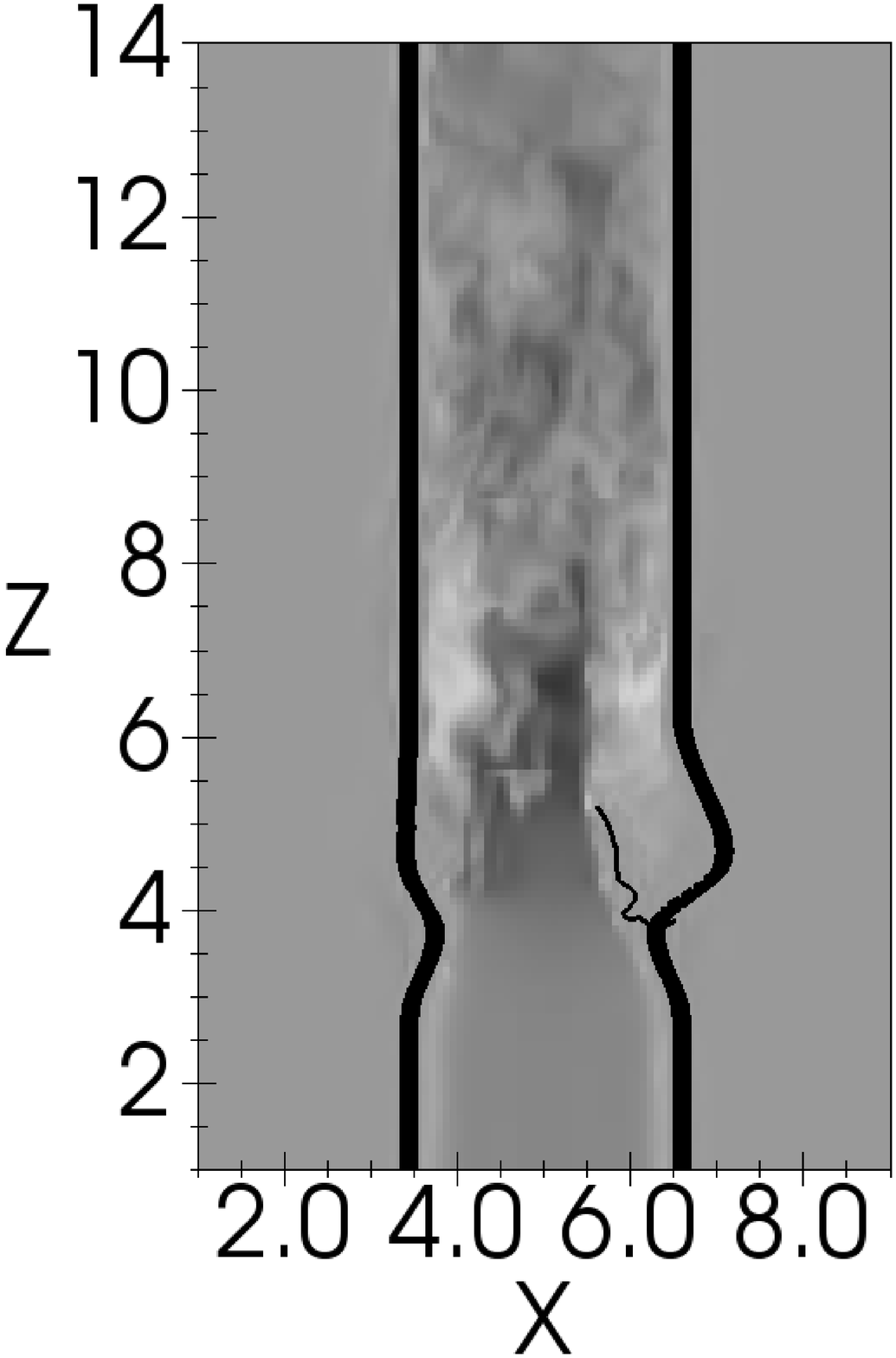}
    & \includegraphics[height=91.75pt]{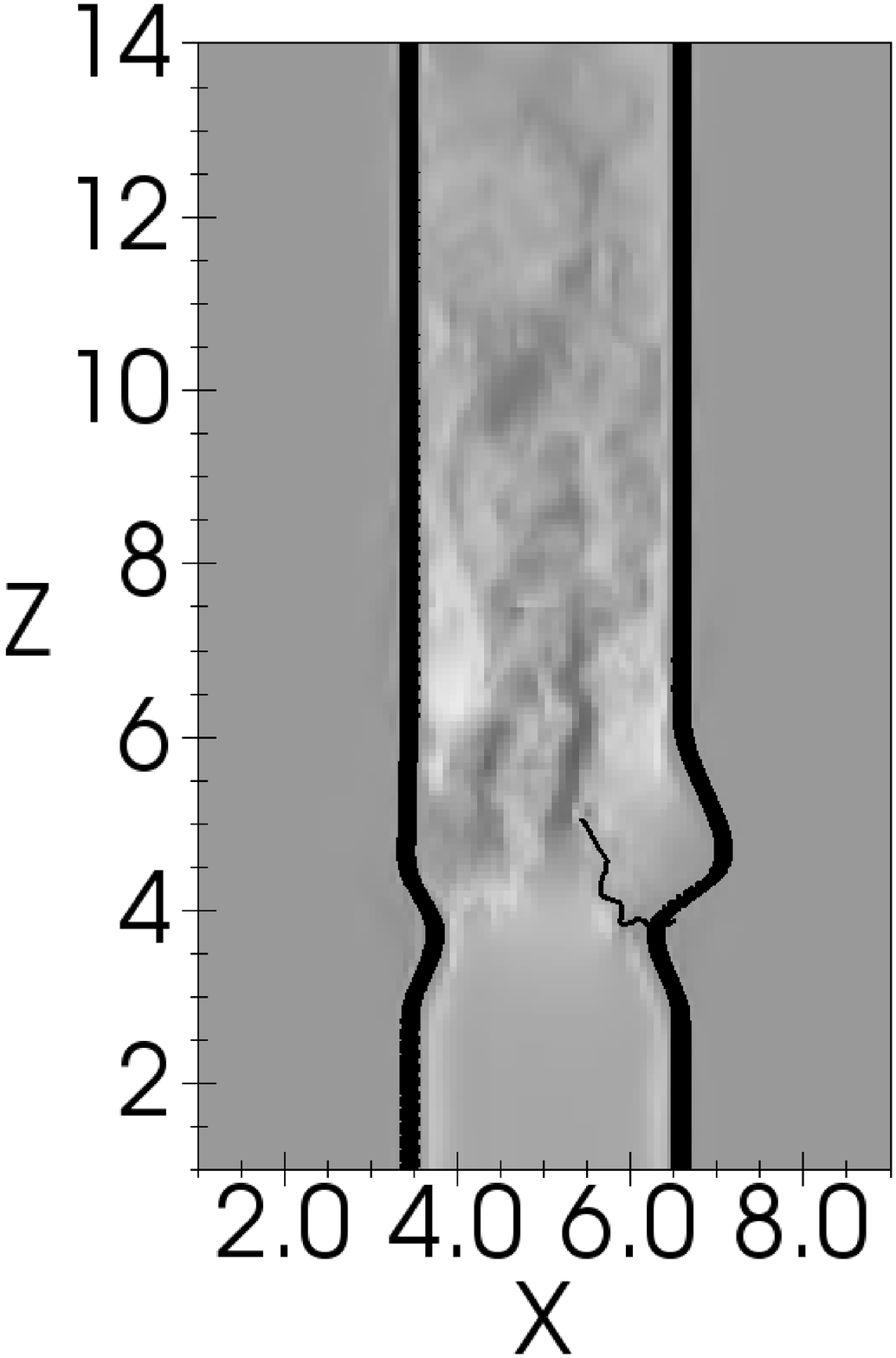}
    & \includegraphics[height=91.75pt]{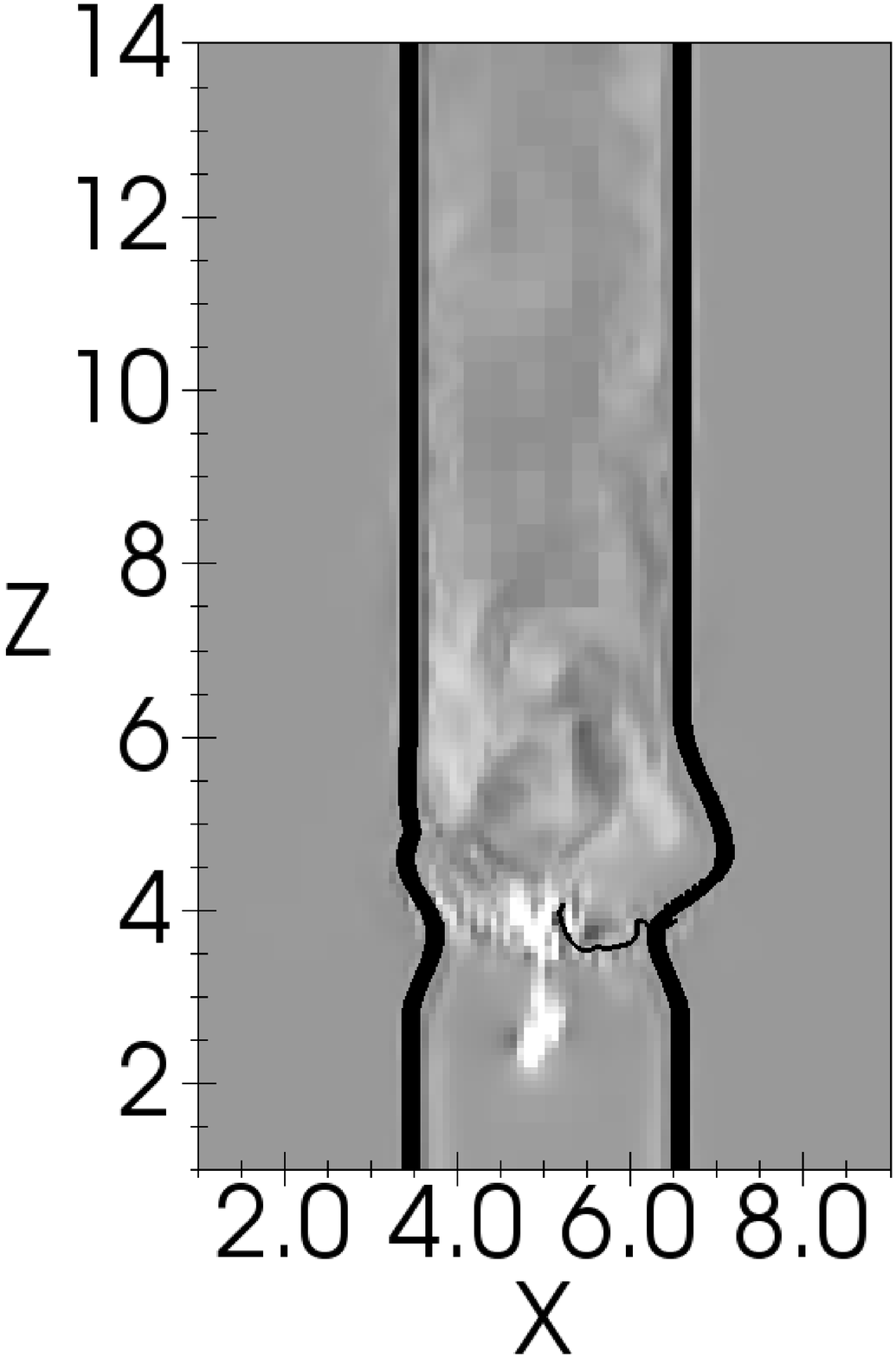}
    & \includegraphics[height=91.75pt]{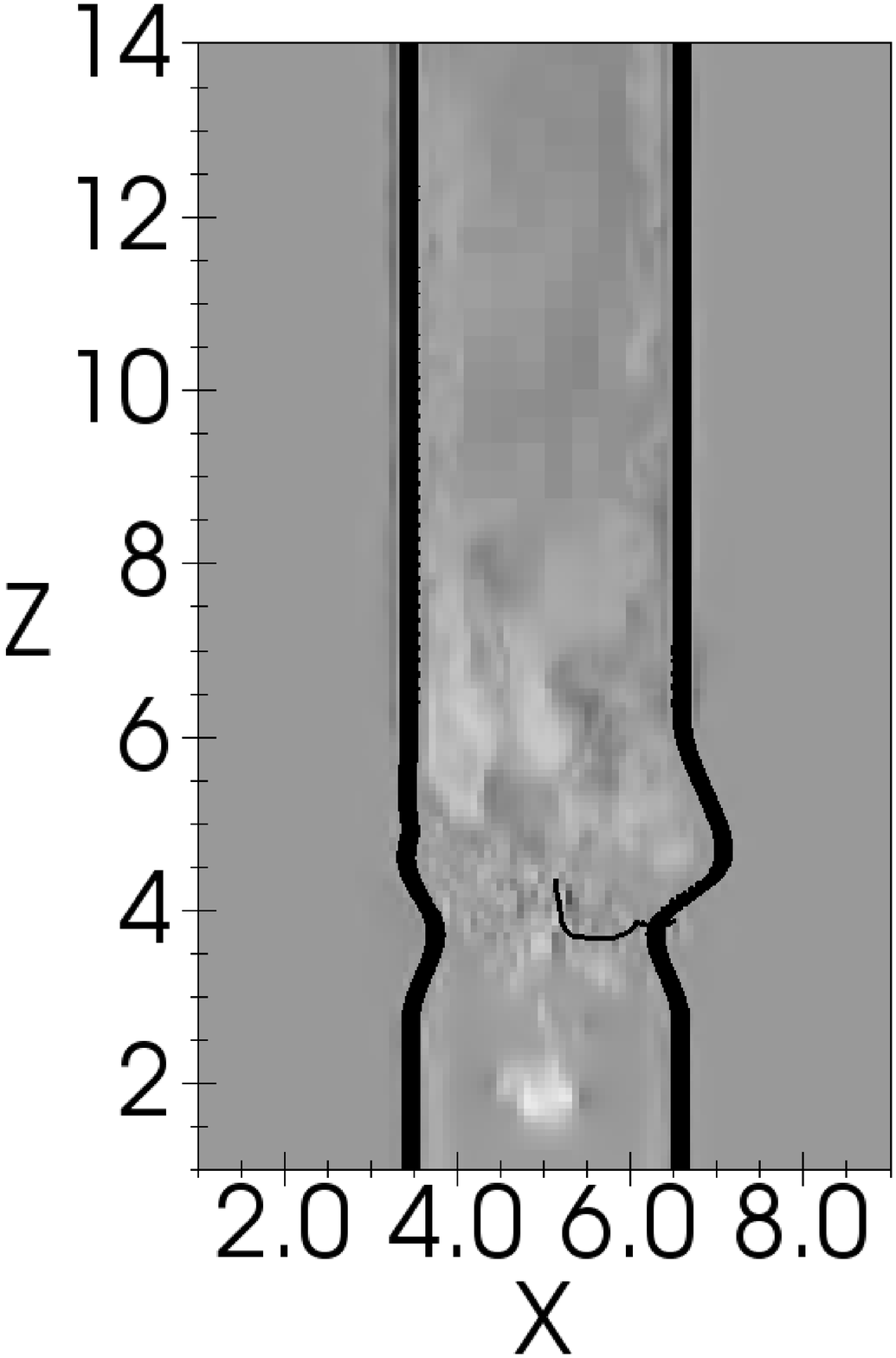} \\
    \vspace{-10pt} {\bf C.} & & & & & \\
    & \includegraphics[height=91.75pt]{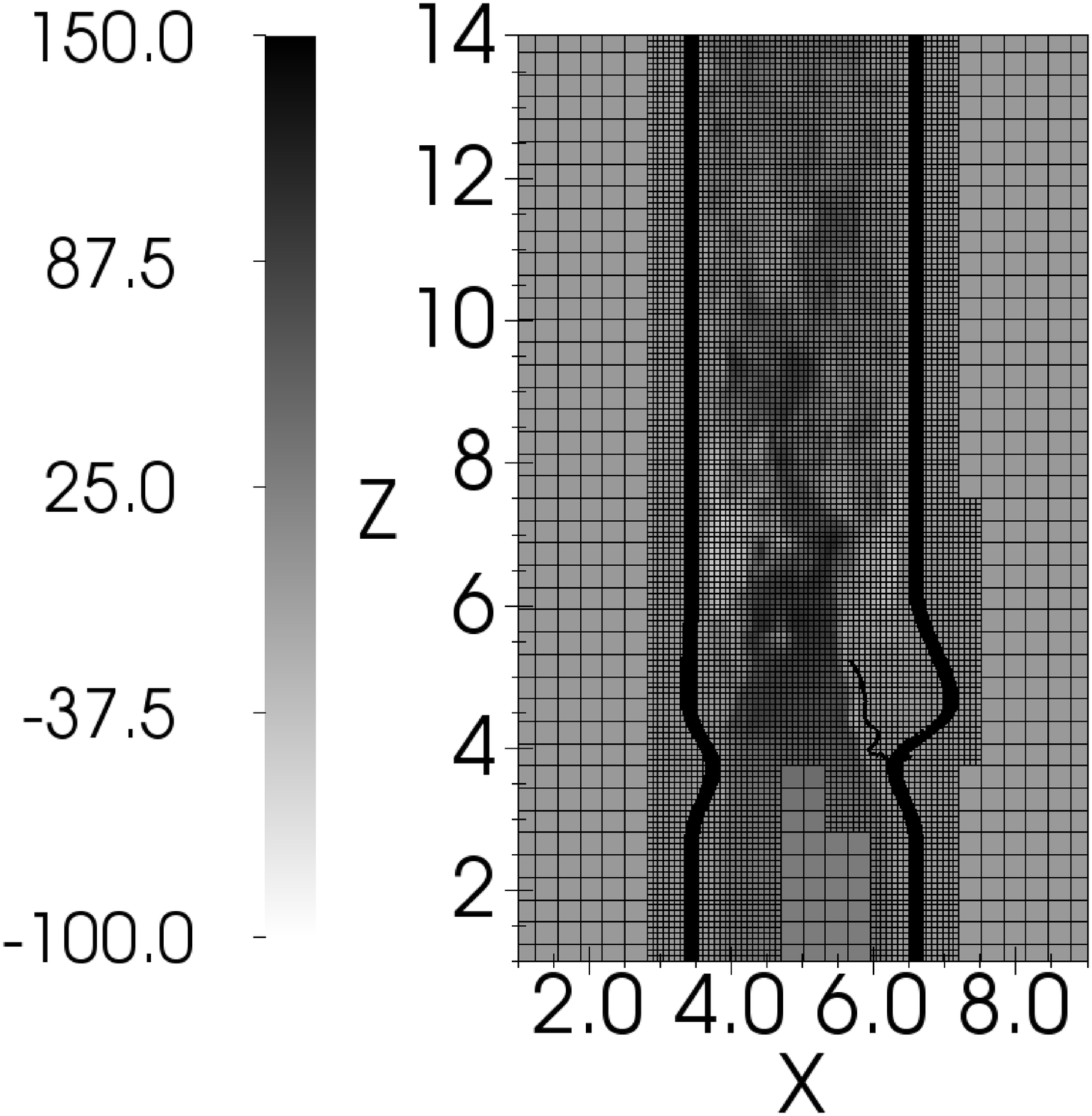}
    & \includegraphics[height=91.75pt]{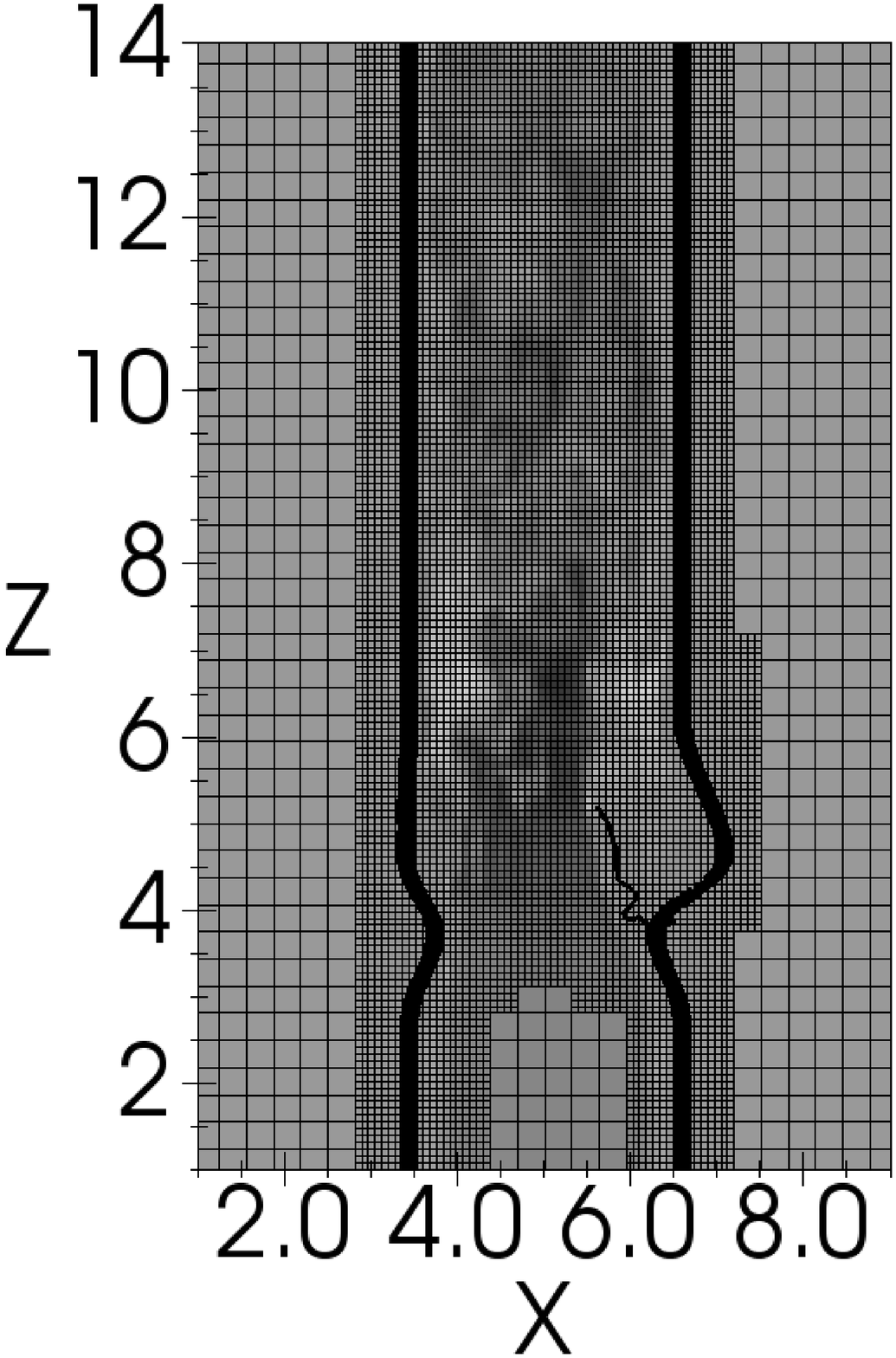}
    & \includegraphics[height=91.75pt]{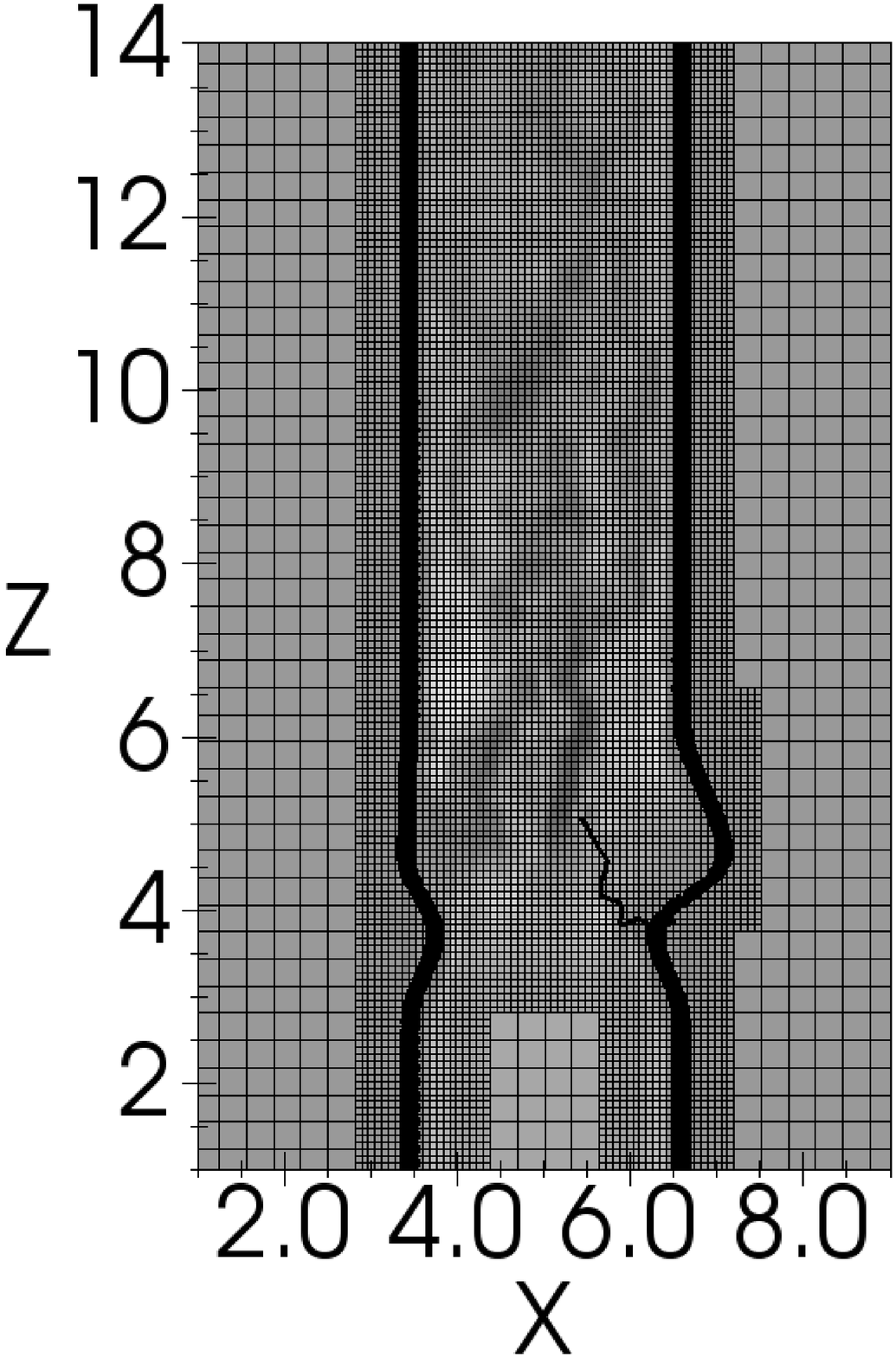}
    & \includegraphics[height=91.75pt]{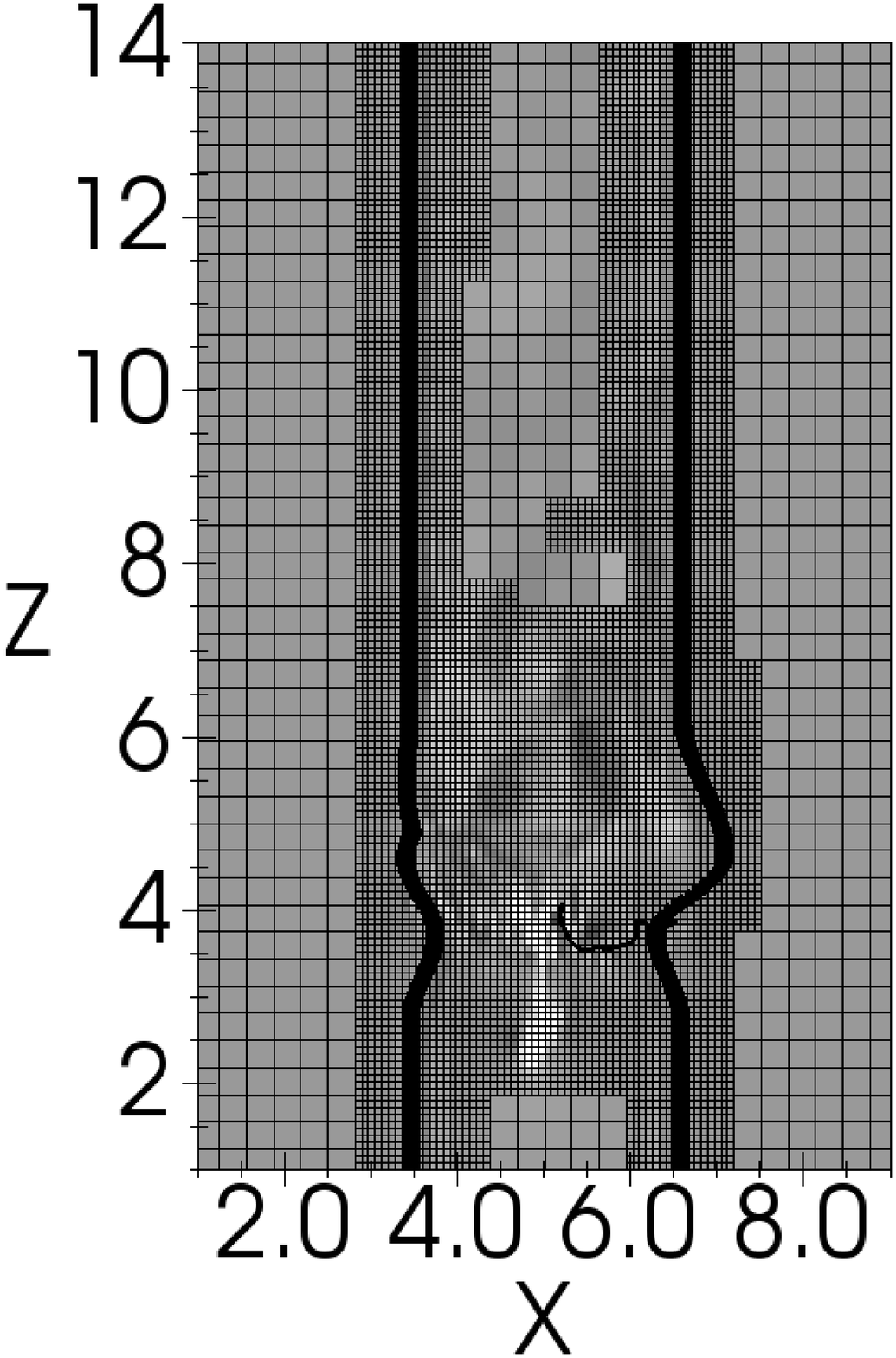}
    & \includegraphics[height=91.75pt]{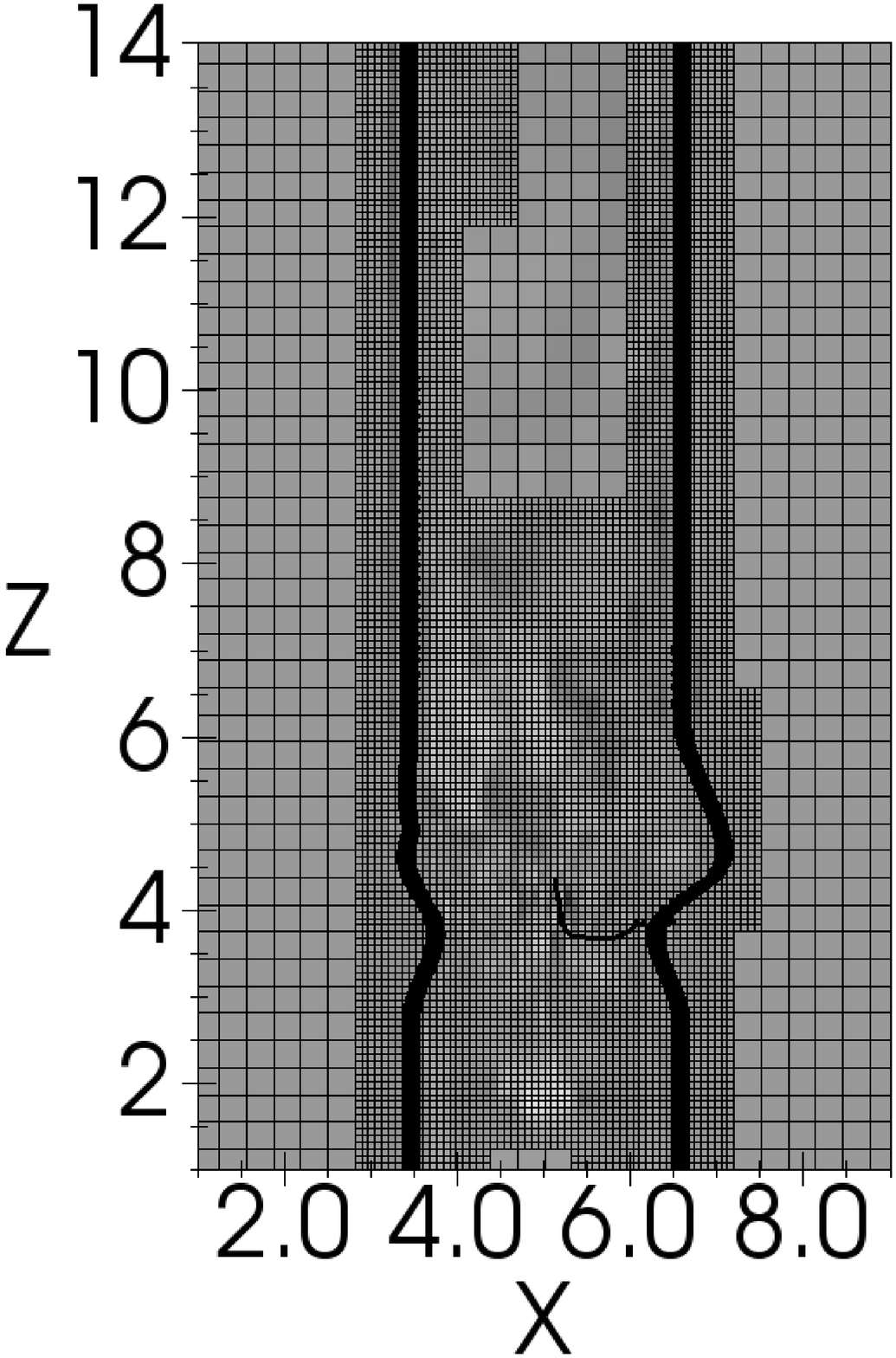}
  \end{tabular}
  \caption{Similar to fig.~\ref{f:opening_dynamics}, but here showing
    the closing dynamics of the valve during the second cardiac cycle.
    Notice that although there is minor regurgitation during valve
    closure, fig.~\ref{f:bulk_flow_properties}A shows that once
    closed, the valve permits no further leak.}
  \label{f:closing_dynamics}
\end{figure}

\begin{figure}
  \centering
  \begin{tabular}{lccc}
    \vspace{-10pt} {\bf A.} & & & \\
    \vspace{3pt}
    & \includegraphics[height=0.2875\textheight]{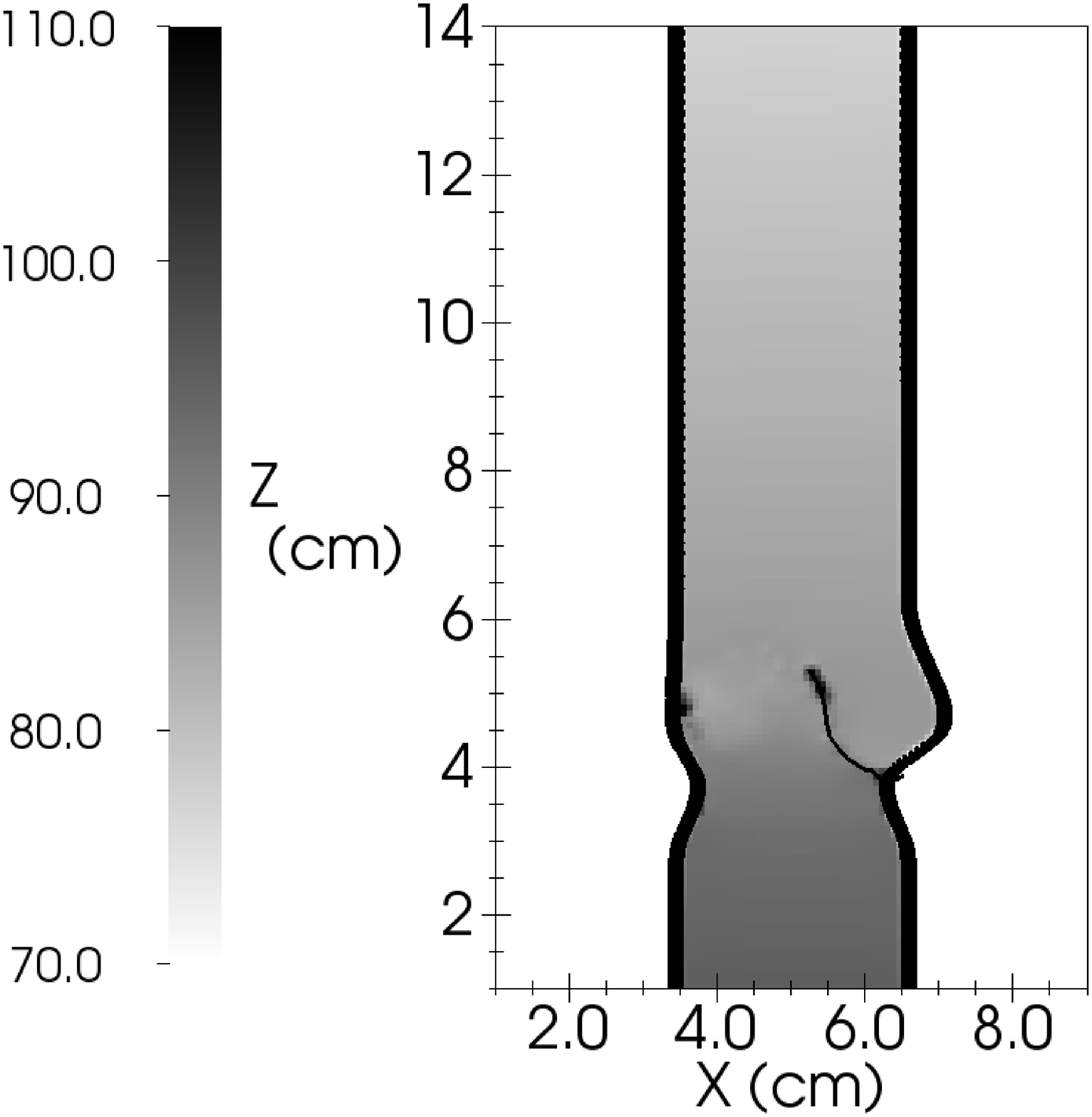} &
    & \includegraphics[height=0.2875\textheight]{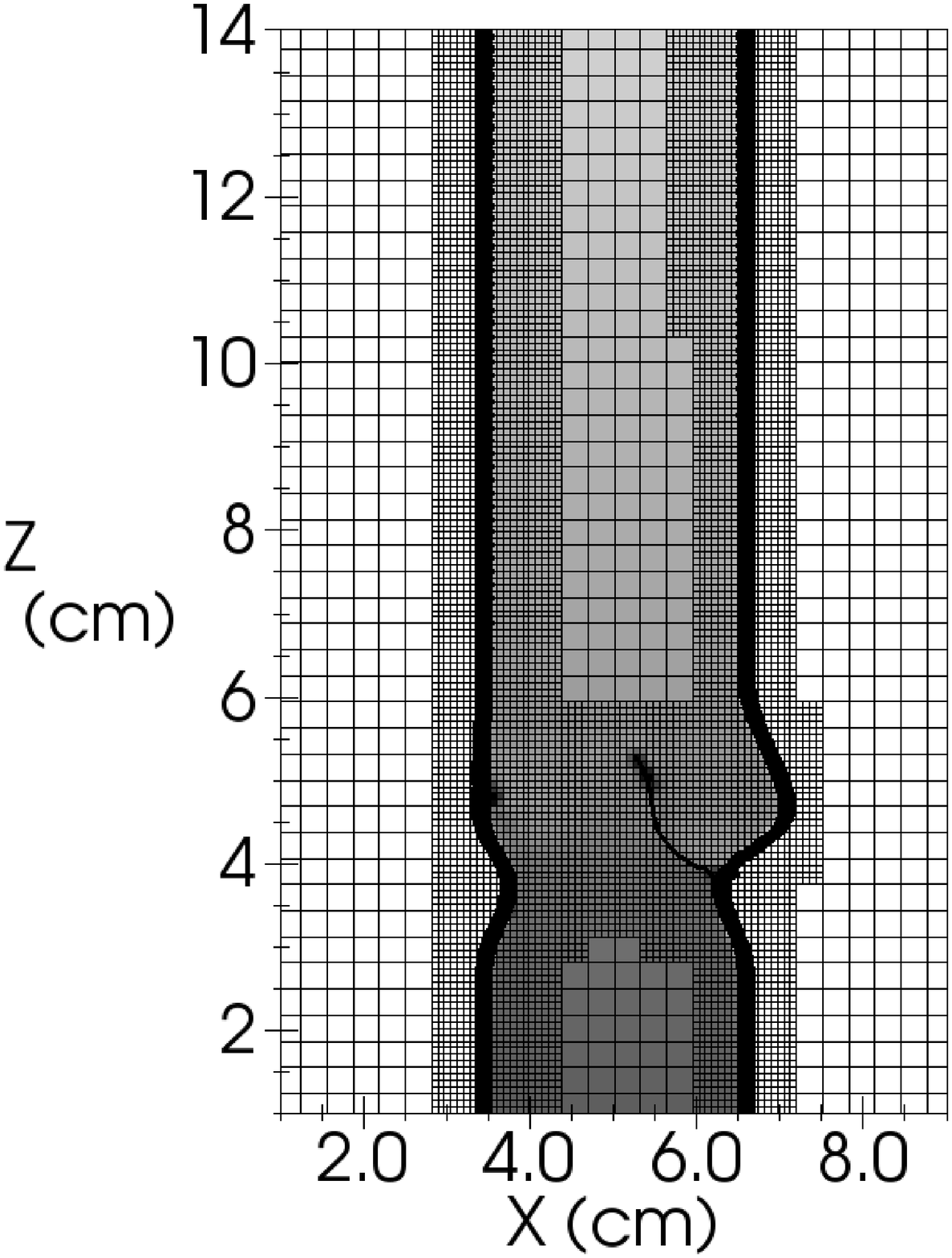} \\
    \vspace{-10pt} {\bf B.} & & & \\
    & \includegraphics[height=0.2875\textheight]{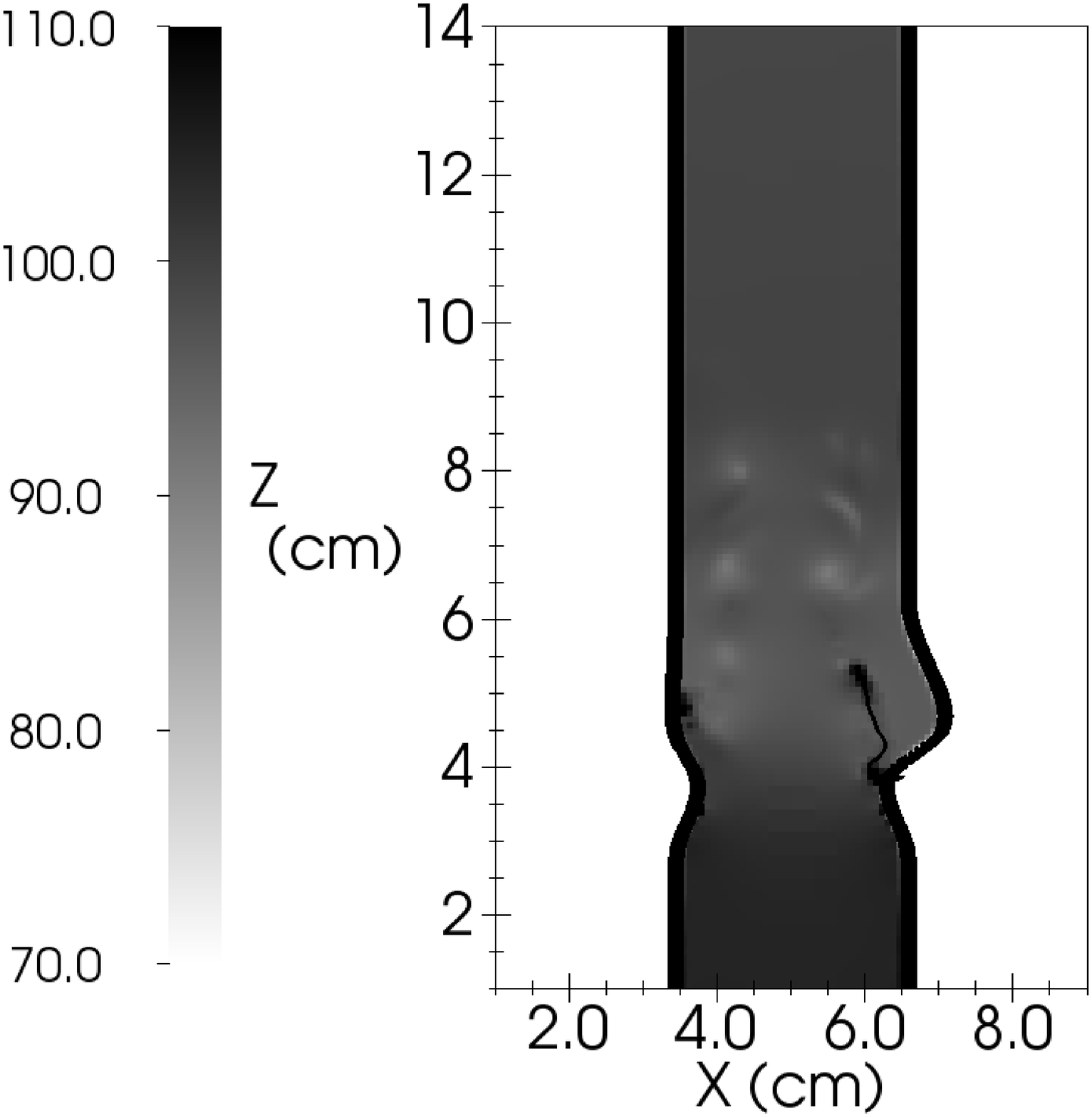} & \hspace{24pt}
    & \includegraphics[height=0.2875\textheight]{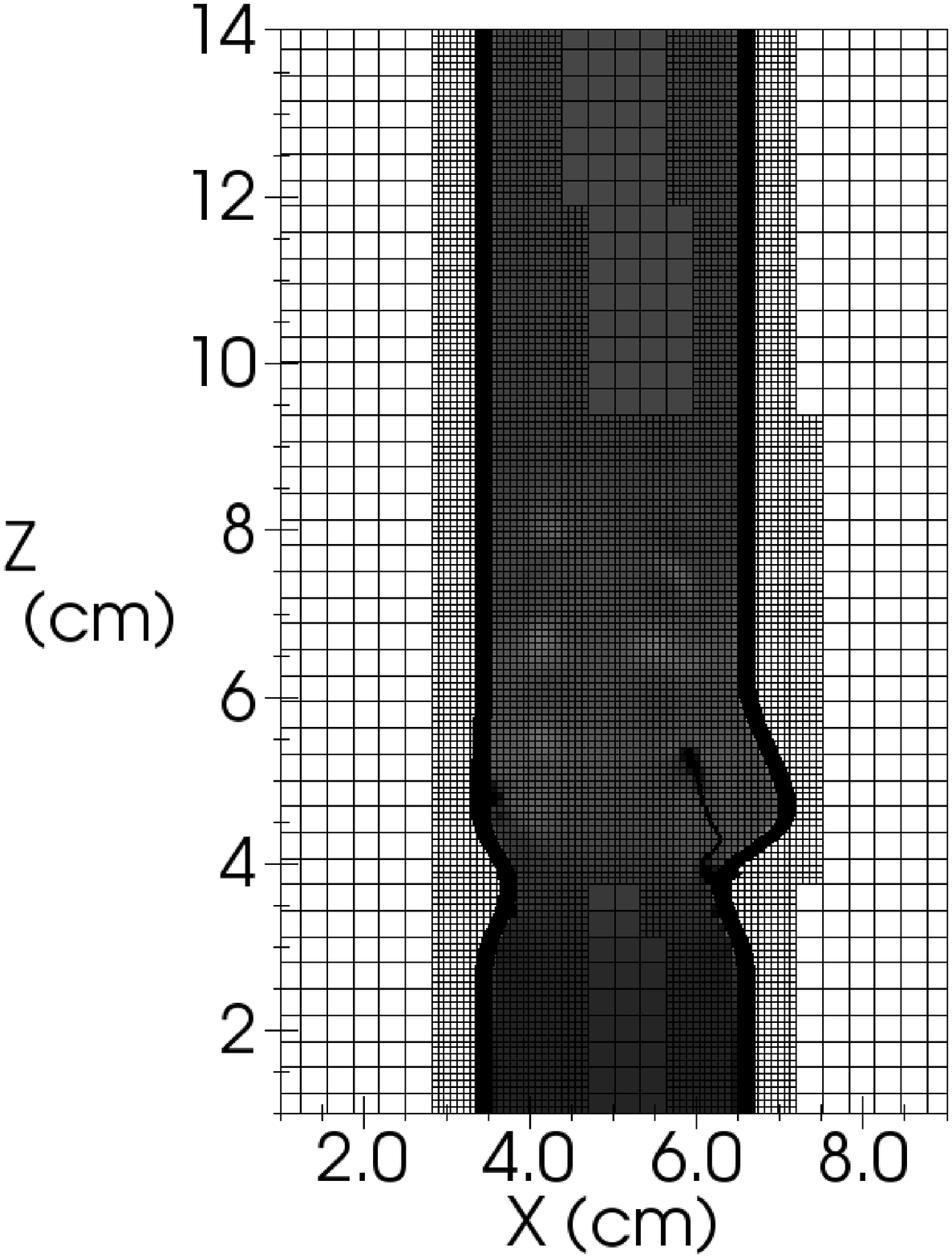}
  \end{tabular}
  \caption{The configuration of the model valve and vessel, the
    pressure field, and, in the right panels, the locally refined
    Cartesian grid are shown along a plane that bisects one of the
    valve leaflets at ({\bf A}) the time of maximum systolic pressure
    difference across the model valve and ({\bf B}) the time of peak
    flow rate through the valve during the second simulated cardiac
    cycle.  The maximum systolic pressure difference across the model
    valve, which is computed as the difference between the left
    ventricular pressure and the pressure approximately $\mbox{1.5
      cm}$ downstream of the valve, is $\mbox{12.2 mmHg}$ during the
    second beat, and the pressure difference across the valve at the
    time of peak flow is $\mbox{7.9 mmHg}$.}
  \label{f:pressure_slice}
\end{figure}

\begin{figure}
  \centering
  \includegraphics[width=0.225\textwidth]{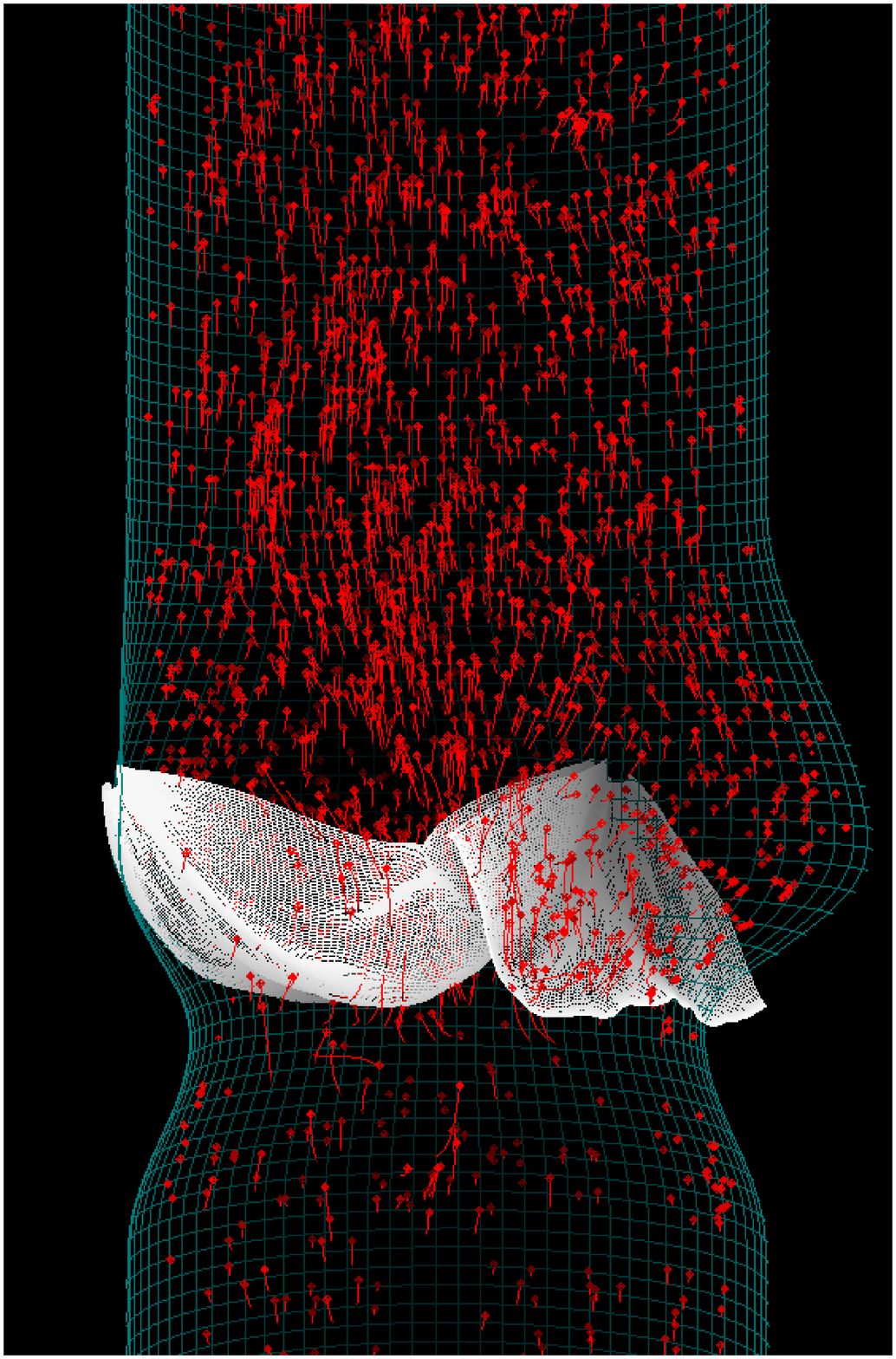}
  \hspace{8pt}
  \includegraphics[width=0.225\textwidth]{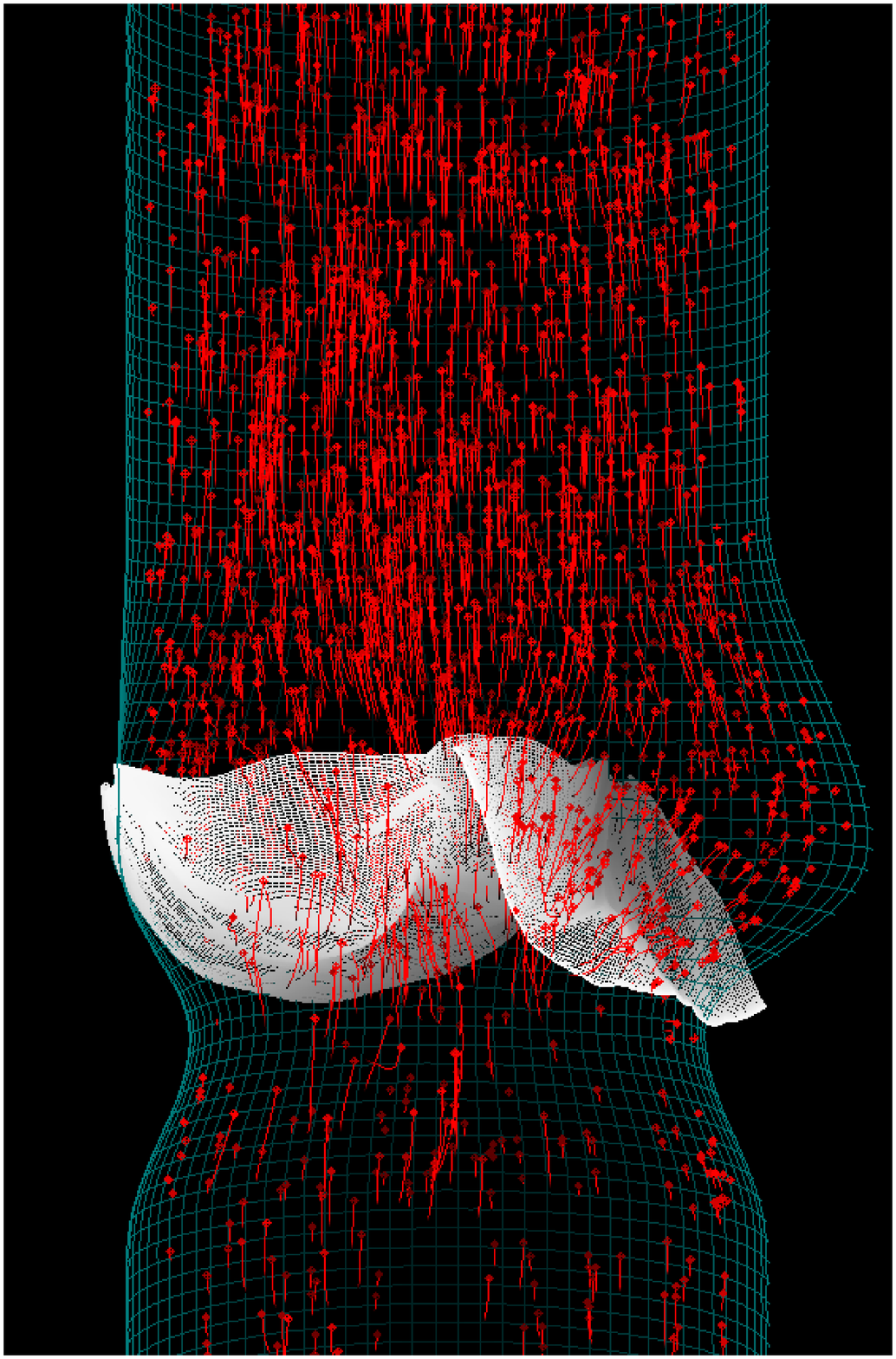}
  \hspace{8pt}
  \includegraphics[width=0.225\textwidth]{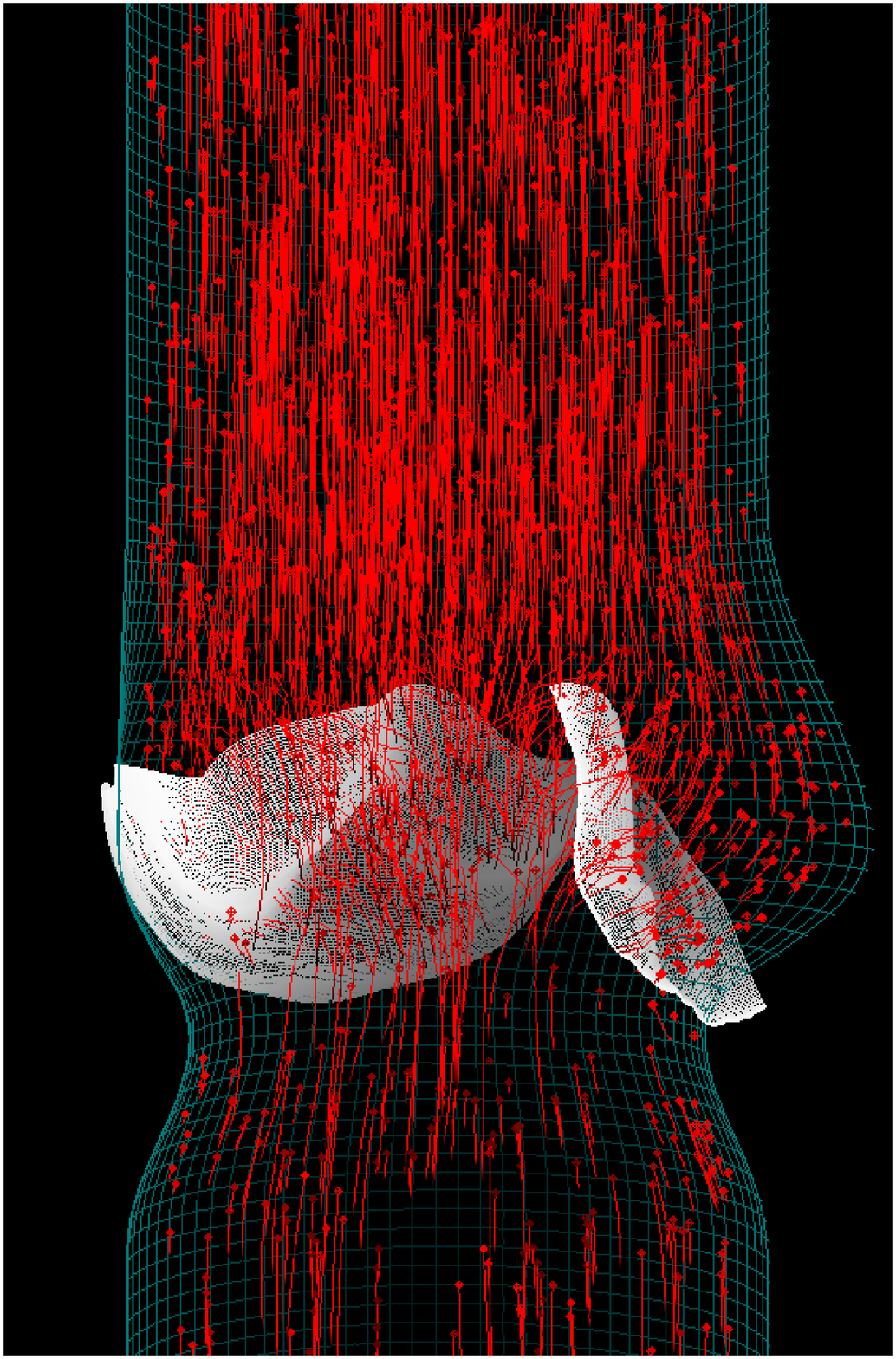}
  \hspace{8pt}
  \includegraphics[width=0.225\textwidth]{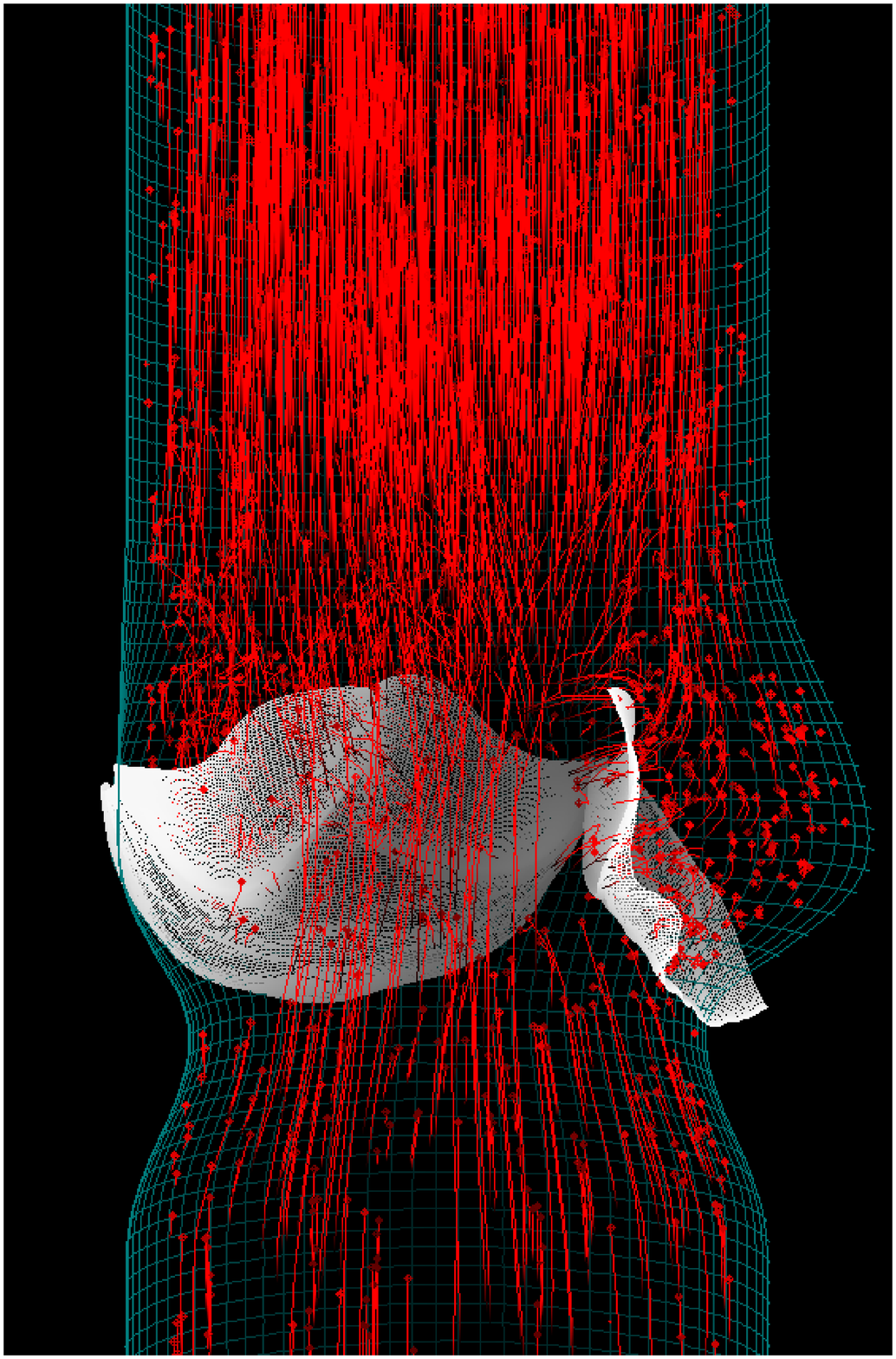}
  \caption{Flow through the model valve during valve opening as
    indicated by passively advected markers particles, shown at
    equally spaced time intervals.}
  \label{f:valve_flow_patterns}
\end{figure}

Figs.~\ref{f:bulk_flow_properties}--\ref{f:valve_flow_patterns}
display representative results from a multibeat simulation using this
model with left-ventricular driving pressures adapted from the human
clinical data of Murgo et al.~\cite{JPMurgo80}, and with loading
conditions provided by the three-element Windkessel model of
Stergiopulos et al.~\cite{Stergiopulos99}.  Recall that pressure
boundary conditions are imposed at both the upstream and downstream
boundaries of the model vessel.  This is necessary to obtain a
simulation in which the model valve supports a realistic pressure load
during diastole.  We remark that the realistic, nearly periodic flow
rates produced by the model are not specified; instead, they emerge
from the fluid-structure interaction simulation.

During the second and third beats of the simulation, mean stroke
volume is approximately $\mbox{65 ml}$, which is within the
physiological range \cite{GuytonHall00}, whereas the peak flow rate is
approximately $\mbox{420 ml s$^{-1}$}$, which is somewhat lower than
the peak flow rate reported by Murgo et al.~\cite{JPMurgo80}.  Because
the pressure field appears to be smoothly resolved in our simulation,
we speculate that this difference is primarily related to the
unphysiological rigidity of our vessel model and is not a consequence
of numerical underresolution; however, this has not yet been
demonstrated.  The maximum systolic pressure difference across the
model valve, which is computed as the difference between the left
ventricular pressure and the pressure approximately $\mbox{1.5 cm}$
downstream of the valve, is $\mbox{12.2 mmHg}$ during the second beat
and $\mbox{12.0 mmHg}$ during the third beat, values that are in good
agreement with the corresponding experimentally obtained value of
$\mbox{12.8 mmHg}$ reported by Driscol and Eckstein
\cite{TEDriscol65}.  The pressure difference across the valve at the
time of peak flow is $\mbox{7.9 mmHg}$ and $\mbox{7.2 mmHg}$ during
the second and third beats, respectively.  The mean transvalvular
pressure difference of the model is $\mbox{5.5 mmHg}$ and $\mbox{5.1
  mmHg}$ during the second and third beats, values that are somewhat
higher than the experimental range of $\mbox{1.2--2.9 mmHg}$
\cite{TEDriscol65}.  Notice that the familiar S$_{\text{2}}$ (``dup'')
heart sounds, which correspond to the reverberations of the aortic
valve leaflets upon the closure of the valve, are clearly visible in
both the computed flow rate (fig.~\ref{f:bulk_flow_properties} panel
A) and also the aortic loading pressure $\PAo(t)$
(fig.~\ref{f:bulk_flow_properties} panels B and C).

\subsection{Performance Analysis}

\begin{table}
  \centering
  \begin{tabular}{r|rrr}
    & \multicolumn{3}{c}{number of cores} \\
    & \multicolumn{1}{c}{24} & \multicolumn{1}{c}{48} & \multicolumn{1}{c}{96} \\ \hline
    uniform grid & 13.02 & 7.62 & 7.24 \\
    2-level grid &  9.09 & 6.21 & 4.54 \\
    3-level grid &  9.19 & 6.24 & 4.54 \\
  \end{tabular}
  \caption{Average wallclock time (in seconds) per timestep during the
    first 128 timesteps of the simulation for different numerical and
    computational parameters.}
  \label{t:wallclock_times}
\end{table}

To gauge the computational performance of our adaptive method, we
record the wallclock time required to perform the first 128 timesteps
of the simulation.  To obtain representative results, we average
timings obtained for five successive runs of these first 128
timesteps.  We compare the performance of the method when using a
uniform Cartesian grid, a two-level Cartesian grid with refinement
ratio $\nref=4$ between levels, and a three-level Cartesian grid with
refinement ratio $\nref=4$ between levels.  In all cases, the
effective resolution of the finest level of the Cartesian grid
hierarchy corresponds to that of a uniform $128 \times 128 \times 192$
grid.  Timings were performed on the \texttt{cardiac} cluster at New
York University, which is comprised of 80 Sun Microsystems, Inc., Sun
Blade X8440 server modules interconnected by an InfiniBand network.
Each compute server is equipped with four 2.3 GHz quad-core AMD
Barcelona 8356 processors with 2 GB memory per core.  We perform
timings on two, four, or eight compute nodes, using four processors
per node and three cores per processor for a total of 24, 48, or 96
cores per simulation.  Timing results are summarized in
table~\ref{t:wallclock_times}.

Notice that for the present model at the present effective fine-grid
spatial resolution, the efficiency of the adaptive scheme is not
improved by using a Cartesian grid hierarchy consisting of more than
two levels.  Specifically, when we use a three-level grid, most of the
coarsest level is tagged for refinement and covered by the grid
patches that comprise the next finer level of the grid hierarchy.
Consequently, the total number of degrees of freedom is not reduced by
an amount that is sufficient to overcome the increased computational
overhead associated with the increased number of grid levels.  In
general, the optimal number of levels of refinement is problem
dependent.  For instance, by increasing the size of the physical
domain $\Omega$ by a sufficient amount, it would eventually become
computationally beneficial to use a Cartesian grid hierarchy comprised
of three or more levels.  Similarly, higher-resolution versions of the
present aortic valve model would almost certainly benefit from
additional levels of refinement.  Because our semi-implicit
timestepping scheme requires that $\dt =
O\left(\left(\hellmax\right)^4\right)$ for problems involving
bending-resistant elastic elements, however, it appears likely that a
prerequisite for such a higher-resolution adaptive model is the
development of an efficient implementation of an implicit version of
this adaptive IB method.

\section{Conclusions}
\label{s:conclusions}

In this work, we have described an adaptive, staggered-grid version of
the IB method, and we have presented multibeat simulation results
obtained by applying this method to a fluid-structure interaction
model of the aortic heart valve that includes realistic driving and
loading conditions.  In our simulations, physiological flows are
obtained with physiological pressure differences over multiple cardiac
cycles.  Only pressure boundary conditions are prescribed at the
upstream and downstream boundaries of the model vessel.  Nonetheless,
realistic, time-periodic flow rates emerge from the coupled
fluid-structure interaction model.

The primary differences between the simulations described herein and
those of our earlier simulation study of cardiac valve dynamics
\cite{GriffithLuoMcQueenPeskin09} are that, in the present
simulations, we use a physiological driving pressure waveform; we
consider multiple cardiac cycles; and we employ an adaptive
staggered-grid version of the IB method.  The staggered-grid version
of the IB method used in this work yields dramatic improvements in
accuracy compared to our earlier cell-centered scheme
\cite{BEGriffithXX-ib_volume_conservation}, but despite these
improvements in accuracy, work remains to obtain fully resolved
simulations of the three-dimensional fluid dynamics of the aortic
valve.  Because of the severe restriction on the timestep size imposed
by our semi-implicit timestepping scheme, however, it is difficult to
deploy higher spatial resolution, even with the benefit of adaptive
mesh refinement.  We expect that obtaining higher-resolution IB
simulations of aortic valve dynamics will require the development of
an efficient parallel implementation of an implicit version of the IB
method.  Such methods promise to overcome the severe stability
restrictions of semi-implicit IB methods, like that used in the
present work, and we and others are actively working to develop
efficient implicit IB methods.

Higher spatial resolution is not the only factor limiting the realism
of our model.  Another limitation of the present model is its simple
description of the elasticity of the aortic valve and root.  The
present model could be improved by replacing these simple models with
experimentally based constitutive models.  Fiber-based elasticity
models like those used in this work provide a convenient description
of anisotropic structures commonly encountered in biological
applications, and are well-suited for modeling the thin aortic valve
leaflets, but combining realistic constitutive models with the
fiber-based approach traditionally used with the IB method is
difficult.  The IB method is not restricted to fiber models, however,
and several recent extensions of the IB method allow for more general
elasticity models that permit finite element discretizations
\cite{LZhang04,WKLiu06,LTZhang07,DBoffi08,BEGriffithXX-ib_fem}.  Using
one such extension of the IB method \cite{BEGriffithXX-ib_fem}, we aim
to develop realistic IB models of aortic valve mechanics that will use
experimentally characterized models of the aortic valve leaflets
\cite{HKim06,HKim07,KMay-Newman09}, the sinuses \cite{Gundiah08}, and
the ascending aorta \cite{Gundiah08b}, and to use such models to study
the fluid dynamics of the aortic valve in both health and disease.

\appendix

\section{Composite-Grid Discretization}

Here, we provide additional details of the composite-grid finite
difference discretizations used in the adaptive staggered-grid IB
method.  We use $(I,J,K)$ to index grid cells of a coarse level $\ell$
of the AMR grid hierarchy, and we use $(i,j,k)$ to index grid cells of
the next finer level of the grid, level $\ell+1$.  Throughout this
appendix, $(i,j,k) = (\nref I, \nref J, \nref K)$, i.e., fine grid
cell $(i,j,k)$ and coarse grid cell $(I,J,K)$ share the vertex
$\x_{i-\half,j-\half,k-\half} = \left(i \hellplus, j \hellplus, k
  \hellplus\right) = \left(I \hell, J \hell, K \hell\right) =
\x_{I-\half,J-\half,K-\half}$.  We use the notation
$u\!\left(I-\half,J,K\right)$, $v\!\left(I,J-\half,K\right)$,
$w\!\left(I,J,K-\half\right)$, and $p(I,J,K)$ to denote the values of
$\u = (u,v,w)$ and $p$ that are stored on the coarse level $\ell$ of
the grid.  Similar notation is used to denote values stored on the
fine level $\ell+1$ of the grid.

\subsection{Restriction}

The composite-grid finite difference discretizations used in this work
require a cell-centered cubic restriction procedure along with
conservative and cubic face-centered restriction operators.  To define
these restriction procedures, we consider a single coarse grid cell
$\mbox{$(I,J,K)$}$ on level $\ell$ along with the overlying $\nref
\times \nref \times \nref$ grid cells on level $\ell+1$, namely grid
cells $(i+\alpha,j+\beta,k+\gamma)$ for $\alpha,\beta,\gamma =
0,\ldots,\nref-1$.  Notice that
\begin{equation}
  \c_{I,J,K}^{\ell} = \bigcup_{\alpha,\beta,\gamma=0,\ldots,\nref-1} \c_{i+\alpha,j+\beta,k+\gamma}^{\ell+1}.
\end{equation}

\subsubsection{Cell-centered cubic restriction}

The cell-centered cubic restriction procedure, which requires $\nref$
to be even and at least four, defines a cell-centered quantity
$p(I,J,K)$ on coarse level $\ell$ in terms of the closest overlying $4
\times 4 \times 4$ fine-grid values stored on level $\ell+1$ via
\begin{equation}
  p(I,J,K) := \sum_{\alpha,\beta,\gamma=-2,\ldots,1} \omega(\alpha) \, \omega(\beta) \, \omega(\gamma) \, p\!\left(i+\frac{\nref}{2}+\alpha,j+\frac{\nref}{2}+\beta,k+\frac{\nref}{2}+\gamma\right),
\end{equation}
in which $\omega(-2) = \omega(1) = -\frac{1}{16}$ and $\omega(-1) =
\omega(0) = \frac{9}{16}$.  In our implementation, if $\nref=2$, we
revert to linear interpolation, which results in a reduction of the
formal order of accuracy of our composite-grid discretization at
coarse-fine interfaces.  We do not permit $\nref$ to be odd because
our recursive implementation of the divergence-~and curl-preserving
interpolation scheme \cite{TothRoe02} used to transfer the
staggered-grid velocity field from one Cartesian grid hierarchy to
another during adaptive regridding requires $\nref$ to be a power of
two.

\subsubsection{Face-centered conservative restriction}

The face-centered conservative restriction procedure defines a
face-centered quantity $\left(u\!\left(I-\half,J,K\right)\!,
  v\!\left(I,J-\half,K\right)\!, w\!\left(I,J,K-\half\right)\right)$
on coarse level $\ell$ in terms of values stored in the overlying
$\nref \times \nref \times \nref$ fine-grid cells on level $\ell+1$
via
\begin{align}
  u\!\left(I-\half,J,K\right) &:= \frac{1}{\nref^2} \sum_{\beta,\gamma=0,\ldots,\nref-1} u\!\left(i-\half,j+\beta,k+\gamma\right),\\
  v\!\left(I,J-\half,K\right) &:= \frac{1}{\nref^2} \sum_{\alpha,\gamma=0,\ldots,\nref-1} v\!\left(i+\alpha,j-\half,k+\gamma\right),\\
  w\!\left(I,J,K-\half\right) &:= \frac{1}{\nref^2} \sum_{\alpha,\beta=0,\ldots,\nref-1} w\!\left(i+\alpha,j+\beta,k-\half\right).
\end{align}
This procedure is conservative in the sense of a finite volume scheme,
i.e.,
\begin{align}
  u\!\left(I-\half,J,K\right) \left(\hell\right)^2 &= \sum_{\beta,\gamma=0,\ldots,\nref-1} u\!\left(i-\half,j+\beta,k+\gamma\right) \left(\hellplus\right)^2,\\
  v\!\left(I,J-\half,K\right) \left(\hell\right)^2 &= \sum_{\alpha,\gamma=0,\ldots,\nref-1} v\!\left(i+\alpha,j-\half,k+\gamma\right) \left(\hellplus\right)^2,\\
  w\!\left(I,J,K-\half\right) \left(\hell\right)^2 &= \sum_{\alpha,\beta=0,\ldots,\nref-1} w\!\left(i+\alpha,j+\beta,k-\half\right) \left(\hellplus\right)^2.
\end{align}

\subsubsection{Face-centered cubic restriction}

The face-centered cubic restriction procedure, which requires $\nref$
to be even and at least four, defines a face-centered quantity
$u\!\left(I-\half,J,K\right)$ on coarse level $\ell$ in terms of the
closest overlying $4 \times 4$ fine-grid values stored on level
$\ell+1$ via
\begin{equation}
  u\!\left(I-\half,J,K\right) := \sum_{\beta,\gamma=-2,\ldots,1} \omega(\beta) \, \omega(\gamma) \, u\!\left(i-\half,j+\frac{\nref}{2}+\beta,k+\frac{\nref}{2}+\gamma\right),
\end{equation}
in which $\omega(-2) = \omega(1) = -\frac{1}{16}$ and $\omega(-1) =
\omega(0) = \frac{9}{16}$.  Similar formulae define
$v\!\left(I,J-\half,K\right)$ and $w\!\left(I,J,K-\half\right)$.  In
our implementation, if $\nref=2$, we revert to linear interpolation,
which results in a reduction of the formal order of accuracy of our
composite-grid discretization at coarse-fine interfaces.  We do not
permit $\nref$ to be odd because our recursive implementation of the
divergence-~and curl-preserving interpolation scheme \cite{TothRoe02}
used to transfer the staggered-grid velocity field from one Cartesian
grid hierarchy to another during adaptive regridding requires $\nref$
to be a power of two.

\subsection{Interpolation at coarse-fine interfaces}

\begin{figure}
  \centering
  \input{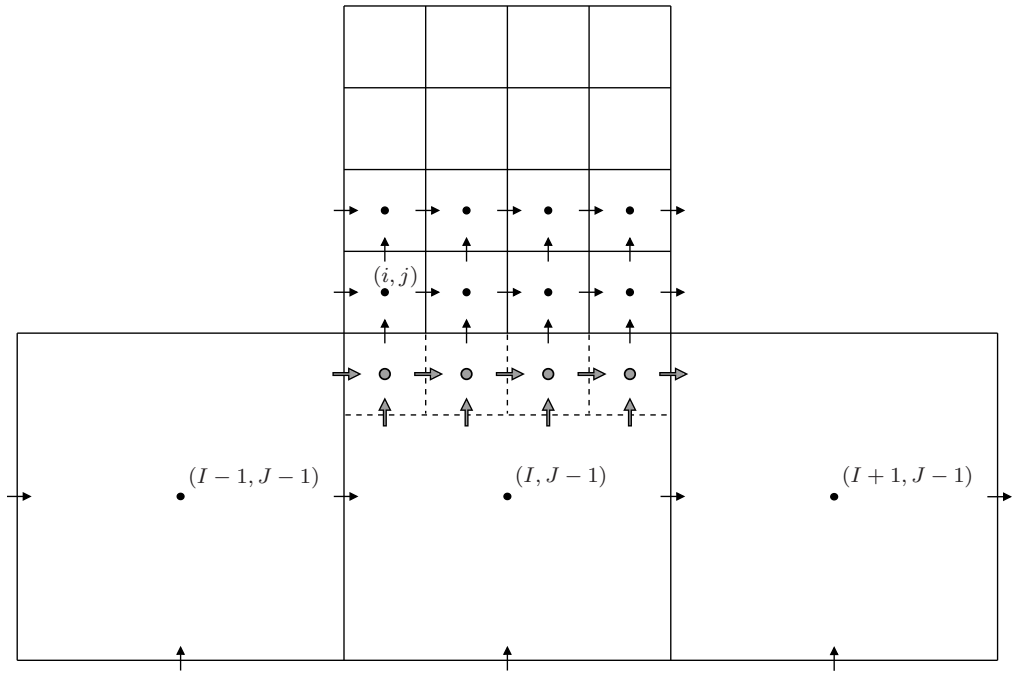}
  \caption{Two-dimensional grid configuration used to describe the
    coarse-fine interface interpolation scheme for cell-~and
    face-centered quantities.  In this configuration, a coarse level
    $\ell$ grid cell $(I,J-1)$ abuts $\nref=4$ fine level $\ell+1$
    grid cells $(i+\alpha,j)$ for $\alpha=0,\ldots,\nref-1$.  The
    coarse-fine interface ghost cells are indicated by dashed lines,
    and the values stored in these ghost cells are shown in gray with
    a black border.  The cell-~and face-centered quantities indicated
    in black are used to compute the ghost-cell values.}
  \label{f:cf_discretization}
\end{figure}

To describe the specialized interpolation scheme used to define values
in the ghost cells abutting a coarse-fine interface, we temporarily
restrict our discussion to two spatial dimensions.  The extension of
this interpolation scheme to three spatial dimensions is
straightforward but difficult to visualize and cumbersome to describe,
and therefore is not presented in detail.

We specifically consider the two-dimensional configuration shown in
figs.~\ref{f:cf_discretization}--\ref{f:cf_discretization_fc_t}, in
which a coarse level $\ell$ grid cell $(I,J-1)$ abuts $\nref$ fine
level $\ell+1$ grid cells $(i+\alpha,j)$ for
$\alpha=0,\ldots,\nref-1$.  (In the figures, as in our computations,
we set $\nref=4$; however, the formulae presented in this subsection
are valid for any even value of $\nref$.  Different formulae would be
required to treat the case in which $\nref$ is odd.)  Formulae similar
to those presented hold for other coarse-fine interface orientations.
To define the values in the ghost cells located along this coarse-fine
interface, we interpolate values stored in the coarse grid cells
$(I+\alpha,J-1)$ for $\alpha = -1,0,1$ along with values stored in the
two layers of fine grid cells that are adjacent to the coarse-fine
interface, i.e., grid cells $(i+\alpha,j)$ and $(i+\alpha,j+1)$ for
$\alpha=0,\ldots,\nref-1$.  The values stored in coarse grid cells
$(I-1,J-1)$ and $(I+1,J-1)$ may be either valid or invalid values,
i.e., grid cells $(I-1,J-1)$ or $(I+1,J-1)$ could be located within
the refined region of level $\ell$.  If the values in coarse grid
cells $(I-1,J-1)$ or $(I+1,J-1)$ are invalid values, those values are
defined to be the cubic restriction of the overlying fine-grid values.

Our approach extends the cell-centered approach of Minion
\cite{Minion96b}, Martin and Colella \cite{MartinColella00}, and
Martin et al.~\cite{MartinColellaGraves08} to treat both cell-centered
and face-centered quantities.  We first interpolate coarse-grid values
in the direction tangential to the coarse-fine interface, so as to
obtain interpolated values at locations that are aligned with the
valid fine-grid values.  We then define the values in the ghost cells
by interpolating in the direction normal to the coarse-fine interface,
using the interpolated coarse-grid values along with the valid
fine-grid values.

\subsubsection{Cell-centered coarse-fine interpolation}
\label{s:cc_cf_interpolation}

\begin{figure}
  \centering
  \input{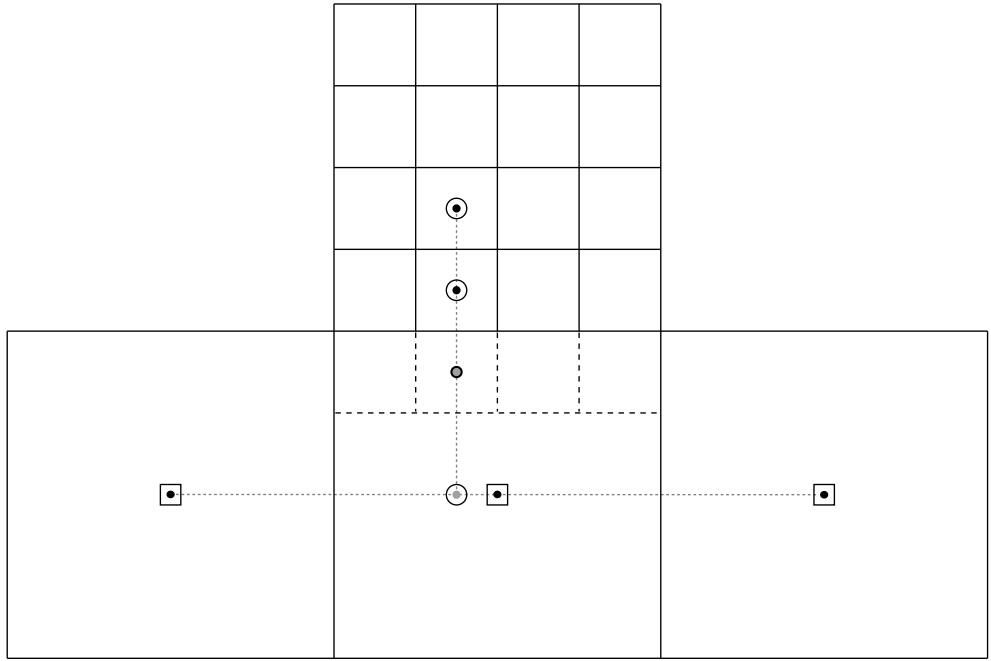}
  \caption{Cell-centered coarse-fine interface interpolation.  First,
    an intermediate interpolated value, indicated in gray, is defined
    via quadratic interpolation in the direction tangential to the
    coarse-fine interface.  The coarse-grid values that are used in
    this initial interpolation step are indicated by black boxes.
    Notice that the intermediate quantity is aligned with the
    fine-grid values, whereas the original coarse-grid values are not.
    Second, the coarse-fine interface ghost-cell value is defined via
    quadratic interpolation in the direction normal to the coarse-fine
    interface.  The values that are used in this second interpolation
    step are indicated by black circles.}
  \label{f:cf_discretization_cc}
\end{figure}

In reference to fig.~\ref{f:cf_discretization}, the cell-centered
quantities that we wish to compute are denoted $p(i+\alpha,j-1)$,
$\alpha = 0,\ldots,\nref-1$.  To define these values, we first compute
intermediate values that are defined by performing quadratic
interpolation in the direction tangential to the coarse-fine
interface, using the coarse-grid values $p(I-1,J-1)$, $p(I,J-1)$, and
$p(I+1,J-1)$.  Specifically, we compute
\begin{align}
  p\!\left(i+\alpha,j-\frac{\nref+1}{2}\right)
  &:= \frac{\left(2 (i+\alpha) + 1 - \nref\right) \left(2 (i+\alpha) + 1 - 3 \nref\right)}{8 \nref^2} p(I-1,J-1) \nonumber \\
  &\quad \mbox{} - \frac{\left(2 (i+\alpha) + 1 + \nref\right) \left(2 (i+\alpha) + 1 - 3 \nref\right)}{4 \nref^2} p(I,J-1) \nonumber \\
  &\quad \mbox{} + \frac{\left(2 (i+\alpha) + 1 + \nref\right) \left(2 (i+\alpha) + 1 - \nref\right)}{8 \nref^2} p(I+1,J-1)
\end{align}
for $\alpha = 0,\ldots,\nref-1$.

We then compute $p(i+\alpha,j-1)$, $\alpha = 0,\ldots,\nref-1$, by
performing quadratic interpolation in the direction normal to the
coarse-fine interface, using the fine-grid values $p(i+\alpha,j)$ and
\mbox{$p(i+\alpha,j+1)$} along with the intermediate interpolated
values.  Specifically, we compute
\begin{align}
  p(i+\alpha,j-1)
  &:= \frac{2 (\nref - 1)}{1 + \nref} p(i+\alpha,j) - \frac{\nref - 1}{3 + \nref} p(i+\alpha,j+1) \nonumber \\
  & \quad \mbox{} + \frac{8}{(1 + \nref) (3 + \nref)} p\!\left(i+\alpha,j-\frac{\nref+1}{2}\right)
\end{align}
for $\alpha = 0,\ldots,\nref-1$.

This two-step interpolation procedure is summarized in
fig.~\ref{f:cf_discretization_cc} for $\nref=4$ and $\alpha=1$.

\subsubsection{Face-centered coarse-fine interpolation of components
  normal to the coarse-fine interface}
\label{s:fc_n_cf_interpolation}

\begin{figure}
  \centering
  \input{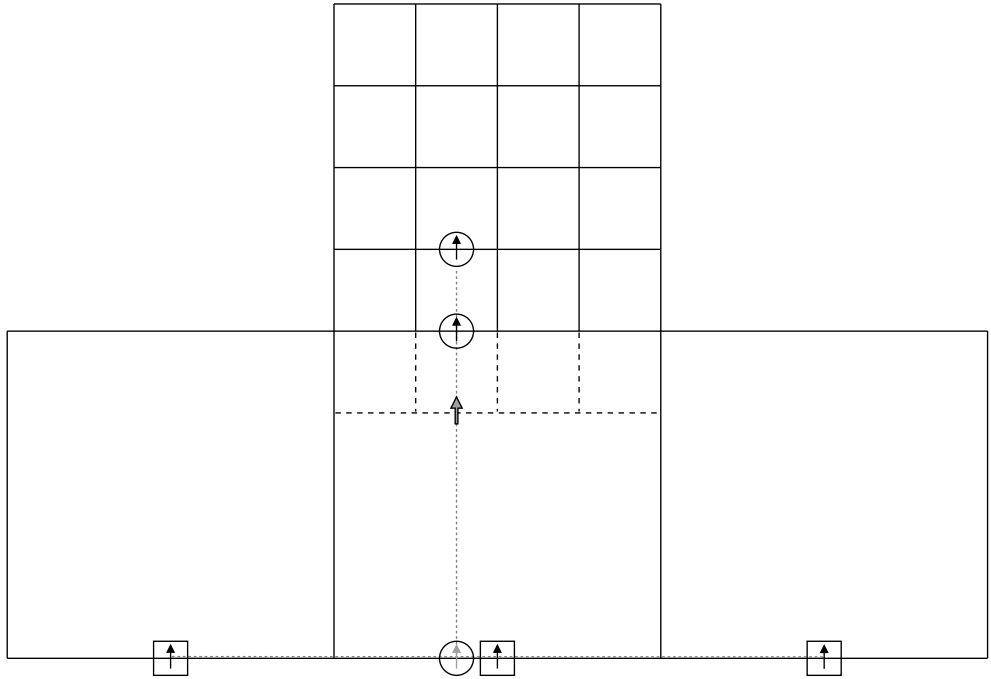}
  \caption{Face-centered coarse-fine interface interpolation of
    components normal to the coarse-fine interface.  First, an
    intermediate interpolated value, indicated in gray, is defined via
    quadratic interpolation in the direction tangential to the
    coarse-fine interface.  The coarse-grid values that are used in
    this initial interpolation step are indicated by black boxes.
    Second, the coarse-fine interface ghost-cell value is defined via
    quadratic interpolation in the direction normal to the coarse-fine
    interface.  The values that are used in this second interpolation
    step are indicated by black circles.}
  \label{f:cf_discretization_fc_n}
\end{figure}

The components of the staggered-grid velocity field that are normal to
the coarse-fine interface are treated in a manner that is similar to
the cell-centered coarse-fine interface interpolation scheme described
in sec.~\ref{s:cc_cf_interpolation}.  In reference to
fig.~\ref{f:cf_discretization}, the face-centered quantities that we
wish to compute are denoted $v\!\left(i+\alpha,j-\frac{3}{2}\right)$,
$\alpha = 0,\ldots,\nref-1$.  To define these values, we first compute
intermediate values that are defined by performing quadratic
interpolation in the direction tangential to the coarse-fine
interface, using the coarse-grid values
$v\!\left(I-1,J-\frac{3}{2}\right)$,
$v\!\left(I,J-\frac{3}{2}\right)$, and
$v\!\left(I+1,J-\frac{3}{2}\right)$.  Specifically, we compute
\begin{align}
  v\!\left(i+\alpha,j-\nref-\half\right)
  &:= \frac{\left(2 (i+\alpha) + 1 - \nref\right) \left(2 (i+\alpha) + 1 - 3 \nref\right)}{8 \nref^2} v\!\left(I-1,J-\frac{3}{2}\right) \nonumber \\
  &\quad \mbox{} - \frac{\left(2 (i+\alpha) + 1 + \nref\right) \left(2 (i+\alpha) + 1 - 3 \nref\right)}{4 \nref^2} v\!\left(I,J-\frac{3}{2}\right) \nonumber \\
  &\quad \mbox{} + \frac{\left(2 (i+\alpha) + 1 + \nref\right) \left(2 (i+\alpha) + 1 - \nref\right)}{8 \nref^2} v\!\left(I+1,J-\frac{3}{2}\right)
\end{align}
for $\alpha = 0,\ldots,\nref-1$.

We then compute $v\!\left(i+\alpha,j-\frac{3}{2}\right)$, $\alpha =
0,\ldots,\nref-1$, by performing quadratic interpolation in the
direction normal to the coarse-fine interface, using the fine-grid
values $v\!\left(i+\alpha,j-\half\right)$ and
$v\!\left(i+\alpha,j+\half\right)$ along with the intermediate
interpolated values.  Specifically, we compute
\begin{align}
  v\!\left(i+\alpha,j-\frac{3}{2}\right)
  &:= \frac{2 (\nref - 1)}{\nref} v\!\left(i+\alpha,j-\half\right)
  - \frac{\nref - 1}{1 + \nref} v\!\left(i+\alpha,j+\half\right) \nonumber \\
  & \quad \mbox{} + \frac{2}{\nref (1 + \nref)} v\!\left(i+\alpha,j-\nref-\half\right)
\end{align}
for $\alpha = 0,\ldots,\nref-1$.

This two-step interpolation procedure is summarized in
fig.~\ref{f:cf_discretization_fc_n} for $\nref=4$ and $\alpha=1$.

\subsubsection{Face-centered coarse-fine interpolation of components
  tangential to the coarse-fine interface}
\label{s:fc_t_cf_interpolation}

\begin{figure}
  \centering
  \input{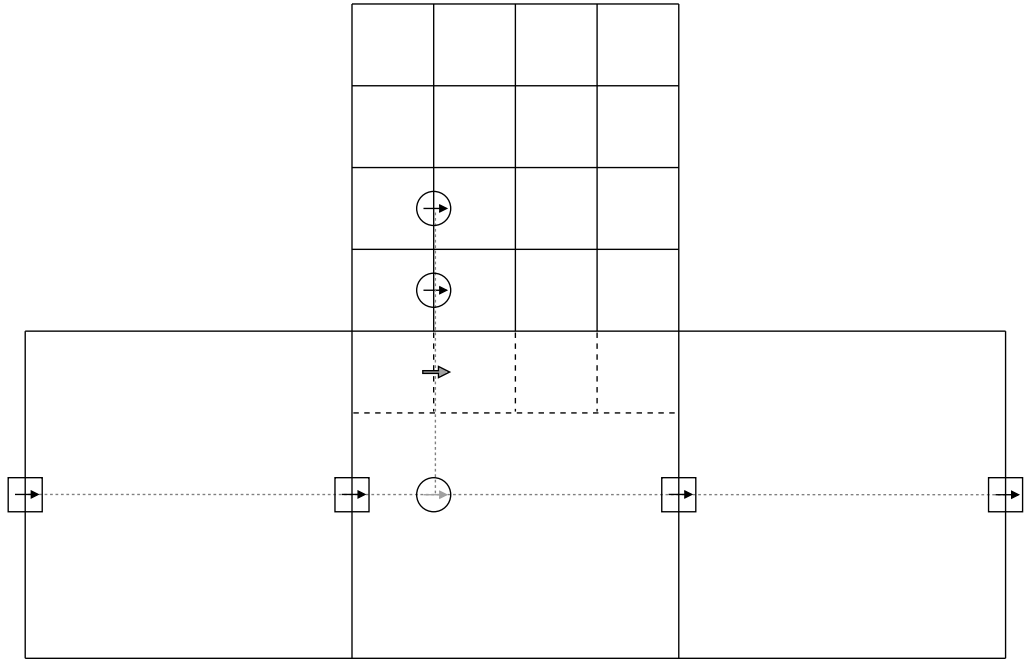}
  \caption{Face-centered coarse-fine interface interpolation of
    components tangential to the coarse-fine interface.  First, an
    intermediate interpolated value, indicated in gray, is defined via
    cubic interpolation in the direction tangential to the coarse-fine
    interface.  The coarse-grid values that are used in this initial
    interpolation step are indicated by black boxes.  Second, the
    coarse-fine interface ghost-cell value is defined via quadratic
    interpolation in the direction normal to the coarse-fine
    interface.  The values that are used in this second interpolation
    step are indicated by black circles.}
  \label{f:cf_discretization_fc_t}
\end{figure}

The components of the staggered-grid velocity field that are
tangential to the coarse-fine interface are treated in a manner that
is similar to the interpolation schemes described in
secs.~\ref{s:cc_cf_interpolation} and \ref{s:fc_n_cf_interpolation},
except that in this case, we perform cubic interpolation in the
tangential direction instead of quadratic interpolation.  In reference
to fig.~\ref{f:cf_discretization}, the face-centered quantities that
we wish to compute are denoted $u\!\left(i+\alpha-\half,j-1\right)$,
$\alpha = 0,\ldots,\nref$.  To define these values, we first compute
intermediate values that are defined by performing cubic interpolation
in the direction tangential to the coarse-fine interface, using the
coarse-grid values $u\!\left(I-\frac{3}{2},J-1\right)$,
$u\!\left(I-\frac{1}{2},J-1\right)$,
$u\!\left(I+\frac{1}{2},J-1\right)$, and
$u\!\left(I+\frac{3}{2},J-1\right)$.  Specifically, we compute
\begin{align}
  u\!\left(i+\alpha-\half,j-\frac{\nref+1}{2}\right)
  &:= - \frac{(i+\alpha) \left(i+\alpha - \nref\right) \left(i+\alpha - 2 \nref\right)}{6 \nref^3} u\!\left(I-\frac{3}{2},J-1\right) \nonumber \\
  & \quad \mbox{} + \frac{\left(i+\alpha - \nref\right) \left(i+\alpha - 2 \nref\right) \left(i+\alpha + \nref\right)}{2 \nref^3} u\!\left(I-\frac{1}{2},J-1\right) \nonumber \\
  & \quad \mbox{} - \frac{(i+\alpha) \left(i+\alpha + \nref\right) \left(i+\alpha - 2 \nref\right)}{2 \nref^3} u\!\left(I+\frac{1}{2},J-1\right) \nonumber \\
  & \quad \mbox{} + \frac{(i+\alpha) \left(i+\alpha - \nref\right) \left(i+\alpha + \nref\right)}{6 \nref^3} u\!\left(I+\frac{3}{2},J-1\right)
\end{align}
for $\alpha = 0,\ldots,\nref$.

We then compute $u\!\left(i+\alpha-\half,j-1\right)$, $\alpha =
0,\ldots,\nref$, by performing quadratic interpolation in the
direction normal to the coarse-fine interface, using the fine-grid
values $u\!\left(i+\alpha-\half,j\right)$ and
$u\!\left(i+\alpha-\half,j+1\right)$ along with the intermediate
interpolated values.  Specifically, we compute
\begin{align}
  u\!\left(i+\alpha-\half,j-1\right)
  &:= \frac{2 (\nref - 1)}{1 + \nref} u\!\left(i+\alpha-\half,j\right) - \frac{\nref - 1}{3 + \nref} u\!\left(i+\alpha-\half,j+1\right) \nonumber \\
  & \quad \mbox{} + \frac{8}{(1 + \nref) (3 + \nref)} u\!\left(i+\alpha-\half,j-\frac{\nref+1}{2}\right)
\end{align}
for $\alpha = 0,\ldots,\nref$.

This two-step interpolation procedure is summarized in
fig.~\ref{f:cf_discretization_fc_t} for $\nref=4$ and $\alpha=1$.

\subsubsection{Extension to three spatial dimensions}

The essential difference between the two-dimensional coarse-fine
interface interpolation scheme and its extension to three spatial
dimensions is that, in the three-dimensional case, the initial
interpolation of coarse-grid values in the direction tangential to the
coarse-fine interface must employ a two-dimensional interpolation
scheme to align the intermediate values with the fine-grid values.  We
use tensor-product interpolation rules that combine the tangential
interpolation schemes described in
secs.~\ref{s:cc_cf_interpolation}--\ref{s:fc_t_cf_interpolation} with
a quadratic interpolation rule along the additional tangential
direction.  In the direction normal to the coarse-fine interface, the
same interpolation procedure is used in both two and three spatial
dimensions to compute the final ghost values from the fine-grid values
and the intermediate interpolated quantities.  Implementations of both
the two-~and three-dimensional versions of this scheme are available
online \cite{IBAMR-web-page}.

\subsection{Composite-grid approximations to the divergence, gradient,
  and Laplace operators}

Finally, we summarize the manner in which we compute finite difference
approximations to $\grad \cdot \u$, $\grad p$, and $\grad^2 \u$ on the
AMR grid hierarchy.  To compute $\Div \u$, we \begin{inparaenum}[(1)]
\item use the conservative face-centered restriction procedure to
  coarsen $\u$ from finer levels of the grid to coarser levels
  of the grid; and
\item use eq.~\eqref{e:div_h} to compute a discrete approximation to
  $\grad \cdot \u$ in each cell of the grid hierarchy.
\end{inparaenum}
To compute $\Grad p$, we \begin{inparaenum}[(1)]
\item use the cubic cell-centered restriction procedure to coarsen $p$
  from finer levels of the grid to coarser levels of the grid;
\item use the cell-centered coarse-fine interface interpolation
  procedure to compute values of $p$ stored in the coarse-fine
  interface ghost cells; and
\item use eqs.~\eqref{e:grad_h_1}--\eqref{e:grad_h_3} to compute a
  discrete approximation to the face-normal components of $\grad p$ on
  each cell face of the grid hierarchy.
\end{inparaenum}
To compute $\Lap \u$, we \begin{inparaenum}[(1)]
\item use the cubic face-centered restriction procedure to coarsen
  $\u$ from finer levels of the grid to coarser levels of the grid;
\item use the face-centered coarse-fine interface interpolation
  procedure to compute values of $\u$ stored in the coarse-fine
  interface ghost cells; and
\item use eq.~\eqref{e:Lap_h} and its analogues to compute a discrete
  approximation to the face-normal components of $\grad^2 \u$ on each
  cell face of the grid hierarchy.
\end{inparaenum}
Additional ghost-cell values are determined, where needed, either by
copying values from neighboring grid patches, or by employing a
discrete approximation to the physical boundary conditions.

\acks The author gratefully acknowledges discussion of this work with
Charles Peskin and David McQueen of the Courant Institute of
Mathematical Sciences, New York University.  The author also
gratefully acknowledges research support from the American Heart
Association (Scientist Development Grant 10SDG4320049) and the
National Science Foundation (DMS Award 1016554 and OCI Award 1047734).
Computations were performed at New York University using computer
facilities funded in large part by a generous donation by St.~Jude
Medical, Inc.

\bibliography{/Users/griffith/documents/me/bibtex/cardiac,/Users/griffith/documents/me/bibtex/num_analysis,/Users/griffith/documents/me/bibtex/ib}

\end{document}